\documentclass[journal]{IEEEtran}
\pdfoutput=1
\usepackage[noadjust]{cite}
\usepackage{amsmath}
\usepackage{array}
\usepackage{mdwmath}
\usepackage{amssymb}
\usepackage{mathtools}
\usepackage{algorithm}
\usepackage[noend]{algpseudocode}
\usepackage{eqparbox}
\usepackage{stfloats}
\usepackage{url}
\usepackage[pdftex]{graphicx}
\graphicspath{{../pdf/}{../jpeg/}}
\DeclareGraphicsExtensions{.pdf,.jpeg,.png}
\usepackage{epstopdf}
\usepackage{amsfonts}

\usepackage{color}

\usepackage{tikz}
\usepackage{pgfplots}
\usepackage{subcaption}
\usetikzlibrary{spy}

\newtheorem{lemma}{Lemma}
\newtheorem{theorem}{Theorem}
\newtheorem{remark}{Remark}
\newtheorem{corollary}{Corollary}

\begin{document}

\title{  Analysis of One-Bit Quantized Linear Precoding Schemes in Multi-Cell Massive MIMO Downlink}
\author{Qurrat-Ul-Ain~Nadeem,~\IEEEmembership{Member,~IEEE,} and Anas~Chaaban,~\IEEEmembership{Senior Member,~IEEE}
\thanks{Q.-U.-A. Nadeem and A. Chaaban are with School of Engineering, The University of British Columbia, Kelowna, Canada  (e-mail:  \{qurrat.nadeem, anas.chaaban\}@ubc.ca). Part of this work has been presented in \cite{BSC}.}
}

\maketitle

\begin{abstract}

This work studies a multi-cell one-bit massive multiple-input multiple-output (MIMO) system  that employs one-bit analog-to-digital converters (ADCs) and digital-to-analog converters (DACs) at each base station (BS). We utilize Bussgang decomposition to derive  downlink signal-to-quantization-plus-interference-plus-noise ratio (SQINR) and ergodic achievable  rate expressions  under one-bit quantized  maximum ratio transmission (MRT) and zero-forcing (ZF) precoding schemes considering scenarios with and without pilot contamination (PC) in the derived channel estimates. The results are also simplified for the mixed architecture that employs full resolution (FR) ADCs and one-bit DACs, and the conventional architecture that employs FR ADCs and DACs.   The SQINR is shown to decrease by a factor of $2/\pi$ and $4/\pi^2$ in the one-bit  setting compared to that achieved in the mixed setting  and conventional setting respectively under MRT precoding without PC. Interestingly, the decrease in SQINR is less when we consider PC, which is shown to adversely impact the conventional  system more than the one-bit  system. Similar insights are obtained under ZF precoding with the decrease in the SQINR with the use of one-bit ADCs and DACs being more pronounced. We utilize the derived expressions to yield performance insights related to power efficiency, the numbers of antennas needed by the three architectures to achieve the same sum-rate, and energy efficiency.

\end{abstract} \vspace{-.05in}

\section{Introduction}
\label{Sec:Intro}

Massive multiple-input multiple-output (MIMO) systems utilize a large number of antennas at each base station (BS) to yield spatial multiplexing and beamforming gains  \cite{massive1, massivenew}. In the downlink, these gains can  be achieved with simple linear precoding schemes, like maximum ratio transmission (MRT) and zero-forcing (ZF), that are asymptotically optimal \cite{ZFref, addref}. While the benefits of massive MIMO scale with the number of antennas, so does the  power consumption associated with the active components, like power amplifiers, analogue-to-digital converters (ADCs) and digital-to-analogue converters (DACs), that constitute the radio frequency (RF) chain connected to each antenna \cite{refe}.

 In practice, the real and imaginary parts of the transmit and receive signal at each RF chain are generated by a pair of DACs and processed by a pair of ADCs  in the downlink and uplink respectively. The power consumption of these ADCs and DACs increases exponentially with their resolution (in bits) and linearly with the sampling frequency \cite{addref,ADC1}, with commercially available data converters with resolution of $12$ to $16$ bits consuming on the order of several watts \cite{ref2}.   Hence, the resolution of each ADC and DAC must be limited to keep the power consumption at the massive MIMO BSs within tolerable levels. Moreover utilizing low-resolution ADCs and DACs relaxes the requirement of employing highly linear power amplifiers in the RF chains, which further reduces cost and power consumption.

 Motivated by these observations, we will consider the simplest possible scenario of a  one-bit massive MIMO cellular network with BSs  employing one-bit ADCs and DACs, that consist of a simple comparator and consume negligible power.   For this setting, we will study the impact of one-bit quantization on the sum-rate and energy efficiency performance of classical linear precoders. To this end, we first outline the related literature and the research gap, and then present our contributions. 
 
 \subsection{Related Literature and Research Gap}

Preliminary works on this subject  focused on the impact of using one-bit ADCs in the uplink of single-cell MIMO systems, with the work in \cite{lit4} analyzing the capacity of a point-to-point MIMO system  with one-bit ADCs, and that in  \cite{addref2}  studying receiver designs for  MIMO communication scenarios with a limited number of one-bit ADCs. More recently, some works have studied the impact of one-bit ADCs on the performance of  linear combiners \cite{lit8, lit9}. Specifically, the authors in \cite{lit8} and \cite{lit9} derived the channel estimates and uplink achievable  rates in wide-band and narrow-band one-bit massive MIMO  systems respectively  under maximum-ratio (MR) and ZF combining.

The current literature studying the impact of one-bit DACs on the downlink performance of massive MIMO systems is limited. For the quantization-free case, dirty paper coding (DPC) \cite{costa} is known to achieve the sum-rate capacity but is computationally very demanding to implement for large numbers  of  antennas. Linear precoding schemes like MRT and ZF, on the other hand, are attractive low-complexity  approaches  that offer competitive performance to DPC for large antenna arrays \cite{addref, ZFref}. Given the advantages offered by linear precoders in the conventional massive MIMO systems where the BSs employ full resolution (FR) ADCs and DACs, it is important to analyze how their performance is impacted by the use of  one-bit data converters.
 
In this context, the downlink performance of one-bit quantized linear precoders has been analyzed in \cite{lit12, lit10, lit11}. The authors in \cite{lit12}  utilized Bussgang decomposition to derive asymptotic closed-form expressions of the signal-to-quantization-plus-interference-plus-noise ratio (SQINR) and  symbol error rate at each user under one-bit quantized ZF precoding assuming perfect channel state information (CSI) and a single-cell. The authors in \cite{lit10} and \cite{lit11}  derived the  downlink ergodic achievable rate expressions considering MRT precoding and imperfect CSI, in single-cell  and cell-free one-bit massive MIMO systems respectively.  There a very few works that incorporate the effect of low-resolution data converters in the design of multi-cell massive MIMO systems \cite{lit15}, \cite{addref3}. In this context, the authors in \cite{lit15}   solved  uplink and downlink  transmit power minimization problems  to design uplink combining and downlink precoding in a multi-cell massive MIMO setting where BSs employ low resolution data converters. In another work \cite{addref3}, the authors   considered a full-duplex cellular network with low-resolution ADCs and DACs, and utilized the additive quantization noise model (AQNM) to derive SQINR and spectral efficiency expressions under MR precoding and combining.

 To the best of the authors' knowledge, the downlink performance of one-bit quantized MRT and ZF precoding  in a multi-cell one-bit massive MIMO setting has  not been analyzed before, and is the subject of this work. In this setting, the impact of one-bit quantization on the pilot contamination (PC) in the channel estimates, the intra-cell and inter-cell interference, and the resulting sum-rate  becomes  important, and is thoroughly analyzed. Our channel estimation and achievable rate analysis will encompass the one-bit architecture where the BSs employ one-bit ADCs and DACs, the mixed architecture where the BSs employ FR ADCs and one-bit DACs, and the conventional architecture where the BSs employ FR ADCs and DACs, in the RF chain associated with each antenna. The analysis will yield  interesting  insights related to   power scaling laws, energy efficiency, and the numbers of antennas needed by the three architectures to achieve the same sum-rate. 

\subsection{Contributions}

We analyze the downlink sum ergodic  rate performance of a  one-bit massive MIMO cellular system, in which the RF chains at each BS are equipped with one-bit ADCs and DACs. The analysis is done for one-bit quantized MRT and ZF precoding schemes implemented using imperfect CSI considering the scenarios with and without PC.   For this framework, our detailed contributions are summarized below.

\begin{itemize}
\item  We derive the  minimum mean squared error (MMSE) channel estimates at each BS for the users in its cell based on the quantized received  training signals, by utilizing  Bussgang decomposition \cite{Bussgang} to represent the non-linear quantizer  as a statistically equivalent linear system. In contrast to the channel estimates in \cite{lit9, lit10} and \cite{lit11}, our estimates account for both quantization noise and PC due to the re-use of pilot sequences. While the normalized mean squared error (NMSE) in   estimates is larger when the BSs deploy one-bit ADCs instead of FR ADCs due to the quantization noise, the impact of PC on the NMSE  is shown to reduce by a factor of $\frac{2}{\pi}$ under one-bit ADCs. 
\item We derive closed-form expressions of the downlink SQINR and ergodic achievable  rate at each user  under one-bit quantized MRT and ZF precoding. In contrast to the expressions in \cite{lit9, lit12} and \cite{lit10} that are derived for single-cell systems, our derivations are for a multi-cell system and explicitly show the impact of quantization noise, PC, and  intra-cell and inter-cell interference.   The expressions are simplified for the mixed BS architecture employing FR ADCs and one-bit DACs, and the conventional BS architecture employing FR ADCs and DACs, as well as for the scenarios without PC.  The SQINR is shown to decrease by a factor of $\frac{2}{\pi}$ and $\frac{4}{\pi^2}$ in the one-bit architecture compared to the  mixed  and conventional architectures respectively under MRT precoding without PC. Interestingly, the decrease in SQINR is less when we consider PC, which is shown to adversely impact the conventional  system more than the one-bit  system. Similar insights are obtained under ZF precoding with the decrease in the SQINR with the use of one-bit ADCs and DACs being more pronounced at high signal-to-noise ratios (SNRs). As the number of antennas at the BSs increases, the desired signal to PC ratio is shown to be the limiting factor in the SQINR expressions in all scenarios with PC, and is shown to be unaffected by one-bit quantization. 
\item We utilize the derived results to study the power efficiency of the one-bit massive MIMO cellular system, which is defined in \cite{lit9} as a measure of the reduction in the transmit and training powers that can be achieved with an increase in the number of  antennas $M$ at each BS while maintaining a given sum-rate. We show that  1) for fixed  uplink training power, the transmit power at each BS  can be reduced proportionally to $1/M$, and 2) the uplink training and downlink transmit powers together can be  reduced proportionally to  $1/\sqrt{M}$ as $M$ increases such that the sum-rate converges to fixed values under both precoders. We further show that a one-bit massive MIMO system inherits the power efficiency advantage of  a conventional massive MIMO system that employs FR ADCs and DACs.   
\item We study the ratio of the number of antennas at each BS in the one-bit massive MIMO cellular system that employs one-bit ADCs and DACs to that at each BS in the conventional massive MIMO cellular system that employs FR ADCs and DACs, required for both systems to achieve the same sum-rate. The ratio turns out to be $2.5$ under MRT precoding at all SNR values, while it is $2.5$ at low SNR and increases with the SNR  under ZF precoding. When comparing the mixed architecture that employs FR ADCs and one-bit DACs with the conventional architecture, the ratio of the number of antennas to achieve the same sum-rate turns out to be $1.57$ under MRT, while it  is $1.57$ at low SNR and increases with SNR  under ZF. The ratio in all cases decreases to one as the number of antennas grows large since the rate loss due to one-bit quantization vanishes asymptotically.
\item Numerical results verify the performance analysis, and show that the one-bit massive MIMO cellular architecture is more energy-efficient than the mixed and conventional cellular architectures, especially at high sampling frequencies. 
\end{itemize}

The rest of the paper is organized as follows. In Sec. \ref{Sec:Sys}, the system model is outlined, and in Sec. III the channel estimates are derived. The achievable rates under one-bit quantized MRT and ZF precoding are derived in Sec. \ref{Sec:Asym}, and a detailed performance analysis is presented in Sec. V. Simulation results and conclusions are provided in Sec. \ref{Sec:Sim} and Sec. \ref{Sec:Con} respectively.

\section{System Model}
\label{Sec:Sys}
\begin{figure}
\centering
\includegraphics[scale=.34]{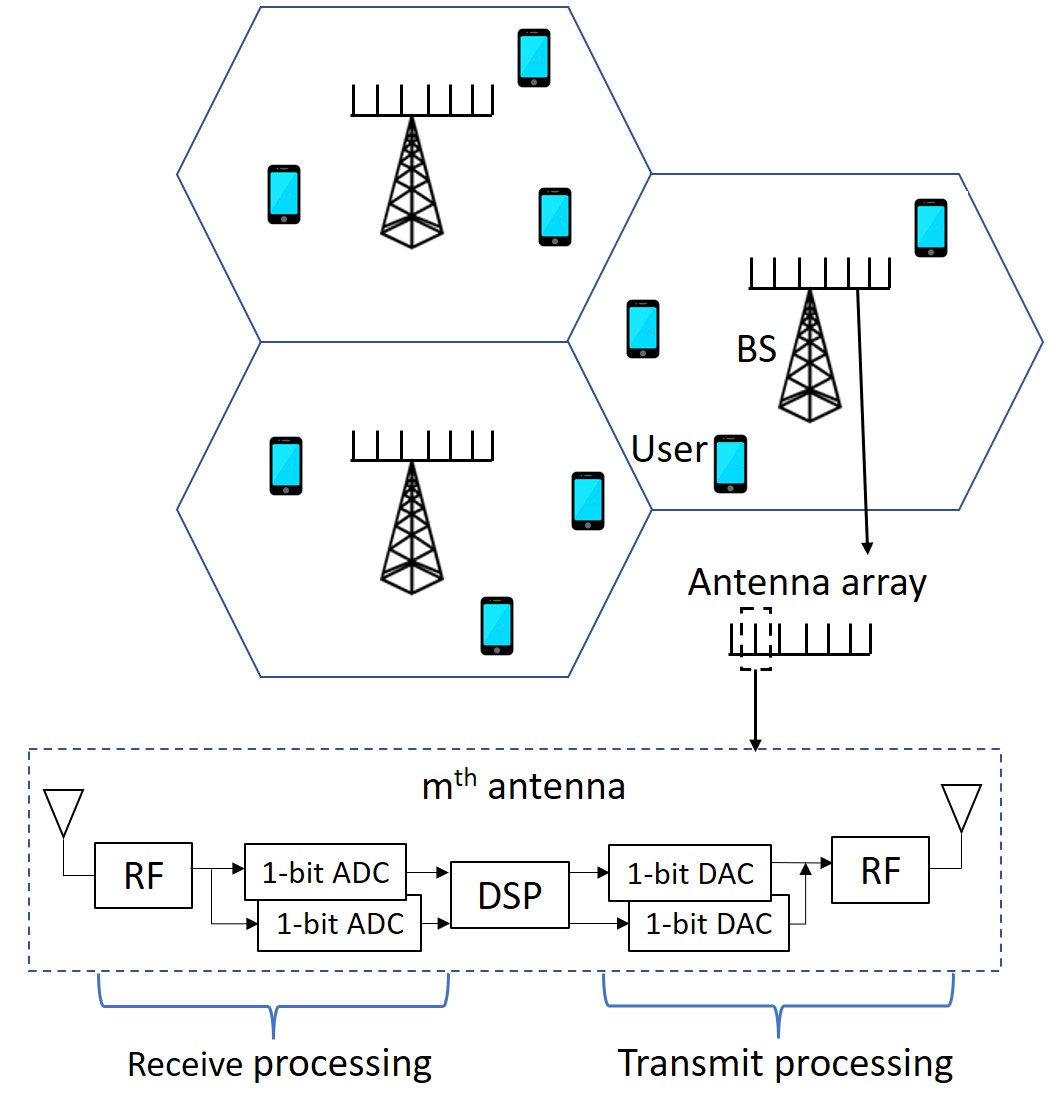}
\caption{One-bit massive MIMO cellular system model. }
\label{model}
\end{figure}

We consider a multi-cell  massive MIMO system with $ L > 1$ cells, each having an $M$-antenna BS and $K\leq M$ single-antenna users. The RF chain associated with each antenna at each BS is equipped with a pair of one-bit ADCs and DACs as shown in Fig. \ref{model}, that quantize the  real and imaginary parts of the received and transmitted signals respectively.  We denote the block-fading  channel between BS $j$ and user $k$ in cell $l$  as $\mathbf{h}_{jlk} \in \mathbb{C}^{M\times 1}$, and describe the signal models in the uplink and downlink for  channel estimation and downlink transmission respectively next. 

\subsection{Uplink Signal Model}

The BSs  need CSI to design the transmit signals, which is obtained through uplink training by exploiting channel reciprocity in a time division duplex framework. Specifically BS $j$ obtains an estimate of the channel  $\mathbf{H}_{jj}=[\mathbf{h}_{jj1},\dots, \mathbf{h}_{jjK}] \in \mathbb{C}^{M\times K}$  in an uplink training phase of length $\tau_p$ symbols at the start of each coherence block. In this training phase, the $K$ users in  cell $j$  transmit mutually orthogonal pilot sequences, represented as $\boldsymbol{\Phi}_j=[\boldsymbol{\phi}_{j1},\dots, \boldsymbol{\phi}_{jK}] \in \mathbb{C}^{\tau_p \times K}$,  satisfying  $\boldsymbol{\Phi}_j^H\boldsymbol{\Phi}_j=\mathbf{I}_K$. The same set of  pilot sequences is transmitted by the $K$ users in every cell resulting in the channel estimates at each BS to be corrupted by PC. The received training signal $\mathbf{Y}_j^p\in \mathbb{C}^{M \times \tau_p}$ at BS $j$ is  given as  $\mathbf{Y}_j^p=\sum_{l=1}^L\sqrt{\rho_p \tau_p} \mathbf{H}_{jl} \boldsymbol{\Phi}_l^T + \mathbf{N}^p_{j}$,   where $\rho_p$ is the training SNR, $\mathbf{H}_{jl}=[\mathbf{h}_{jl1}, \dots, \mathbf{h}_{jlK}]$,  and $\mathbf{N}_j^p$ has i.i.d. $\mathcal{CN}(\mathbf{0}, \mathbf{I}_M)$ columns representing   noise. We vectorize  $\mathbf{Y}_j^P$ as 
\begin{align}
\label{URx1}
&\mathbf{y}_j^p=\text{vec}(\mathbf{Y}_j^P)=\sum_{l=1}^L \bar{\boldsymbol{\Phi}}_l \mathbf{h}_{jl} +\mathbf{n}_j^p,
\end{align}
where $\bar{\boldsymbol{\Phi}}_l=\boldsymbol{\Phi}_l \otimes \sqrt{\rho_p \tau_p} \mathbf{I}_M$,  $ \mathbf{h}_{jl}=\text{vec}(\mathbf{H}_{jl})$, $\mathbf{n}_j^p=\text{vec}(\mathbf{N}_{j}^p)$, and $\otimes$ is the Kronecker product. The RF chain associated with each antenna is equipped with a pair of one-bit ADCs (see Fig. \ref{model}), that separately quantize the real and imaginary parts of the received  signal  to one-bit representation based on their sign. The quantized  training signal after one-bit ADCs is  given as 
\begin{align}
\label{URx2}
&\mathbf{r}_j^p =Q(\mathbf{y}_j^p)=Q\left(\sum_{l=1}^L  \bar{\boldsymbol{\Phi}}_l \mathbf{h}_{jl} +\mathbf{n}_j^p\right),
\end{align}
where  $Q(.)$ is the one-bit quantization operation defined as 
\begin{align}
\label{Q}
&Q(\mathbf{a}) = \frac{1}{\sqrt{2}}\text{sign}(\Re(\mathbf{a})) + j \frac{1}{\sqrt{2}} \text{sign}(\Im(\mathbf{a})),
\end{align}
where $\Re(\mathbf{a})$ and  $\Im(\mathbf{a})$ represent the real and imaginary parts of $\mathbf{a}$, and $\text{sign}(t)$ is the sign of $t$. The elements of  ${\mathbf{r}}_j^p$ will  belong to the set  $\mathcal{R}=\frac{1}{\sqrt{2}}\{1+j, 1-j, -1+j, -1-j  \}$.  Based on the quantized training signal model in \eqref{URx2}, we will derive  MMSE  estimates of $\mathbf{h}_{jjk}$ denoted as $\hat{\mathbf{h}}_{jjk}$, $\forall j, k$ in Sec. III.

\subsection{Downlink Signal Model}
In the downlink,  BS $j$ wants to send information at rate $R_{jk}$ to user $k$ in  cell $j$. To do this, it constructs codewords with symbols $s_{jk} \in \mathbb{C}$ and combines them in a transmit signal vector  $\mathbf{x}_j=\mathbf{W}_j\mathbf{s}_j \in \mathbb{C}^{M\times 1}$, where $\mathbf{W}_j \in \mathbb{C}^{M \times K}$ is the linear precoder and $\mathbf{s}_j=[s_{j1},\dots, s_{jK}]^T\in \mathbb{C}^{K\times 1}$  satisfies $\mathbb{E}[\mathbf{s}_j\mathbf{s}_j^H]=\mathbf{I}_K$. The RF chain associated with each antenna is equipped with a pair of one-bit DACs. The real and imaginary parts of the  transmit signal $\mathbf{x}_j$ are thus converted into one-bit representation (element-wise) based on the sign of each component as shown in \eqref{Q}, and then converted to analog using   one-bit DACs. The final transmit signal from BS $j$ is  written as 
\begin{align}
\label{Tx}
&\tilde{\mathbf{x}}_j=Q(\mathbf{x}_j)=Q(\mathbf{W}_j\mathbf{s}_j).
\end{align}
Note that the elements of  $\tilde{\mathbf{x}}_j$ will  belong to the set $\mathcal{R}=\frac{1}{\sqrt{2}}\{1+j, 1-j, -1+j, -1-j  \}$. The received signal  at all users in cell $j$ is then given as
\begin{align}
\label{Rx1}
&\mathbf{y}_j= \sum_{l=1}^L \sqrt{\eta_l} \mathbf{H}_{lj}^H \tilde{\mathbf{x}}_l+\mathbf{n}_j,
\end{align}
where $\mathbf{H}_{lj}=[\mathbf{h}_{lj1}, \dots, \mathbf{h}_{ljK}] \in \mathbb{C}^{M\times K}$, $\mathbf{h}_{ljk}$ is the channel between BS $l$ and user $k$ in cell $j$, $\mathbf{n}_j=[n_{j1}, \dots, n_{jK}]^T\in \mathbb{C}^{K \times 1}$, and $n_{jk} \sim \mathcal{CN}(0,\sigma^2)$ is the noise at user $k$ in cell $j$.  Moreover $\eta_l$ is a normalization constant chosen to satisfy the average transmit power constraint at BS $l$ as $\mathbb{E}[||\sqrt{\eta_l}\tilde{\mathbf{x}}_l||^2]=P_t$. Since $\mathbb{E}[||\tilde{\mathbf{x}}_l||^2]=M$ due to the quantization operation defined in \eqref{Q}, we  obtain $\eta_l=\frac{P_t}{M}$.  The channel  $\mathbf{H}_{lj}$ is modeled as 
\begin{align}
\label{channel}
&\mathbf{H}_{lj}=\mathbf{G}_{lj}\mathbf{D}_{lj}^{1/2},
\end{align}
 where $\mathbf{G}_{lj}=[\mathbf{g}_{lj1}, \dots, \mathbf{g}_{ljK}]\in \mathbb{C}^{M\times K}$  and $\mathbf{D}_{lj}=\text{diag}(\beta_{lj1}\dots,  \beta_{ljK}) \in \mathbb{C}^{K\times K}$. The entries of $\mathbf{g}_{ljk}$ are i.i.d. complex Gaussian RVs with zero mean and unit variance, and the coefficients  $\beta_{ljk}$ represent the channel attenuation factors.

 We consider two  linear precoders of practical interest in this work, namely MRT  and ZF precoders, defined respectively as\vspace{-.05in}
\begin{align}
\label{MRT}
&\mathbf{W}_j^{\rm m}=\hat{\mathbf{H}}_{jj}, \\
\label{ZF}
&\mathbf{W}_j^{\rm z}=\hat{\mathbf{H}}_{jj}(\hat{\mathbf{H}}_{jj}^H \hat{\mathbf{H}}_{jj})^{-1},
\end{align}
where $\hat{\mathbf{H}}_{jj}=[\hat{\mathbf{h}}_{jj1},\dots, \hat{\mathbf{h}}_{jjK}]\in \mathbb{C}^{M\times K}$, and $\hat{\mathbf{h}}_{jjk}$ is the channel estimate that will be derived in Sec. III. Linear precoding schemes are   attractive in multi-user massive MIMO downlink because (i) they have much lower computational complexity compared to DPC, and (ii) they achieve optimal performance as $M$ grows large \cite{massive1,massiveMIMOO}. While  quantization does not imply additional computational complexity, it is important to study the performance of one-bit quantized linear precoders in \eqref{Tx} in massive MIMO settings to see whether they inherit the second advantage of conventional linear precoders. To do this, we will derive closed-form expressions of $R_{jk}$ under one-bit quantized MRT and ZF precoding, and study their  behaviour with $M$.

 Note that MRT and ZF are single-cell processing schemes implemented at each BS  using local CSI of  the $K$ users in its own cell. Recently, there has been an interest in  multi-cell zero-forcing (M-ZF) \cite{MZF} and multi-cell MMSE (M-MMSE) precoding \cite{MMMSE} schemes, that can mitigate inter-cell interference especially when the system utilizes larger pilot re-use factors (longer training sequences and overhead) to allow for a larger  number  ($\gg K$) of independent channel directions to be estimated. Our work focuses on the practically relevant scenario where  $K$ orthogonal pilot sequences are re-used in every cell resulting in a relatively small training overhead \cite{massiveMIMOO}. For such a scenario, M-ZF and M-MMSE schemes do not offer  noticeable gains over their single-cell counterparts since only $K$ independent user directions can be estimated due to PC. Therefore, in this work we focus on analyzing the performance of MRT and ZF precoders in \eqref{MRT} and \eqref{ZF} in a multi-cell one-bit massive MIMO setting, which has not been the subject of a work yet. The extension of the analysis to M-ZF and M-MMSE precoders with different pilot reuse factors is left for future work.

\section{Uplink Channel Estimation}
\label{Sec:Sys2}

In this section, we derive the estimate of the channel $\mathbf{h}_{jjk}$  based on the quantized received training signal model in \eqref{URx2}.

\subsection{Bussgang Decomposition}

Quantizing the training signal in  \eqref{URx2} causes a distortion $Q(\mathbf{y}_j^p)-\mathbf{y}_j^p$ that is correlated with the input $\mathbf{y}_j^p$ to the ADCs. For Gaussian inputs, Bussgang’s theorem \cite{Bussgang} allows us to decompose the quantized signal into a linear function of the input to the quantizer and a distortion term that is uncorrelated with the input  \cite{addref, lit9}. The resulting linear representation of the  quantization operation is statistically equivalent up to the second moments of  data.  Specifically, the Bussgang decomposition of the quantized training signal $\mathbf{r}_j^p$ in \eqref{URx2}, which involves the quantization of a Gaussian input  $\mathbf{y}_j^p$, is given as \cite{lit9}, \cite{Bussgang} 
\begin{align}
\label{URx3}
&\mathbf{r}_j^p =\mathbf{A}_j^p\mathbf{y}_j^p+\mathbf{q}_j^p ,
\end{align}
 where $\mathbf{A}_j^p$ is a linear operator chosen to satisfy $\mathbb{E}[\mathbf{y}_j^p \mathbf{q}_j^{p^H}]=\mathbf{0}$ as 
 \begin{align}
 \label{A}
&\mathbf{A}_j^p= \sqrt{\frac{2}{\pi}}\text{diag}(\mathbf{R}_{\mathbf{y}_j^p \mathbf{y}_j^p})^{-1/2},
\end{align}
where $\text{diag}(\mathbf{C})$ denotes a diagonal matrix with diagonal elements   equal to those of $\mathbf{C}$. Further, using the arcsin law for a hard limiting one-bit quantizer, we have the result $\mathbf{R}_{\mathbf{r}_j^p \mathbf{r}_j^p}=\frac{2}{\pi} (\arcsin(\mathbf{B})+j \arcsin(\mathbf{C})  )$ \cite{Bussgang3, lit9}, where $\mathbf{B}=\text{diag}(\mathbf{R}_{\mathbf{y}_j^p \mathbf{y}_j^p})^{-1/2} \Re(\mathbf{R}_{\mathbf{y}_j^p \mathbf{y}_j^p})  \text{diag}(\mathbf{R}_{\mathbf{y}_j^p \mathbf{y}_j^p})^{-1/2}$ and  $\mathbf{C}=\text{diag}(\mathbf{R}_{\mathbf{y}_j^p \mathbf{y}_j^p})^{-1/2} \Im(\mathbf{R}_{\mathbf{y}_j^p \mathbf{y}_j^p})  \text{diag}(\mathbf{R}_{\mathbf{y}_j^p \mathbf{y}_j^p})^{-1/2}$. This yields 
\begin{align}
\label{C_q}
&\mathbf{R}_{\mathbf{q}_j^p \mathbf{q}_j^p}=\frac{2}{\pi} (\arcsin(\mathbf{B})+j \arcsin(\mathbf{C})  )-\frac{2}{\pi}(\mathbf{B}+j\mathbf{C}).
\end{align}

Next we utilize these results to complete the Bussgang decomposition of  $\mathbf{r}_j^p$. Utilizing \eqref{URx1} in \eqref{URx3}, we write $\mathbf{r}_j^p$ as
\begin{align}
\label{URx4}
&\mathbf{r}_j^p=\sum_{l=1}^L\mathbf{A}_j^p \bar{\boldsymbol{\Phi}}_l   \mathbf{h}_{jl} +\mathbf{A}_j^p\mathbf{n}_j^p+\mathbf{q}_j^p.
\end{align}

In order to find $\mathbf{A}_j^p$  using \eqref{A}, we compute  $\mathbf{R}_{\mathbf{y}_j^p \mathbf{y}_j^p}$ as $\mathbf{R}_{\mathbf{y}_j^p \mathbf{y}_j^p}=\mathbb{E}\left[\mathbf{y}_j^p\mathbf{y}_j^{p^H}\right]=\sum_{l=1}^L\bar{\boldsymbol{\Phi}}_l \bar{\mathbf{D}}_{jl} \bar{\boldsymbol{\Phi}}_l^H+\mathbf{I}_{M\tau_p}$, where  $\bar{\mathbf{D}}_{jl}=\mathbf{D}_{jl}\otimes \mathbf{I}_M\in \mathbb{C}^{MK \times MK}$ and $\mathbf{D}_{jl}= \text{diag}(\beta_{jl1}\dots,  \beta_{jlK})$ as defined in \eqref{channel}. The expression of $\mathbf{R}_{\mathbf{y}_j^p \mathbf{y}_j^p}$  indicates that the choice of $\boldsymbol{\Phi}_l$'s will affect the linear operator $\mathbf{A}_j^p$ as well as the quantization noise. In order to obtain analytically tractable expressions for $\mathbf{A}_j^p$ and the   estimates, we consider $\tau_p=K$ and choose the $K$-dimensional identity matrix as the pilot matrix  as done in  \cite{ lit11}.   As a result, $\mathbf{R}_{\mathbf{y}_j^p \mathbf{y}_j^p}=\sum_{l=1}^L K\rho_p \bar{\mathbf{D}}_{jl}+\mathbf{I}_{MK}$. Thus we can write $\mathbf{A}_j^p$ using \eqref{A} as a diagonal matrix given as 
\begin{align}
\label{A_fin}
&\hspace{-.1in}\mathbf{A}_j^p=\sqrt{\frac{2}{\pi}}\text{diag}\left(\sum_{l=1}^L K\rho_p \bar{\mathbf{D}}_{jl}+\mathbf{I}_{MK}\right)^{-1/2}\hspace{-.15in} =\bar{\mathbf{A}}_j^p\otimes \mathbf{I}_M,
\end{align}
where  $[\bar{\mathbf{A}}_{j}^p]_{k,k}=\bar{a}_{jk}=\sqrt{\frac{2}{\pi(\sum_{l=1}^L K\rho_p \beta_{jlk}+1)}}$.  Further using the expression of $\mathbf{R}_{\mathbf{y}_j^p \mathbf{y}_j^p}$ in \eqref{C_q}, we get 
\begin{align}
\label{C_qp}
\mathbf{R}_{\mathbf{q}_j^p\mathbf{q}_j^p}=\left(1-\frac{2}{\pi}  \right) \mathbf{I}_{MK}.
\end{align}
This completes the Bussgang decomposition with  \eqref{URx4} being statistically equivalent to  \eqref{URx2} under the definition \hspace{-.02in}of \hspace{-.02in}$\mathbf{A}_j^p$ \hspace{-.02in}in \hspace{-.02in}\eqref{A_fin}.

\subsection{MMSE Channel Estimation}

The MMSE estimate of $\mathbf{h}_{jj}=[\mathbf{h}_{jj1}^T, \dots, \mathbf{h}_{jjK}^T]^T \in \mathbb{C}^{MK\times 1}$, where $\mathbf{h}_{jjk}$ undergoes independent Rayleigh fading  as outlined in \eqref{channel}, is presented next.

\begin{lemma} \label{L1} The MMSE  estimate of  $\mathbf{h}_{jj}$  is computed based on  the quantized received  training signal $\mathbf{r}_j^p$ at  BS $j$  in \eqref{URx4}  as
\begin{align}
\label{est1}
&\hat{\mathbf{h}}_{jj}=\sqrt{\rho_p K} \bar{\mathbf{D}}_{jj} {\mathbf{A}}_j^{p^H}  \mathbf{r}_j^p, \hspace{.15in} j=1,\dots, L
\end{align}
where $\bar{\mathbf{D}}_{jj}=\mathbf{D}_{jj}\otimes \mathbf{I}_M$, and $\mathbf{A}_j^p$ is defined in \eqref{A_fin}.
\end{lemma}
\begin{IEEEproof}
The proof follows by utilizing the definition of  MMSE estimate given as $\hat{\mathbf{h}}_{jj}=\mathbf{R}_{\mathbf{h}_{jj}\mathbf{r}_j^p}\mathbf{R}^{-1}_{\mathbf{r}_{j}^p\mathbf{r}_j^p} \mathbf{r}_{j}^p$, and computing $\mathbf{R}_{\mathbf{h}_{jj}\mathbf{r}_j^p}$ and $\mathbf{R}_{\mathbf{r}_{j}^p\mathbf{r}_j^p}$  using  the derived Bussgang decomposition of $\mathbf{r}_{j}^p$ in \eqref{URx4} and the channel model in \eqref{channel}.
\end{IEEEproof}

 Although the  estimate in \eqref{est1} is not Gaussian in general due to the quantization noise $\mathbf{q}_j^p$ that appears in $\mathbf{r}_j^p$ in \eqref{URx4}, we can approximate it as Gaussian using  the Cramer's central limit theorem for large $M$ \cite{CLT, lit10, lit9}. Thus  $\hat{\mathbf{h}}_{jj}\sim \mathcal{CN}(\mathbf{0},\mathbf{R}_{\hat{\mathbf{h}}_{jj}\hat{\mathbf{h}}_{jj}} )$, where $\mathbf{R}_{\hat{\mathbf{h}}_{jj}\hat{\mathbf{h}}_{jj}} =\mathbf{T}_{jj} \otimes  \mathbf{I}_{M}$ and $\mathbf{T}_{jj}\in \mathbb{C}^{K\times K}$ is a diagonal matrix with entries
\begin{align}
\label{TT}
[\mathbf{T}_{jj}]_{k,k}=t_{jjk}=\frac{2 \beta_{jjk}^2 \rho_p K}{\pi (\sum_{l=1}^L K\rho_p \beta_{jlk}+1)}.
\end{align}
Under the orthogonality property of  MMSE estimate, the channel estimate and the estimation error defined as $\tilde{\mathbf{h}}_{jj}={\mathbf{h}}_{jj}-\hat{\mathbf{h}}_{jj}$, are uncorrelated, with  $\tilde{\mathbf{h}}_{jj}\sim \mathcal{CN}(\mathbf{0},\bar{\mathbf{D}}_{jj}-\mathbf{R}_{\hat{\mathbf{h}}_{jj}\hat{\mathbf{h}}_{jj}} )$.

The MMSE estimate of the channel $\mathbf{h}_{jjk}$ from BS $j$ to user $k$ in cell $j$  under one-bit ADCs  can be extracted from \eqref{est1} as
\begin{align}
&\hat{\mathbf{h}}_{jjk}= \sqrt{\rho_p K} \beta_{jjk} \bar{a}_{jk} \mathbf{r}_{jk}^p,
\end{align}
 where $\mathbf{r}_{jk}^p=\sum_{l=1}^L  \sqrt{\rho_p K}  \bar{a}_{jk} \mathbf{h}_{jlk}+\bar{a}_{jk}\mathbf{n}_{jk}^p+\mathbf{q}_{jk}^p$, $\bar{a}_{jk}$ is defined after \eqref{A_fin}, and $\mathbf{n}_{jk}^p$ and $\mathbf{q}_{jk}^p$ are vectors of $(k-1)M+1$ to $kM$ entries of $\mathbf{n}_{j}^p$ and $\mathbf{q}_{j}^p$ in \eqref{URx4}. It then follows that $\hat{\mathbf{h}}_{jjk}\sim \mathcal{CN}(\mathbf{0},t_{jjk}\mathbf{I}_M)$ with $t_{jjk}$ given in \eqref{TT}, and the estimation error $\tilde{\mathbf{h}}_{jjk} \sim \mathcal{CN}(\mathbf{0},\tilde{t}_{jjk}\mathbf{I}_M)$, where $\tilde{t}_{jjk}=\beta_{jjk}-t_{jjk}$. The normalized mean squared error (NMSE) in $\hat{\mathbf{h}}_{jjk}$ is given as
\begin{align}
\label{NMSE1}
&\Gamma_{jk}=\frac{\tilde{t}_{jjk}}{\beta_{jjk}}=1-\frac{2\beta_{jjk}\rho_p K}{\pi(\sum_{l=1}^L \rho_pK \beta_{jlk}+1)}
\end{align}
 and represents the effects of thermal noise, PC, as well as  quantization noise due to the use of one-bit ADCs.

 The NMSE  under one-bit ADCs increases as $\rho_p$ decreases, and as $L$ increases due to an increase in PC from other cells.   Further, we have $\underset{\rho_p \rightarrow \infty}{\text{lim}}\Gamma_{jk}=1-\frac{2\beta_{jjk}}{\pi\sum_{l=1}^L \beta_{jlk}}$, which represents the effects of quantization noise and PC on the NMSE. 

Next we present the  estimate of $\mathbf{h}_{jjk}$ for the conventional massive MIMO system where the BSs employ FR ADCs, and the received training signal $\mathbf{y}_j^p$ in \eqref{URx1} is  not quantized. 
\begin{lemma}
\label{Cor1}
\hspace{-.03in}The  MMSE estimate of $\mathbf{h}_{jjk}$ when the BS employs FR ADCs is computed based on the received training signal in \eqref{URx1} as 
\begin{align}
\label{est12}
&\hat{\mathbf{h}}^{\rm FR}_{jjk}= \frac{\sqrt{\rho_p K} \beta_{jjk} }{\sum_{l=1}^L K\rho_p \beta_{jlk}+1 }  \mathbf{y}^{p}_{jk}
\end{align}
 where $\mathbf{y}_{jk}^{p}\hspace{-.04in}=\hspace{-.04in}\sum_{l=1}^L \hspace{-.04in} \sqrt{\rho_p K}  \mathbf{h}_{jlk}\hspace{-.04in}+\hspace{-.04in}\mathbf{n}_{jk}^p$ and $\mathbf{n}_{jk}^p\hspace{-.04in}\sim \hspace{-.04in} \mathcal{CN}(\mathbf{0},\mathbf{I}_M)$.  We have $\hat{\mathbf{h}}^{\rm FR}_{jjk}\sim \mathcal{CN}(\mathbf{0},t^{\rm FR}_{jjk}\mathbf{I}_M)$ where $t^{\rm FR}_{jjk}=\frac{\beta_{jjk}^2 \rho_p K}{\sum_{l=1}^L K\rho_p \beta_{jlk}+1}$, and the  estimation error $\tilde{\mathbf{h}}^{\rm FR}_{jjk} \sim \mathcal{CN}(\mathbf{0},\tilde{t}_{jjk}^{\rm FR}\mathbf{I}_M)$ where $\tilde{t}^{\rm FR}_{jjk}=\beta_{jjk}-t^{\rm FR}_{jjk}$.  The NMSE in $\hat{\mathbf{h}}^{\rm FR}_{jjk}$ is given as
\begin{align}
\label{NMSE3}
&\Gamma^{\rm FR}_{jk}=\frac{\tilde{t}^{\rm FR}_{jjk}}{\beta_{jjk}}=1-\frac{\beta_{jjk}\rho_p K}{\sum_{l=1}^L \rho_pK \beta_{jlk}+1}.
\end{align} 
\end{lemma}
\begin{IEEEproof}
This lemma follows by correlating $\mathbf{Y}^p_j$ with user $k$'s pilot sequence to get $\mathbf{y}_{jk}^p$, and utilizing the definition of the MMSE estimate $\hat{\mathbf{h}}_{jjk}=\mathbf{R}_{\mathbf{h}_{jjk}\mathbf{y}_{jk}^p}\mathbf{R}^{-1}_{\mathbf{y}_{jk}^p\mathbf{y}_{jk}^p} \mathbf{y}_{jk}^p$, with the channel $\mathbf{h}_{jjk}$ given by \eqref{channel}. The  estimate in \eqref{est12} is also presented in [20, Sec. II-C] that studies a conventional massive MIMO system.
\end{IEEEproof}

Note that  $\underset{\rho_p \rightarrow \infty}{\text{lim}}\Gamma_{jk}^{\rm FR}=1-\frac{\beta_{jjk}}{\sum_{l=1}^L \beta_{jlk}}$, which represents the effect of PC  on the NMSE under FR ADCs. Comparing \eqref{NMSE1} and \eqref{NMSE3}, it is evident that using one-bit ADCs increases the NMSE in the estimates since $\Gamma_{jk}-\Gamma^{\rm FR}_{jk}=(1-\frac{2}{\pi})\frac{\beta_{jjk}\rho_p K}{(\sum_{l=1}^L \rho_pK \beta_{jlk}+1)}>0$.   We also observe that $\underset{\rho_p \rightarrow \infty}{\text{lim}} (\Gamma_{jk}-\Gamma^{\rm FR}_{jk})=(1-\frac{2}{\pi})\frac{\beta_{jjk}}{\sum_{l=1}^L \beta_{jlk}}>0$, implying that that the NMSE difference under  one-bit and FR ADCs  not only depends on quantization noise represented by the $(1-\frac{2}{\pi})$ factor, but also on the amount of PC, represented by the $\frac{\beta_{jjk}}{\sum_{l=1}^L \beta_{jlk}}$ factor, as will be made explicit in Remark 1 later.

Next we present the channel estimates considering the scenario where the training sequences $\boldsymbol{\phi}_{jk}$ have length $\tau_p=KL$ symbols, and can therefore be chosen to be orthogonal for all the users in the multi-cell system, circumventing the problem of PC.

\begin{lemma}\label{NoPCest}
When $\tau_p=KL$, $\boldsymbol{\Phi}=[\boldsymbol{\Phi}_1, \dots, \boldsymbol{\Phi}_L]\in \mathbb{C}^{\tau_p\times KL}=\mathbf{I}_{KL}$, and the BS employs one-bit ADCs,  the estimate of the channel from BS $j$ to user $k$ in cell $j$ is obtained as 
\begin{align}
&\hat{\mathbf{h}}^{\rm s}_{jjk}= \sqrt{\rho_p KL} \beta_{jjk} \bar{a}^{ p,s}_{jjk} \mathbf{r}^{p}_{jk},
\end{align}
 where $\mathbf{r}_{jk}^p= \sqrt{\rho_p KL}  \bar{a}^{ p,s}_{jjk} \mathbf{h}_{jjk}+\bar{a}^{p, s}_{jjk}\mathbf{n}_{jk}^{p}+\mathbf{q}_{jk}^{p}$, $\bar{a}^{ p,s}_{jjk}=\sqrt{\frac{2}{\pi(\rho_p KL\beta_{jjk}+1)}}$, and $\mathbf{n}_{jk}^{p}$ and $\mathbf{q}_{jk}^{p}$ are vectors of $(j-1)MK+(k-1)M+1$ to $(j-1)MK+(k-1)M+M$ entries of $\mathbf{n}_{j}^{p}$ and $\mathbf{q}_{j}^{p}$ defined in \eqref{URx4}. It then follows that $\hat{\mathbf{h}}^s_{jjk}\sim \mathcal{CN}(\mathbf{0},t^{\rm s}_{jjk}\mathbf{I}_M)$ where $t^{\rm s}_{jjk}=\frac{2\beta_{jjk}^2 \rho_p KL}{\pi(\rho_p KL\beta_{jjk}+1)}$, and  $\tilde{\mathbf{h}}^{\rm s}_{jjk} \sim \mathcal{CN}(\mathbf{0},\tilde{t}^{\rm s}_{jjk}\mathbf{I}_M)$, where $\tilde{t}^{\rm s}_{jjk}=\beta_{jjk}-t^{\rm s}_{jjk}$. The NMSE in $\hat{\mathbf{h}}^s_{jjk}$ is given as
\begin{align}
\label{NMSE5}
&\Gamma^{\rm s}_{jk}=1-\frac{2\beta_{jjk}\rho_p KL}{\pi( \rho_p KL \beta_{jjk}+1)},
\end{align}
\end{lemma}
\begin{IEEEproof}
The proof follows by writing  $\mathbf{r}_j^{p} $  as \eqref{URx4} for $\tau_p=KL$, where $\mathbf{A}_j^{p}=\bar{\mathbf{A}}_j^{ p,s} \otimes \mathbf{I}_M $ with $\bar{\mathbf{A}}_j^{ p,s}\in \mathbb{C}^{KL\times KL}$ being a diagonal matrix with entries $[\bar{\mathbf{A}}_j^{ p,s}]_{(l-1)K+k, (l-1)K+k}=\bar{a}^{ p,s}_{jlk}=\sqrt{\frac{2}{ \pi(\rho_p KL\beta_{jlk}+1)}}$. The result then follows from the definition of the MMSE estimate and the orthogonality of pilot sequences. 
\end{IEEEproof}

Note that $\underset{\rho_p \rightarrow \infty}{\text{lim}}\Gamma^{\rm s}_{jk}=1-\frac{2}{\pi}$, which represents the effect of quantization noise only on the NMSE, and also follows from $\mathbf{R}_{\mathbf{q}_j^p\mathbf{q}_j^p}=\left(1-\frac{2}{\pi}  \right) \mathbf{I}_{M\tau_p}$.  Comparing the NMSE results with and without PC in \eqref{NMSE1} and \eqref{NMSE5} respectively, PC is observed to increase the NMSE with the error becoming larger with $L$. 

Finally we present the channel estimate $\hat{\mathbf{h}}^{\rm FR^s}_{jjk}$ when the BS has FR ADCs and the pilot matrix is set as that in Lemma \ref{NoPCest}.
\begin{lemma}
\label{Cor1noPC}
The   estimate of $\mathbf{h}_{jjk}$  when $\tau_p=KL$, $\boldsymbol{\Phi}=[\boldsymbol{\Phi}_1, \dots, \boldsymbol{\Phi}_L]=\mathbf{I}_{KL}$, and   BS employs FR ADCs  is given as  
\begin{align}
&\hat{\mathbf{h}}^{\rm FR^s}_{jjk}=  \frac{\sqrt{\rho_p KL} \beta_{jjk}}{ \rho_p KL\beta_{jjk}+1 } \mathbf{y}^{p}_{jk},
\end{align}
 where $\mathbf{y}_{jk}^{p}\hspace{-.04in}=\hspace{-.04in}  \sqrt{\rho_p KL}  \mathbf{h}_{jjk}\hspace{-.04in}+\hspace{-.04in}\mathbf{n}_{jk}^p$. Also $\hat{\mathbf{h}}^{\rm FR^s}_{jjk}\sim \mathcal{CN}(\mathbf{0},t^{\rm FR^s}_{jjk}\mathbf{I}_M)$ where $t^{\rm FR^s}_{jjk}=\frac{\beta_{jjk}^2 \rho_p K L}{\rho_p KL \beta_{jjk}+1}$ and   $\tilde{\mathbf{h}}^{\rm FR^s}_{jjk} \sim \mathcal{CN}(\mathbf{0},\tilde{t}_{jjk}^{\rm FR^s}\mathbf{I}_M)$ where $\tilde{t}^{\rm FR^s}_{jjk}=\beta_{jjk}-t^{\rm FR^s}_{jjk}$. The NMSE is given as
\begin{align}
\label{NMSE7}
&\Gamma^{\rm FR^s}_{jk}=1-\frac{\beta_{jjk}\rho_p K L}{\rho_p K L \beta_{jjk}+1}.
\end{align}
\end{lemma}
  Comparing \eqref{NMSE3} and \eqref{NMSE7}, PC is observed to increase the NMSE in the  estimates in the conventional setting as well.  Moreover, $\underset{\rho_p \rightarrow \infty}{\text{lim}}\Gamma^{\rm FR^s}_{jk}=0$ since there is no PC in this setting.

\begin{remark} Using \eqref{NMSE1}, \eqref{NMSE3}, \eqref{NMSE5} and \eqref{NMSE7}, we obtain
\begin{align}
\label{PCfin}
\frac{\Gamma_{jk}-\Gamma^{\rm s}_{jk}}{\Gamma^{\rm FR}_{jk}-\Gamma^{\rm FR^s}_{jk}}=\frac{2}{\pi}, \hspace{.1in} \text{for } j=1,\dots, L, k=1,\dots, K,
\end{align}

which represents the ratio of the impact of PC on the NMSE when the BS has one-bit ADCs to when the BS has FR ADCs. We see that the impact of PC on the NMSE in the one-bit case is reduced by a factor of $\frac{2}{\pi}$ compared to the conventional case. This can be explained as follows. We know that quantization reduces the desired signal energy and introduces quantization noise as evident in \eqref{URx3}. The PC in the estimate of the channel of user $k$  at BS $j$ is caused by the received training signals at this  BS from every user $k$ in  cell $l\neq j$ that is using the same  pilot sequence. When the BS has one-bit ADCs, not only is the received training signal from the desired user quantized, but the received training signals from the interfering users from other cells are also quantized reducing their impact on the estimation error, compared to the case with FR ADCs where the interfering signals from the contaminating users  are stronger.  
\end{remark}

\section{Downlink Achievable Rate Analysis}
\label{Sec:Asym}

In this section, we present the Bussgang decomposition of the quantized transmit signal and analyze the  achievable  rates. 

\subsection{Bussgang Decomposition of Transmit Signal}

We utilize the Bussgang theorem to obtain a linear representation of the quantized transmit signal in \eqref{Tx} as  \cite{lit10}, \cite[Th. 2]{addref}  \begin{align}
\label{Tx2}
&\tilde{\mathbf{x}}_j=Q(\mathbf{x}_j)=\mathbf{A}_j \mathbf{x}_j+\mathbf{q}_j,
\end{align}
where $\mathbf{A}_j$ is the linear operator  given as \cite{addref, lit10} 
\begin{align}
\label{Ad}
&\mathbf{A}_j= \sqrt{\frac{2}{\pi}}\text{diag}\left( \mathbf{R}_{\mathbf{x}_j \mathbf{x}_j} \right)^{-1/2},
\end{align}
 and $\mathbf{R}_{\mathbf{x}_j \mathbf{x}_j}$ is  computed  for a given precoder $\mathbf{W}_j$ as $\mathbf{R}_{\mathbf{x}_j \mathbf{x}_j}=\mathbb{E}_{\mathbf{s}_j}[\mathbf{W}_{j}\mathbf{s}_j \mathbf{s}_j^H\mathbf{W}_{j}^H]=\mathbf{W}_j\mathbf{W}_j^H$  \cite[Theorem 2]{addref}. Moreover  the covariance matrix of the  quantization noise is given as \cite{lit10} \vspace{-.05in}
\begin{align}
\label{C_qd}
&\mathbf{R}_{\mathbf{q}_j \mathbf{q}_j}=\frac{2}{\pi} (\arcsin(\bar{\mathbf{B}})+j \arcsin(\bar{\mathbf{C}})  )-\frac{2}{\pi}(\bar{\mathbf{B}}+j\bar{\mathbf{C}}),
\end{align}
where $\bar{\mathbf{B}}=\text{diag}( \mathbf{R}_{\mathbf{x}_j \mathbf{x}_j} )^{-1/2} \Re( \mathbf{R}_{\mathbf{x}_j \mathbf{x}_j} )  \text{diag}( \mathbf{R}_{\mathbf{x}_j \mathbf{x}_j} )^{-1/2}$ and $\bar{\mathbf{C}}=\text{diag}( \mathbf{R}_{\mathbf{x}_j \mathbf{x}_j} )^{-1/2} \Im( \mathbf{R}_{\mathbf{x}_j \mathbf{x}_j} ) \text{diag}( \mathbf{R}_{\mathbf{x}_j \mathbf{x}_j} )^{-1/2}$.  In order to facilitate subsequent statistical analysis, we approximate $\mathbf{R}_{\mathbf{x}_j \mathbf{x}_j} $ with deterministic quantities  for large $(M,K)$ values and present the Bussgang decompositions for both precoders.
 
\begin{lemma}\label{LemmaMRT}
Under MRT precoding in \eqref{MRT}, the Bussgang decomposition of the quantized transmit signal is given as 
\begin{align}
\label{Tx2MRT}
&\tilde{\mathbf{x}}^{\rm m}_j=\mathbf{A}^{\rm m}_j \mathbf{W}^{\rm m}_j \mathbf{s}_j+\mathbf{q}^{\rm m}_j,
\end{align}
where $\mathbf{A}_j^{\rm m} = \sqrt{\frac{2}{\pi \sum_{k=1}^K t_{jjk}}}\mathbf{I}_M$ and  $\mathbf{R}_{\mathbf{q}^{\rm m}_j \mathbf{q}^{\rm m}_j}= \left(1-\frac{2}{\pi} \right) \mathbf{I}_M$ for  large $(M, K)$ values, with $t_{jjk}$ defined in  \eqref{TT}.
\end{lemma}
\begin{IEEEproof}
 We utilize the channel hardening property of  massive MIMO systems for large $(M,K)$ values  to write  $\mathbf{R}_{\mathbf{x}^{\rm m}_j \mathbf{x}^{\rm m}_j}=\mathbf{W}^{\rm m}_j\mathbf{W}^{\rm m^H}_j=\hat{\mathbf{H}}_{jj}\hat{\mathbf{H}}_{jj}^H\cong\sum_{k=1}^K t_{jjk}\mathbf{I}_M$, and use it to express  $\mathbf{A}^{\rm m}_j$ and $\mathbf{R}_{\mathbf{q}^{\rm m}_j \mathbf{q}^{\rm m}_j}$ using \eqref{Ad} and \eqref{C_qd} respectively. 
\end{IEEEproof}
 
\begin{lemma}\label{LemmaZF}
Under ZF precoding  in \eqref{ZF}, the Bussgang decomposition of the quantized transmit signal   is given as\vspace{-.05in}
\begin{align}
\label{Tx2ZF}
&\tilde{\mathbf{x}}^{\rm z}_j=\mathbf{A}^{\rm z}_j \mathbf{W}^{\rm z}_j \mathbf{s}_j+\mathbf{q}^{\rm z}_j
\end{align}
where $\mathbf{A}_j^{\rm z}= \sqrt{\frac{2 K (c-1)^2}{\pi \zeta_j}}\mathbf{I}_M$,  $\mathbf{R}_{\mathbf{q}^{\rm z}_j \mathbf{q}^{\rm z}_j}= \left(1-\frac{2}{\pi} \right) \mathbf{I}_M$, and $\zeta_j=\frac{1}{K}\sum_{k=1}^K \frac{1}{t_{jjk}}$  for large $(M,K)$  such that $M/K=c < \infty$.
\end{lemma} \vspace{.04in}
\begin{IEEEproof}
 We utilize the asymptotic approximation given in \cite[(34)]{lit12} as  $ \mathbf{R}_{\mathbf{x}^{\rm z}_j \mathbf{x}^{\rm z}_j} =\mathbf{W}^{\rm z}_j\mathbf{W}^{\rm z^H}_j=\hat{\mathbf{H}}_{jj}(\hat{\mathbf{H}}^H_{jj}\hat{\mathbf{H}}_{jj})^{-2}\hat{\mathbf{H}}_{jj}^H\xrightarrow[M,K\rightarrow \infty]{} \frac{\zeta_j}{K(c-1)^2} \mathbf{I}_M$ for $M/K=c < \infty$, where $\zeta_j=\frac{1}{K}\sum_{k=1}^K \frac{1}{t_{jjk}}$, to approximate  $\mathbf{A}^{\rm z}_j$ and $\mathbf{R}_{\mathbf{q}^{\rm z}_j \mathbf{q}^{\rm z}_j}$ using \eqref{Ad} and \eqref{C_qd} respectively. 
\end{IEEEproof}

Note that the asymptotic approximations used in  Lemma \ref{LemmaMRT} and  \ref{LemmaZF} are tight  for moderate values of system dimensions and hence are of practical value \cite{lit12, lit10, lit9, lit11}. 

\subsection{Achievable Rates}

 We now outline the achievable  rates at the users for which we will develop  closed-form expressions.  Utilizing the  decomposition of $\tilde{\mathbf{x}}_j$ in \eqref{Tx2}, we can write the received signal at user $k$ in cell $j$ in \eqref{Rx1} as ${y}_{jk}=\sqrt{\eta_j}\mathbf{h}_{jjk}^H \mathbf{A}_j \mathbf{w}_{jk} \mathbf{s}_{jk}+\sum_{(l,m)\neq (j,k)} \sqrt{\eta_l}\mathbf{h}_{ljk}^H \mathbf{A}_l \mathbf{w}_{lm} \mathbf{s}_{lm}+\sum_{l=1}^L \sqrt{\eta_l}\mathbf{h}_{ljk}^H \mathbf{q}_l+{n}_{jk}$. Although the quantization noise $\mathbf{q}_l$ is not Gaussian, we can obtain a lower bound on the capacity by making the worst-case Gaussian assumption for it to write the achievable rate as $R_{jk}=$\vspace{-.15in}

\small
\begin{align}
\label{rateperfect}
&\mathbb{E}\Big[\log_2 \hspace{-.04in}\Big(1+\frac{{\eta_j}|\mathbf{h}_{jjk}^H \mathbf{A}_j \mathbf{w}_{jk}|^2}{\hspace{-.05in}\sigma^2+\sum_{l=1}^L \eta_l \mathbf{h}_{ljk}^H \mathbf{C}_{\mathbf{q}_l\mathbf{q}_l} \mathbf{h}_{ljk}+\hspace{-.2in}\underset{(l,m)\neq (j,k)}{\sum} \hspace{-.1in} \eta_l |\mathbf{h}_{ljk}^H \mathbf{A}_l \mathbf{w}_{lm}|^2}\Big)  \Big]
\end{align}\normalsize

Note that \eqref{rateperfect} represents an achievable rate for a genie-aided user that perfectly knows the instantaneous effective channel gain $\mathbf{h}_{jjk}^H \mathbf{A}_j \mathbf{w}_{jk}$. In practice the users do not know these gains and need to estimate them  through downlink training \cite{DLp}. Alternatively, a blind channel estimation scheme has been proposed in \cite{No} to estimate the effective channels gains without requiring any downlink pilots, for which the authors then derive and numerically compute a capacity lower bound.

Another achievable rate expression often utilized in massive MIMO literature is based on the idea that by virtue of channel hardening, the instantaneous effective channel gain of user $k$ in cell $j$ approaches its average value $\mathbb{E}[\mathbf{h}_{jjk}^H \mathbf{A}_j \mathbf{w}_{jk} ]$ as $M$ increases and hence asymptotically it is sufficient for each user to only have   statistical CSI \cite{massiveMIMOO, MZF, AR, emilref, append, cff}. The main idea then is to decompose  ${y}_{jk}$  as ${y}_{jk}=\sqrt{\eta_j}\mathbb{E}[\mathbf{h}_{jjk}^H \mathbf{A}_j \mathbf{w}_{jk} ] {s}_{jk}+\sqrt{\eta_j}(\mathbf{h}_{jjk}^H \mathbf{A}_j \mathbf{w}_{jk}\hspace{-.03in} -\hspace{-.03in}\mathbb{E}[\mathbf{h}_{jjk}^H \mathbf{A}_j \mathbf{w}_{jk} ]) {s}_{jk}+\sum_{(l,m)\neq (j,k)} \sqrt{\eta_l}\mathbf{h}_{ljk}^H \mathbf{A}_l \mathbf{w}_{lm} {s}_{lm}+\sum_{l=1}^L \sqrt{\eta_l}\mathbf{h}_{ljk}^H \mathbf{q}_l+{n}_{jk} $, and assume that  $\mathbb{E}[\mathbf{h}_{jjk}^H \mathbf{A}_j \mathbf{w}_{jk} ]$ can be perfectly learned at  user $k$ in cell $j$.  The sum of the last four terms in $y_{jk}$  is  considered as effective additive noise. Treating this noise   as uncorrelated Gaussian as a worst-case,   user $k$ in cell $j$ can achieve the  ergodic  rate  \cite[Theorem 1]{AR}
\begin{align}
\label{rate}
&R_{jk}=\log_2(1+\gamma_{jk})
\end{align}
where $\gamma_{jk}$ is the associated SQINR given as 
\begin{align}
\label{SQINR}
&\gamma_{jk}=\frac{\text{DS}_{jk}}{\text{CU}_{jk}+\text{QN}_{jk}+\text{IUI}_{jk}+\text{TN}_{jk}}
\end{align}
where  $\text{DS}_{jk}={\eta_j}|\mathbb{E}[\mathbf{h}_{jjk}^H \mathbf{A}_j \mathbf{w}_{jk} ]|^2$ is the power of  average desired signal,   $\text{CU}_{jk}=\eta_j\text{Var}[\mathbf{h}_{jjk}^H \mathbf{A}_j \mathbf{w}_{jk} ]$ is the average channel gain uncertainty power,  $\text{QN}_{jk}=\sum_{l=1}^L \eta_l\mathbb{E}[\mathbf{h}_{ljk}^H \mathbf{C}_{\mathbf{q}_l\mathbf{q}_l} \mathbf{h}_{ljk}]$  is the average quantization noise power,  $\text{IUI}_{jk}=\sum_{(l,m)\neq (j,k)} \eta_l \mathbb{E}[|\mathbf{h}_{ljk}^H \mathbf{A}_l \mathbf{w}_{lm}|^2]$ is the average interference power,  and $\text{TN}_{jk}\hspace{-.04in}=\hspace{-.04in}\sigma^2$ is thermal noise power.

The difference in the achievable rate under genie-aided scheme in \eqref{rateperfect}, blind channel estimation scheme in \cite[eq (44)]{No}, and under channel hardening in \eqref{rate} is  negligible for Rayleigh fading channels, and becomes pronounced for  keyhole channels that do not harden. In this work, we resort to the channel hardening bound for analyzing the performance of one-bit massive MIMO systems. Later in the simulations, we compare the achievable rates  in \eqref{rate} with that for the genie-aided users in \eqref{rateperfect}, and show the performance difference to be negligible when the environment has rich scattering which is the scenario considered in this work.  To this end, the  sum average rate  is given as 
\begin{align}
\label{Rsum}
&R_{\rm sum}=\sum_{j=1}^L \sum_{k=1}^K R_{jk}.
\end{align}
Next we   derive the expectations in \eqref{SQINR} and the resulting ergodic rates in closed-form for both one-bit quantized  precoders.
 
\subsection{Achievable Rates under One-Bit Quantized MRT Precoder}
 We first present  results for one-bit quantized MRT precoding described using the Bussgang decomposition in  Lemma \ref{LemmaMRT}, and implemented using the channel estimates  in Lemma \ref{L1}. 

\begin{theorem}
\label{Th_MRT}
Consider a one-bit  massive MIMO cellular network with BSs equipped with one-bit ADCs and DACs. Then under MRT precoding and large $(M,K)$ values, the ergodic   rate in \eqref{rate} is given in a closed-form as $R^{\rm m, 1}_{jk}=\log_2(1+\gamma^{\rm m, 1}_{jk})$, where 
\begin{align}
\label{SQINR_MRT} 
\gamma^{\rm m, 1}_{jk}&=\frac{1}{\overline{\text{QN}}^{\rm m, 1}_{jk}+\overline{\text{IUI}}^{\rm m, 1}_{jk}+\overline{\text{PC}}^{\rm m, 1}_{jk}+\overline{\text{TN}}^{\rm m, 1}_{jk}} 
\end{align}
where   $\overline{\text{QN}}^{\rm m, 1}_{jk}=\sum_{l=1}^L \left(1-\frac{2}{\pi}  \right) \frac{\pi \bar{t}_{j} \beta_{ljk}}{2M t_{jjk}^2}$ is the normalized average  quantization noise power (normalized by the average desired signal power),  $\overline{\text{IUI}}^{\rm m, 1}_{jk}=\sum_{l=1}^L\sum_{m=1}^K \frac{\bar{t}_{j} t_{llm}\beta_{ljk}}{M \bar{t}_{l} t_{jjk}^2}$ is  the normalized average channel gain uncertainty plus interference power, $\overline{\text{PC}}^{\rm m, 1}_{jk}=\sum_{l\neq j}^L \frac{\bar{t}_{j}  t_{llk}^2 \beta_{ljk}^2}{\bar{t}_{l} t_{jjk}^2 \beta_{llk}^2}$ is the normalized average PC introduced interference power, and  $\overline{\text{TN}}^{\rm m, 1}_{jk}=\frac{\pi \sigma^2 \bar{t}_{j}}{2M P_t  t_{jjk}^2}$ is the normalized average thermal noise power,  with  $\bar{t}_{j}=\sum_{k=1}^K t_{jjk}$.
\end{theorem}
\begin{IEEEproof}
The proof is provided in Appendix \ref{App:Th_MRT}.
\end{IEEEproof}
Next we simplify this result for the special case without PC.
\begin{corollary}\label{Th_MRTs}
Under the setting of Theorem \ref{Th_MRT} and considering the special case outlined in Lemma \ref{NoPCest} that assigns orthogonal pilot sequences to all the users to remove PC, the ergodic achievable  rate   at user $k$ in cell $j$ is given as $R^{\rm m, 1^s}_{jk}=\log_2(1+\gamma^{\rm m, 1^s}_{jk})$, where  $\gamma^{\rm m, 1^s}_{jk}=\frac{1}{\overline{\text{QN}}^{\rm m, 1^s}_{jk}+\overline{\text{IUI}}^{\rm m, 1^s}_{jk}+\overline{\text{TN}}^{\rm m, 1^s}_{jk}} $, and all the terms in $\gamma^{\rm m, 1^s}_{jk}$ have the same definition as their counterpart terms in $\gamma^{\rm m, 1}_{jk}$ in Theorem \ref{Th_MRT} with $t_{jjk}$ replaced by $t^{\rm s}_{jjk}$  defined in Lemma \ref{NoPCest}, and $\bar{t}_{j}$ replaced by $\bar{t}_{j}^{\rm s}=\sum_{k=1}^K t^{\rm s}_{jjk}$.
\end{corollary}
\begin{IEEEproof}
The proof is similar to the one in Appendix \ref{App:Th_MRT}  and follows by using the estimate in Lemma \ref{NoPCest}.
\end{IEEEproof}

Next we simplify the result in Theorem \ref{Th_MRT} for a mixed (mix) architecture, where the BSs are equipped with one-bit DACs and FR ADCs. The  estimates \hspace{-.03in}  are \hspace{-.03in} therefore not \hspace{-.03in} contaminated \hspace{-.03in} by \hspace{-.03in} one-bit quantization  and are given in Lemma \ref{Cor1}.
\begin{theorem}
\label{Cor_MRT}
Consider a massive MIMO cellular network with BSs equipped with FR ADCs and one-bit DACs. Then under MRT precoding and large $(M,K)$ values, the ergodic achievable  rate in \eqref{rate}   is given as  $R^{\rm m, mix}_{jk}=\log_2(1+\gamma^{\rm m, mix}_{jk})$, where
\begin{align}
\label{SQINR_MRTFR}
&\hspace{-.15in}\gamma^{\rm m, mix}_{jk}=\frac{1}{\overline{\text{QN}}^{\rm m, mix}_{jk}+\overline{\text{IUI}}^{\rm m, mix}_{jk}+\overline{\text{PC}}^{\rm m, mix}_{jk}+\overline{\text{TN}}^{\rm m, mix}_{jk}}, 
\end{align} 
where $\overline{\text{QN}}^{\rm m, mix}_{jk}\hspace{-.07in}=\hspace{-.03in}\sum_{l=1}^L \left(1-\frac{2}{\pi}  \right) \frac{\pi \bar{t}^{\rm FR}_{j} \beta_{ljk}}{2M t_{jjk}^{{\rm FR}^2}}$, $\overline{\text{IUI}}^{\rm m, mix}_{jk}=\sum_{l=1}^L\sum_{m=1}^K \frac{\bar{t}^{\rm FR}_{j} t^{\rm FR}_{llm}\beta_{ljk}}{M\bar{t}^{\rm FR}_{l} t_{jjk}^{{\rm FR}^2}}$, $\overline{\text{PC}}^{\rm m, mix}_{jk}=\sum_{l\neq j}^L \frac{\bar{t}^{\rm FR}_{j}  t_{llk}^{{\rm FR}^2} \beta_{ljk}^2}{\bar{t}^{\rm FR}_{l} t_{jjk}^{{\rm FR}^2} \beta_{llk}^2}$, and $\overline{\text{TN}}^{\rm m, mix}_{jk}=\frac{\pi \sigma^2 \bar{t}^{\rm FR}_{j}}{2M P_t  t_{jjk}^{{\rm FR}^2}}$, where $t^{\rm FR}_{jjk}$ is defined under FR ADCs in Lemma \ref{Cor1}, and $\bar{t}^{\rm FR}_{j}=\sum_{k=1}^K t^{\rm FR}_{jjk}$.
\end{theorem}
\begin{IEEEproof}
The proof is similar to the one in Appendix \ref{App:Th_MRT}  and follows by using the estimate in Lemma \ref{Cor1}.
\end{IEEEproof}

\begin{corollary}\label{Cor_MRTs}
Under the setting of Theorem \ref{Cor_MRT} and considering the special case outlined in Lemma \ref{Cor1noPC} that  removes PC, the ergodic achievable  rate  at user $k$ in cell $j$ is given as  $R^{\rm m, mix^s}_{jk}=\log_2(1+\gamma^{\rm m, mix^s}_{jk})$, where  $\gamma^{\rm m, mix^s}_{jk}=\frac{1}{\overline{\text{QN}}^{\rm m, mix^s}_{jk}+\overline{\text{IUI}}^{\rm m, mix^s}_{jk}+\overline{\text{TN}}^{\rm m, mix^s}_{jk}} $ and all the terms in $\gamma^{\rm m, mix^s}_{jk}$ have the same definition as their counterpart terms in $\gamma^{\rm m, mix}_{jk}$ in Theorem \ref{Cor_MRT} with $t^{\rm FR}_{jjk}$ replaced by $t^{\rm FR^s}_{jjk}$  defined in Lemma \ref{Cor1noPC}, and $\bar{t}^{\rm FR}_{j}$ replaced by $\bar{t}_{j}^{\rm FR^s}=\sum_{k=1}^K t^{\rm FR^s}_{jjk}$.
\end{corollary}
\begin{IEEEproof}
The proof is similar to the one in Appendix \ref{App:Th_MRT}  and follows by using the estimate in Lemma \ref{Cor1noPC}.
\end{IEEEproof}

Note that $t^{\rm FR}_{jjk}=\frac{\pi}{2}t_{jjk}$ by comparing Lemma \ref{L1} and Lemma \ref{Cor1}, and $t^{\rm FR^s}_{jjk}=\frac{\pi}{2}t^{\rm s}_{jjk}$ by comparing Lemma \ref{NoPCest} and Lemma \ref{Cor1noPC}. Using these relationships, it is straightforward to see that  $\frac{\overline{\text{QN}}^{\rm m,1}_{jk}}{\overline{\text{QN}}^{\rm m, mix}_{jk}}=\frac{\overline{\text{QN}}^{\rm m,1^s}_{jk}}{\overline{\text{QN}}^{\rm m, mix^s}_{jk}}=\frac{\pi}{2} $, $\frac{\overline{\text{TN}}^{\rm m,1}_{jk}}{\overline{\text{TN}}^{\rm m, mix}_{jk}}=\frac{\overline{\text{TN}}^{\rm m,1^s}_{jk}}{\overline{\text{TN}}^{\rm m, mix^s}_{jk}}=\frac{\pi}{2} $, and $\frac{\overline{\text{IUI}}^{\rm m,1}_{jk}}{  \overline{\text{IUI}}^{\rm m, mix}_{jk}}=\frac{\overline{\text{IUI}}^{\rm m,1^s}_{jk}}{  \overline{\text{IUI}}^{\rm m, mix^s}_{jk}}=\frac{\pi}{2}$, resulting in lower achievable rates under the setting of Theorem \ref{Th_MRT} and Corollary \ref{Th_MRTs} where the BS uses one-bit ADCs and DACs than Theorem \ref{Cor_MRT} and Corollary \ref{Cor_MRTs} where the BS uses FR ADCs and one-bit DACs. Interestingly the normalized average PC power is unaffected by the use of one-bit ADCs, i.e. $\overline{\text{PC}}^{\rm m,1}_{jk}=\overline{ \text{PC}}^{\rm m, mix}_{jk}$. This is because the desired signal energy is reduced by an additional factor of $\frac{2}{\pi}$ when we use one-bit ADCs instead of FR ADCs while using one-bit DACs (can be observed by studying the result for $\text{DS}_{jk}$ in Appendix A with $t_{jjk}=\frac{2}{\pi}t^{\rm FR}_{jjk}$), while the interference due to PC is also reduced by a factor of $\frac{2}{\pi}$ due to the explanation in Remark 1, resulting in the net effect of cancel out. Overall we observe that
\begin{align} 
\label{M1}
&\frac{\gamma_{jk}^{\rm m,1^s}}{\gamma_{jk}^{\rm m,mix^s}}=\frac{2}{\pi} \hspace{.06in} \text{ and} \hspace{.06in} \frac{2}{\pi} \leq \frac{\gamma_{jk}^{\rm m,1}}{\gamma_{jk}^{\rm m,mix}} \leq 1,
\end{align}
where $\frac{\gamma_{jk}^{\rm m,1}}{\gamma_{jk}^{\rm m,mix}}=\frac{2}{\pi}$ for small values of SNR, i.e. $\frac{P_t}{\sigma^2}$, and increases as the SNR increases and PC becomes dominant. This is because the introduction of PC has a smaller adverse impact on the performance when we have one-bit ADCs compared to when we have FR ADCs as discussed in Remark 1.  Therefore the ratio of the SQINR in one-bit system to that in the mixed system improves as  SNR increases and PC becomes dominant. 

Next  we simplify Theorem \ref{Th_MRT} for the conventional system.
\begin{theorem}
\label{Cor_MRTconv}
Consider the conventional massive MIMO cellular network with BSs equipped with FR ADCs and DACs. Then under MRT precoding, the ergodic achievable downlink rate at  user $k$ in cell $j$ is given as $R^{\rm m, conv}_{jk}=\log_2(1+\gamma^{\rm m, conv}_{jk})$, where the associated SINR is given in a closed-form  as
\begin{align}
\label{SQINR_MRTconv}
\hspace{-.1in}\gamma^{\rm m, conv}_{jk}&\hspace{-.05in}=\hspace{-.04in}\frac{1}{\overline{\text{IUI}}^{\rm m, conv}_{jk}\hspace{-.05in}+\overline{\text{PC}}^{\rm m, conv}_{jk}\hspace{-.05in}+\overline{\text{TN}}^{\rm m, conv}_{jk}}, 
\end{align}
where $\overline{\text{IUI}}^{\rm m, conv}_{jk}=\sum_{l=1}^L\sum_{m=1}^K\frac{\bar{t}^{\rm FR}_{j} t^{\rm FR}_{llm}\beta_{ljk}}{M\bar{t}^{\rm FR}_{l} t_{jjk}^{{\rm FR}^2}}$, $\overline{\text{PC}}^{\rm m, conv}_{jk}=\sum_{l\neq j} \frac{\bar{t}^{\rm FR}_{j}  t_{llk}^{{\rm FR}^2}\beta_{ljk}^2}{\bar{t}^{\rm FR}_{l} t_{jjk}^{{\rm FR}^2} \beta_{llk}^2}$, and $\overline{\text{TN}}^{\rm m, conv}_{jk}=\frac{\sigma^2 \bar{t}^{\rm FR}_{j}}{M P_t t_{jjk}^{{\rm FR}^2}}$.
\end{theorem} 
\begin{IEEEproof}
The proof  follows that in \cite[App. B]{massiveMIMOO}.
\end{IEEEproof}
Next we simplify this result for the special case without PC.
\begin{corollary}\label{Cor_MRTconvs}
Under the setting of Theorem \ref{Cor_MRTconv} and considering the special case outlined in Lemma \ref{Cor1noPC} that removes PC, the ergodic achievable  rate  at user $k$ in cell $j$ is given as  $R^{\rm m, conv^s}_{jk}=\log_2(1+\gamma^{\rm m, conv^s}_{jk})$, where  $\gamma^{\rm m, conv^s}_{jk}=\frac{1}{\overline{\text{IUI}}^{\rm m, conv^s}_{jk}\hspace{-.05in}+\overline{\text{TN}}^{\rm m, conv^s}_{jk}} $. All the terms in $\gamma^{\rm m, conv^s}_{jk}$ have the same definition as their counterpart terms in Theorem \ref{Cor_MRTconv} with $t^{\rm FR}_{jjk}$ replaced by $t^{\rm FR^s}_{jjk}$  defined in Lemma \ref{Cor1noPC}, and $\bar{t}^{\rm FR}_{j}$ replaced by $\bar{t}_{j}^{\rm FR^s}=\sum_{k=1}^K t^{\rm FR^s}_{jjk}$.
\end{corollary}

 Comparing the results in Theorem \ref{Th_MRT} and Theorem \ref{Cor_MRTconv} as well as those in Corollary \ref{Th_MRTs} and Corollary \ref{Cor_MRTconvs}, we can see that using one-bit ADCs and DACs not only introduces a quantization noise term ${\overline{\text{QN}}}^{\rm m,1}_{jk}$ in the SINR that will reduce the achievable rates, but it also increases the normalized channel gain uncertainty plus inter-user interference power by a factor of $\frac{\pi}{2}$, and  the normalized thermal noise power  by a factor of $\frac{\pi^2}{4}$ under both scenarios with and without PC. Further, we observe that $\overline{\text{PC}}^{\rm m, 1}_{jk}= \overline{\text{PC}}^{\rm m, conv}_{jk}$, because both the desired signal energy and interference due to PC are reduced by a factor of $\frac{4}{\pi^2}$ (can be observed by studying the derivation of $\text{DS}_{jk}$ in Appendix A and the PC term in \eqref{PCeq}) when the BS employs one-bit ADCs and DACs instead of FR ADCs and DACs, resulting in the net effect to cancel out.  Overall straightforward algebraic manipulations yield 
\begin{align}
\label{M3}
&\frac{\gamma_{jk}^{\rm m,1^s}}{\gamma_{jk}^{\rm m,conv^s}}= \frac{4}{\pi^2} \hspace{.05in} \text{ and } \hspace{.05in} \frac{4}{\pi^2}\leq \frac{\gamma_{jk}^{\rm m,1}}{\gamma_{jk}^{\rm m,conv}}\leq 1
\end{align}
where $ \frac{\gamma_{jk}^{\rm m,1}}{\gamma_{jk}^{\rm m,conv}}=\frac{4}{\pi^2}$ in the noise-limited case and improves as SNR increases and PC becomes dominant. This is because  PC introduces a smaller decrease in the performance when the BS employs one-bit ADCs and DACs compared to when it employs  FR ADCs and DACs, as discussed  in Remark 1. 

\subsection{Achievable Rates under One-Bit Quantized ZF Precoder} 

 Next we present the results for one-bit quantized ZF precoding  implemented using  channel estimates  in Lemma \ref{L1}.

\begin{theorem}
\label{Th_ZF}
Consider a massive MIMO cellular network with BSs equipped with one-bit ADCs and DACs. Then under ZF precoding and large $(M,K)$ values such that $\frac{M}{K}=c<\infty$, the ergodic achievable  rate in \eqref{rate}  at user $k$ in cell $j$ is given as $R^{\rm z, 1}_{jk}=\log_2(1+\gamma^{\rm z, 1}_{jk})$, where the SQINR is given  as 
\begin{align}
\label{SQINR_ZF}
\gamma^{\rm z,1}_{jk}&=\frac{1}{\overline{\text{QN}}^{\rm z,1}_{jk}+\overline{\text{IUI}}^{\rm z,1}_{jk}+\overline{\text{PC}}^{\rm z,1}_{jk}+\overline{\text{TN}}^{\rm z,1}_{jk}},
\end{align}
where    $\overline{\text{QN}}^{\rm z,1}_{jk}=\sum_{l=1}^L \left(1-\frac{2}{\pi}  \right) \frac{\pi M   \beta_{ljk} \zeta_j}{2K (c-1)^2}$ is the normalized average  quantization noise power,  $\overline{\text{IUI}}^{\rm  z,1}_{jk}=\sum_{l=1}^L\sum_{m=1}^K\frac{\zeta_j}{\zeta_l }\left(\frac{ \beta_{ljk}}{ t_{llm} (M-K)}-\frac{\beta_{ljk}^2 t_{llk}}{\beta_{llk}^2 t_{llm} (M-K)}\right)$  is the normalized average channel gain uncertainty plus interference power,  $\overline{\text{PC}}^{\rm  z,1}_{jk}=\sum_{l\neq j}^L \frac{\zeta_j \beta_{ljk}^2}{\zeta_l  \beta_{llk}^2}$ is the normalized average PC introduced interference power, and    $\overline{\text{TN}}^{\rm  z,1}_{jk}=\frac{\pi M \sigma^2 \zeta_j}{2K P_t (c-1)^2}$ is the normalized  average thermal noise power, with $\zeta_j$   defined in Lemma \ref{LemmaZF}.
\end{theorem}
\begin{IEEEproof}
The proof is provided in Appendix \ref{App:Th_ZF}.
\end{IEEEproof}
Next we simplify this result for the special case without PC.
\begin{corollary}\label{Th_ZFs}
Under the setting of Theorem \ref{Th_ZF} and considering the special case outlined in Lemma \ref{NoPCest} that removes PC, the  achievable  rate   at user $k$ in cell $j$ is given as $R^{\rm z, 1^s}_{jk}=\log_2(1+\gamma^{\rm z, 1^s}_{jk})$, where  $\gamma^{\rm z, 1^s}_{jk}=\frac{1}{\overline{\text{QN}}^{\rm z, 1^s}_{jk}+\overline{\text{IUI}}^{\rm z, 1^s}_{jk}+\overline{\text{TN}}^{\rm z, 1^s}_{jk}} $, and $\overline{\text{QN}}^{\rm z, 1^s}_{jk}$ and $\overline{\text{TN}}^{\rm z, 1^s}_{jk}$ have the same definitions as their counterpart terms in $\gamma^{\rm z, 1}_{jk}$ in Theorem \ref{Th_ZF} with $t_{jjk}$ replaced by $t^{\rm s}_{jjk}$  defined in Lemma \ref{NoPCest}, and $\zeta_j$ replaced by $\zeta_{j}^{\rm s}=\frac{1}{K}\sum_{k=1}^K\frac{1}{ t^{\rm s}_{jjk}}$. Further $\overline{\text{IUI}}^{\rm z, 1^s}_{jk}=\sum^K_{m =1} \frac{\beta_{jjk}-t_{jjk}^{\rm s}}{t_{jjm}^{\rm s}(M-K)}+\sum_{l\neq j}^L \sum_{m=1}^K \frac{\zeta_j^{\rm s} \beta_{ljk}}{\zeta_l^{\rm s} t_{llm}^{\rm s}(M-K)}$.
\end{corollary}
\begin{IEEEproof}
The proof is similar to the one in Appendix \ref{App:Th_ZF}  and follows by using the estimate in Lemma \ref{NoPCest}.
\end{IEEEproof}

Next we simplify Theorem \ref{Th_ZF} for the mixed architecture. 
\begin{theorem}
\label{Cor_ZF}
Consider a massive MIMO cellular network with BSs equipped with FR ADCs and one-bit DACs.  Then under ZF precoding and large $(M,K)$ values such that $\frac{M}{K}=c<\infty$, the ergodic   rate in \eqref{rate}  at user $k$ in cell $j$ is given as $R^{\rm z, mix}_{jk}=\log_2(1+\gamma^{\rm z, mix}_{jk})$, where the  SQINR is given as 
\begin{align}
\label{SQINR_ZFFR}
&\gamma^{\rm z, mix}_{jk}=\frac{1}{\overline{\text{QN}}^{\rm z, mix}_{jk}+\overline{\text{IUI}}^{\rm z, mix}_{jk}+\overline{\text{PC}}^{\rm z, mix}_{jk}+\overline{\text{TN}}^{\rm z, mix}_{jk}},
\end{align}
where  $\overline{\text{QN}}^{\rm z, mix}_{jk}=\sum_{l=1}^L \left(1-\frac{2}{\pi}  \right) \frac{\pi M   \beta_{ljk} \zeta^{\rm FR}_j}{2K (c-1)^2}$, $\overline{\text{IUI}}^{\rm  z,mix}_{jk}=\sum_{l=1}^L\sum_{m=1}^K\frac{\zeta^{\rm FR}_j}{\zeta^{\rm FR}_l }\left(\frac{ \beta_{ljk}}{ t^{\rm FR}_{llm} (M-K)}-\frac{\beta_{ljk}^2 t^{\rm FR}_{llk}}{\beta_{llk}^2 t^{\rm FR}_{llm} (M-K)}\right)$, $\overline{\text{PC}}^{\rm z, mix}_{jk}=\sum_{l\neq j}^L \frac{\zeta^{\rm FR}_j \beta_{ljk}^2}{\zeta^{\rm FR}_l  \beta_{llk}^2}$, and $\overline{\text{TN}}^{\rm  z, mix}_{jk}=\frac{\pi M \sigma^2 \zeta^{\rm FR}_j}{2K P_t (c-1)^2}$, where $t^{\rm FR}_{jjk}$ is defined in Lemma \ref{Cor1}, and $\zeta^{\rm FR}_j=\frac{1}{K}\sum_{k=1}^K \frac{1}{t^{\rm FR}_{jjk}}$.
\end{theorem}
\begin{IEEEproof}
The proof is similar to the one in Appendix \ref{App:Th_ZF}, and follows by using the  estimate in Lemma \ref{Cor1}.
\end{IEEEproof}

\begin{corollary}\label{Cor_ZFs}
Under the setting of Theorem \ref{Cor_ZF} and considering the special case outlined in Lemma \ref{Cor1noPC} that  removes PC, the ergodic achievable  rate  at user $k$ in cell $j$ is given as  $R^{\rm z, mix^s}_{jk}=\log_2(1+\gamma^{\rm z, mix^s}_{jk})$, where the SQINR  is given    as $\gamma^{\rm z, mix^s}_{jk}=\frac{1}{\overline{\text{QN}}^{\rm z, mix^s}_{jk}+\overline{\text{IUI}}^{\rm z, mix^s}_{jk}+\overline{\text{TN}}^{\rm z, mix^s}_{jk}} $ and  $\overline{\text{QN}}^{\rm z, mix^s}_{jk}$ and $\overline{\text{TN}}^{\rm z, mix^s}_{jk}$ have the same definitions as their counterpart terms in $\gamma^{\rm z, mix}_{jk}$ in Theorem \ref{Cor_ZF} with $t^{\rm FR}_{jjk}$ replaced by $t^{\rm FR^s}_{jjk}$  defined in Lemma \ref{Cor1noPC}, and $\zeta_j^{\rm FR}$ replaced by $\zeta_{j}^{\rm FR^s}=\frac{1}{K}\sum_{k=1}^K\frac{1}{ t^{\rm FR^s}_{jjk}}$. Further $\overline{\text{IUI}}^{\rm z, mix^s}_{jk}=\sum^K_{m=1} \frac{\beta_{jjk}-t_{jjk}^{\rm FR^s}}{t_{jjm}^{\rm FR^s}(M-K)}+\sum_{l\neq j}^L \sum_{m=1}^K \frac{\zeta_j^{\rm FR^s} \beta_{ljk}}{\zeta_l^{\rm FR^s} t_{llm}^{\rm FR^s}(M-K)}$.
\end{corollary}
\begin{IEEEproof}
The proof is similar to the one in Appendix \ref{App:Th_ZF}  and follows by using the estimate in Lemma \ref{Cor1noPC}.
\end{IEEEproof}

Noting that $t^{\rm FR}_{jjk}=\frac{\pi}{2}t_{jjk}$, $t^{\rm FR^s}_{jjk}=\frac{\pi}{2}t^{\rm s}_{jjk}$, $\zeta^{\rm FR}_{j}=\frac{2}{\pi} \zeta_{j}$ and $\zeta^{\rm FR^s}_{j}=\frac{2}{\pi} \zeta_{j}^{\rm s}$, we obtain $\frac{\overline{\text{QN}}^{\rm z,1}_{jk}}{\overline{\text{QN}}^{\rm z,mix}_{jk}}=\frac{\pi}{2}$,  $\frac{\overline{\text{TN}}^{\rm z,1}_{jk}}{\overline{\text{TN}}^{\rm z,mix}_{jk}}=\frac{\pi}{2}$, and $\frac{\overline{\text{IUI}}^{\rm z,1}_{jk}}{\overline{\text{IUI}}^{\rm z,mix}_{jk}}  =\frac{\pi}{2}\frac{(M-K)\overline{\text{IUI}}_{jk}^{\rm z,mix} +\sum_{l=1}^L\sum_{m=1}^K \left(1-\frac{2}{\pi}\right)\frac{\zeta_j^{\rm FR } \beta_{ljk}^2 t_{llk}^{\rm FR}}{\zeta_l^{\rm FR} \beta_{llk}^2t_{llm}^{\rm FR}} }{(M-K)\overline{\text{IUI}}_{jk}^{\rm z,mix}}   \geq \frac{\pi}{2}$, resulting in lower rates under the setting of Theorem \ref{Th_ZF} than Theorem \ref{Cor_ZF}. Similar results can be obtained for the special case without PC by comparing the results in Corollary \ref{Th_ZFs} and Corollary \ref{Cor_ZFs}.  The decrease in achievable rates under ZF  when using one-bit ADCs instead of FR ADCs is more dominant than that under MRT, since the normalized interference power increases by  a factor greater than $\frac{\pi}{2}$ under ZF, while it increased by a factor of $\frac{\pi}{2}$ under MRT.  Further, we observe that $\overline{\text{PC}}^{\rm z,1}_{jk}=\overline{ \text{PC}}^{\rm z, mix}_{jk}$, because both the desired signal energy and interference due to PC are reduced by a factor of $\frac{2}{\pi}$ when we use one-bit ADCs instead of FR ADCs (also discussed earlier after Corollary \ref{Cor_MRTs} for MRT). Overall the ratio of  SQINR in the two cases (without and with PC) follows 
\begin{align}
\label{Z1}
&\frac{\gamma_{jk}^{\rm z,1^s}}{\gamma_{jk}^{\rm z,mix^s}}=\begin{cases} \frac{2}{\pi} & \text{as } \frac{P_t}{\sigma^2}\rightarrow 0 \\
 <\frac{2}{\pi} & \text{otherwise}, \end{cases} \\ \label{Z2}
&\frac{\gamma_{jk}^{\rm z,1^s}}{\gamma_{jk}^{\rm z,mix^s}} \leq \frac{\gamma_{jk}^{\rm z,1}}{\gamma_{jk}^{\rm z,mix}} \leq 1,
\end{align} 
where $\frac{\gamma_{jk}^{\rm z,1}}{\gamma_{jk}^{\rm z,mix}}=\frac{2}{\pi}$ for small values of SNR, i.e. $\frac{P_t}{\sigma^2}$. Further we see that the ratio is better under PC, because the introduction of PC causes a smaller decrease in  performance when we have one-bit ADCs compared to when we have FR ADCs as discussed  in Remark 1. In the simulations, we will see that $ \frac{\gamma_{jk}^{\rm z,1}}{\gamma_{jk}^{\rm z,mix}}$ is $\frac{2}{\pi}$ at low SNR, improves slightly from $\frac{2}{\pi}$ as the SNR increases, and then starts to decrease as interference (that is increased by a factor $>\frac{\pi}{2}$) becomes more dominant. 

Next  we simplify Theorem \ref{Th_ZF} for the  conventional system.
\begin{theorem}
\label{Cor_ZFconv}
Consider the conventional massive MIMO cellular network with BSs equipped with FR ADCs and DACs. Then under ZF precoding, the  achievable  rate at  user $k$ in cell $j$ is given as $R^{\rm z, conv}_{jk}=\log_2(1+\gamma^{\rm z, conv}_{jk})$ with  SINR  given as \vspace{-.05in}
\begin{align}
\label{SQINR_ZFconv}
&\gamma^{\rm z, conv}_{jk}\hspace{-.05in}=\frac{1}{\overline{\text{IUI}}^{\rm z, conv}_{jk}+\overline{\text{PC}}^{\rm z, conv}_{jk}+\overline{\text{TN}}^{\rm z, conv}_{jk}}, \end{align}
where  $\overline{\text{PC}}^{\rm z, conv}_{jk}=\sum_{l\neq j}^L \frac{\zeta^{\rm FR}_j \beta_{ljk}^2}{\zeta^{\rm FR}_l  \beta_{llk}^2}$,  $\overline{\text{TN}}^{\rm z, conv}_{jk}=\frac{ \sigma^2 K \zeta^{\rm FR}_j}{P_t (M-K)}$ and $\overline{\text{IUI}}^{\rm  z,conv}_{jk}=\sum_{l=1}^L\sum_{m=1}^K\frac{\zeta^{\rm FR}_j}{\zeta^{\rm FR}_l }\left(\frac{ \beta_{ljk}}{ t^{\rm FR}_{llm} (M-K)}-\frac{\beta_{ljk}^2 t^{\rm FR}_{llk}}{\beta_{llk}^2 t^{\rm FR}_{llm} (M-K)}\right)$.
\end{theorem}

Next we simplify this result for the special case without PC.
\begin{corollary}\label{Cor_ZFconvs}
Under the setting of Theorem \ref{Cor_ZFconv} and considering the special case outlined in Lemma \ref{Cor1noPC} that removes PC, the  achievable  rate  at user $k$ in cell $j$ is given as  $R^{\rm z, conv^s}_{jk}=\log_2(1+\gamma^{\rm z, conv^s}_{jk})$, where $\gamma^{\rm z, conv^s}_{jk}=\frac{1}{\overline{\text{IUI}}^{\rm z, conv^s}_{jk}\hspace{-.05in}+\overline{\text{TN}}^{\rm z, conv^s}_{jk}} $. The term $\overline{\text{TN}}^{\rm z, conv^s}_{jk}$ has the same definition as its counterpart term in Theorem \ref{Cor_ZFconv} with $t^{\rm FR}_{jjk}$ replaced by $t^{\rm FR^s}_{jjk}$  defined in Lemma \ref{Cor1noPC}, and $\zeta_j^{\rm FR}$ replaced by $\zeta_{j}^{\rm FR^s}=\frac{1}{K}\sum_{k=1}^K\frac{1}{ t^{\rm FR^s}_{jjk}}$. Further $\overline{\text{IUI}}^{\rm z, conv^s}_{jk}=\sum^K_{m =1} \frac{\beta_{jjk}-t_{jjk}^{\rm FR^s}}{t_{jjm}^{\rm FR^s}(M-K)}+\sum_{l\neq j}^L \sum_{m=1}^K \frac{\zeta_j^{\rm FR^s} \beta_{ljk}}{\zeta_l^{\rm FR^s} t_{llm}^{\rm FR^s}(M-K)}$.
\end{corollary}

Comparing the results in Theorem \ref{Th_ZF} and Theorem \ref{Cor_ZFconv}, we can see that using one-bit ADCs and DACs not only introduces a quantization noise term $\overline{\text{QN}}^{\rm z,1}_{jk}$, but it also increases the remaining  terms as  $\frac{\overline{\text{TN}}^{\rm z, 1}_{jk}}{\overline{\text{TN}}^{\rm z, conv}_{jk}}=\frac{\pi^2 M}{4K (c-1)}\approx \frac{\pi^2}{4}$, and $\frac{\overline{\text{IUI}}^{\rm z,1}_{jk}}{\overline{\text{IUI}}^{\rm z,conv}_{jk}}  =\frac{\pi}{2}\frac{(M-K)\overline{\text{IUI}}_{jk}^{\rm z,conv} +\sum_{l=1}^L\sum_{m=1}^K \left(1-\frac{2}{\pi}\right)\frac{\zeta_j^{\rm FR } \beta_{ljk}^2 t_{llk}^{\rm FR}}{\zeta_l^{\rm FR} \beta_{llk}^2t_{llm}^{\rm FR}} }{(M-K)\overline{\text{IUI}}_{jk}^{\rm z,conv}}   \geq \frac{\pi}{2}$. Similar results can be obtained for the special case without PC by comparing the results in Corollary \ref{Th_ZFs} and Corollary \ref{Cor_ZFconvs}.   Compared to MRT, the decrease in SINR under ZF   when the BS uses one-bit ADCs and DACs instead of FR ADCs and DACs is observed to be  more dominant  as the SNR increases and interference becomes dominant. Moreover, similar to previous comparisons, the normalized average PC power is   unaffected by the use of one-bit data converters, i.e. $\overline{\text{PC}}^{\rm z, 1}_{jk}= \overline{\text{PC}}^{\rm z, conv}_{jk}$. Overall we observe  the ratio of the SQINR and SINR to follow
\begin{align}
\label{Z3}
&\frac{\gamma_{jk}^{\rm z,1^s}}{\gamma_{jk}^{\rm z,conv^s}}=\begin{cases} \frac{4}{\pi^2} & \text{as } \frac{P_t}{\sigma^2}\rightarrow 0 \\
 <\frac{4}{\pi^2} & \text{otherwise}, \end{cases} \\ \label{Z4}
&\frac{\gamma_{jk}^{\rm z,1^s}}{\gamma_{jk}^{\rm z,conv^s}} \leq \frac{\gamma_{jk}^{\rm z,1}}{\gamma_{jk}^{\rm z,conv}} \leq 1,
\end{align}
where $ \frac{\gamma_{jk}^{\rm z,1}}{\gamma_{jk}^{\rm z,conv}}=\frac{4}{\pi^2}$ for small values of SNR. The ratio is again seen to be better under PC since PC has a smaller adverse impact on performance when  BS employs one-bit ADCs and DACs, as discussed in Remark 1. The improvement in ratio will depend on the relative amount of PC compared to interference and noise, which depends on $M$ as we see next. 

\begin{theorem}
\label{Cor_MRTlimit}
The ergodic achievable downlink rates  under MRT precoding converge as $R^{\rm m}_{jk}\xrightarrow[M\rightarrow \infty]{}R^{\rm m, \infty}_{jk}=\log_2\left(1+\frac{1}{\overline{\text{PC}}^{\rm m}_{jk}}\right)$ for all three settings  in Theorems \ref{Th_MRT},   \ref{Cor_MRT} and   \ref{Cor_MRTconv}. Similarly under ZF precoding, they converge as $R^{\rm z}_{jk}\xrightarrow[M\rightarrow \infty]{}R^{\rm z, \infty}_{jk}=\log_2\left(1+\frac{1}{\overline{\text{PC}}^{\rm z}_{jk}}\right)$ for all three settings  in Theorems \ref{Th_ZF},  \ref{Cor_ZF} and  \ref{Cor_ZFconv}. Here $\overline{\text{PC}}^{\rm m}_{jk}=\sum_{l\neq j} \frac{\bar{c}_{j}c_{llk}^{2}  \beta_{ljk}^2 }{\bar{c}_{l} c_{jjk}^{2} \beta_{llk}^2}$ and $\overline{\text{PC}}^{\rm z}_{jk}=\sum_{l\neq j}^L \frac{\bar{\zeta}_j \beta_{ljk}^2}{\bar{\zeta}_l  \beta_{llk}^2}$ are the average PC to average desired signal power ratios under MRT and ZF precoding respectively, where $c_{jjk}=\frac{\beta_{jjk}^2}{\sum_{l=1}^L K \rho_p \beta_{jlk}+1}$, $\bar{c}_{j}=\sum_{k=1}^K c_{jjk}$, and  $\bar{\zeta}_{j}=\sum_{k=1}^K \frac{1}{c_{jjk}}$.
\end{theorem} \vspace{.04in}
\begin{IEEEproof}
The proof follows by letting $M\rightarrow \infty$ in the ergodic achievable rate expressions in Theorems \ref{Th_MRT},   \ref{Cor_MRT} and   \ref{Cor_MRTconv} for MRT, and in Theorems \ref{Th_ZF},  \ref{Cor_ZF} and  \ref{Cor_ZFconv} for ZF. 
\end{IEEEproof}

We see that the effects of channel gain uncertainty, quantization noise, thermal noise, and inter-user interference vanish as $M\rightarrow \infty$ under both precoders, while interference due to PC remains the only performance limitation with the average PC to average desired signal power ratio being the same for one-bit, mixed and conventional architectures. Theorem \ref{Cor_MRTlimit} therefore implies that using a larger number of antennas equipped with one-bit ADCs and DACs can compensate for the effect of quantization noise introduced by them even under simple linear precoders. Therefore one-bit quantized linear precoding schemes can ultimately approach the performance these schemes achieve \hspace{-.02in}in \hspace{-.02in}conventional \hspace{-.02in} \hspace{-.02in}MIMO \hspace{-.02in}systems, with $\frac{\gamma_{jk}^{\rm m,1}}{\gamma_{jk}^{\rm m,conv}} $, $\frac{\gamma_{jk}^{\rm m,1}}{\gamma_{jk}^{\rm m,mix}}$, $\frac{\gamma_{jk}^{\rm z,1}}{\gamma_{jk}^{\rm z,conv}}$   and $\frac{\gamma_{jk}^{\rm z,1}}{\gamma_{jk}^{\rm z,mix}} $ approaching $1$ as $M\rightarrow \infty$.

\section{Performance Analysis}

 In this section, we utilize the derived achievable rate expressions to yield several performance insights.

\subsection{Power Efficiency}

First, we study the power efficiency achieved by  the one-bit massive MIMO cellular system, defined as the decrease in transmit power that can be achieved with an increase in the number of antennas to maintain a given asymptotic sum average rate \cite{lit9,lit10}. In this context, we  consider two cases as follows. 

\subsubsection{Case I} In the first case, we assume that the training SNR $\rho_p=\frac{P_p}{\sigma^2}$ is fixed and independent of $M$, while the transmit power at each BS is given by $P_t = \frac{E_t}{M^c}$ for a given $c$, where $E_t$ is independent of $M$. We want to find the largest value of $c$ such that decreasing the transmit power proportionally to $\frac{1}{M^c}$ maintains a given  sum average rate as $M$ grows large. Substituting  $P_t = \frac{E_t}{M^c}$ into the SQINR expressions under MRT and ZF precoding in \eqref{SQINR_MRT} and \eqref{SQINR_ZF} respectively and assuming $M\rightarrow \infty$, we can readily see that we should choose $c = 1$. This implies that when the users' training powers are fixed, the transmit power at each BS  can be reduced proportionally to $1/M$ such that we achieve the following asymptotic sum average rates under MRT and ZF precoding respectively. 
\begin{align}
\label{limMRTPE}
&\underset{M\rightarrow \infty}{\text{lim}} R_{\rm sum}^{\rm m,1}|_{P_t=\frac{E_t}{M}}=\sum_{j=1}^L \sum_{k=1}^K \log_2\left(1+\tilde{\gamma}^{\rm m,1}_{jk} \right) \\
\label{limZFPE}
&\underset{M\rightarrow \infty}{\text{lim}} R_{\rm sum}^{\rm z,1}|_{P_t=\frac{E_t}{M}}=\sum_{j=1}^L \sum_{k=1}^K \log_2\left(1+\tilde{\gamma}^{\rm z,1}_{jk} \right)
\end{align}
where $\tilde{\gamma}^{\rm m,1}_{jk}= \frac{1}{\sum_{l\neq j}^L \frac{\bar{t}_{j} t_{llk}^2 \beta_{ljk}^2}{\bar{t}_{l} t_{jjk}^2 \beta_{llk}^2}+\frac{\pi \sigma^2 \bar{t}_{j}}{2 E_t t_{jjk}^2}}$ and $\tilde{\gamma}^{\rm z,1}_{jk}= \frac{1}{\sum_{l\neq j}^L \frac{\zeta_j \beta_{ljk}^2}{\zeta_l  \beta_{llk}^2}+\frac{\pi \sigma^2 K \zeta_j}{2E_t }}$. Note that the terms $\overline{\text{QN}}_{jk}$ and $\overline{\text{IUI}}_{jk}$  in \eqref{SQINR_MRT} and \eqref{SQINR_ZF}  decrease proportionally to $\frac{1}{M}$ as $M\rightarrow \infty$, while $\overline{\text{PC}}_{jk}$ an $\overline{\text{TN}}_{jk}$ remain the limiting factors in this case. 

For the mixed and conventional architectures,  the transmit power at each BS can also be reduced proportionally to $1/M$ while maintaining a given  sum average rate as $M$ grows large. The asymptotic limits are larger than those for the one-bit MIMO system and are given for the mixed case as $\underset{M\rightarrow \infty}{\text{lim}} R_{\rm sum}^{\rm m, mix}|_{P_t=\frac{E_t}{M}}=\sum_{j=1}^L \sum_{k=1}^K  \log_2\left(1+ \tilde{\gamma}_{jk}^{\rm m, mix} \right) $ and $\underset{M\rightarrow \infty}{\text{lim}} R_{\rm sum}^{\rm z, mix}|_{P_t=\frac{E_t}{M}}=\sum_{j=1}^L \sum_{k=1}^K \log_2\left(1+ \tilde{\gamma}_{jk}^{\rm z, mix} \right)$, where $\tilde{\gamma}_{jk}^{\rm m, mix}  = \frac{1}{\sum_{l\neq j}^L \frac{\bar{t}^{\rm FR}_{j} t^{\rm FR^2}_{llk} \beta_{ljk}^2}{\bar{t}^{\rm FR}_{l} t_{jjk}^{\rm FR^2} \beta_{llk}^2}+\frac{ \pi \sigma^2 \bar{t}^{\rm FR}_{j}}{2 E_t t_{jjk}^{\rm FR^2}}}$ and $\tilde{\gamma}_{jk}^{\rm z, mix}\hspace{-.1in}= \frac{1}{\sum_{l\neq j}^L \frac{\zeta_j^{\rm FR} \beta_{ljk}^2}{\zeta_l^{\rm FR}  \beta_{llk}^2}+\frac{\pi \sigma^2 K \zeta_j^{\rm FR}}{2 E_t }}$. The asymptotic limits for the conventional system are given as $\underset{M\rightarrow \infty}{\text{lim}} R_{\rm sum}^{\rm m, conv}|_{P_t=\frac{E_t}{M}}=\sum_{j=1}^L \sum_{k=1}^K  \log_2\left(1+ \tilde{\gamma}_{jk}^{\rm m, conv} \right) $ and $\underset{M\rightarrow \infty}{\text{lim}} R_{\rm sum}^{\rm z, conv}|_{P_t=\frac{E_t}{M}}=\sum_{j=1}^L \sum_{k=1}^K \log_2\left(1+ \tilde{\gamma}_{jk}^{\rm z, conv} \right)$, where $\tilde{\gamma}_{jk}^{\rm m, conv}  = \\ \frac{1}{\sum_{l\neq j}^L \frac{\bar{t}^{\rm FR}_{j} t^{\rm FR^2}_{llk} \beta_{ljk}^2}{\bar{t}^{\rm FR}_{l} t_{jjk}^{\rm FR^2} \beta_{llk}^2}+\frac{ \sigma^2 \bar{t}^{\rm FR}_{j}}{E_t t_{jjk}^{\rm FR^2}}}$ and $\tilde{\gamma}_{jk}^{\rm z, conv}\hspace{-.1in}= \frac{1}{\sum_{l\neq j}^L \frac{\zeta_j^{\rm FR} \beta_{ljk}^2}{\zeta_l^{\rm FR}  \beta_{llk}^2}+\frac{ \sigma^2 K \zeta_j^{\rm FR}}{E_t }}$.

Therefore all three implementations achieve the same power efficiency, i.e. the same order of reduction in $P_t$ can be achieved by increasing $M$ while maintaining given sum average rates in their respective cases. The ultimately achievable sum average rates in Case I are smaller for the one-bit setting than the mixed and conventional settings due to the quantization of the uplink training  and downlink transmit signals.

\subsubsection{Case II} Next we consider that the training power at the users and transmit  power at the BSs are reduced at the same rate, i.e. $P_p= \frac{E_p}{M^c}$ and $P_t=\frac{E_t}{M^c}$. Substituting these values into the SQINR expressions under MRT and ZF precoding in \eqref{SQINR_MRT} and \eqref{SQINR_ZF} and assuming $M\rightarrow \infty$, the value of $c = 1/2$ can be seen to  result in the sum rate converging to a fixed value. Therefore the training and  transmit powers together cannot be reduced as aggressively as the transmit power alone in Case I where the accuracy of channel estimates was fixed. The asymptotic sum average rates as we decrease $P_p$ and $P_t$ proportionally to $\frac{1}{\sqrt{M}}$  are given under one-bit quantized MRT and ZF precoders  as
\begin{align}
\label{limMRTPEII}
&\underset{M\rightarrow \infty}{\text{lim}} R_{\rm sum}^{\rm m,1}|_{P_t=\frac{E_t}{\sqrt{M}}, P_p=\frac{E_p}{\sqrt{M}}}=\sum_{j=1}^L \sum_{k=1}^K \log_2(1+ \bar{\gamma}_{jk}^{\rm m,1}) \\ 
\label{limZFPEII}
&\underset{M\rightarrow \infty}{\text{lim}} R_{\rm sum}^{\rm z,1}|_{P_t=\frac{E_t}{\sqrt{M}}, P_p=\frac{E_p}{\sqrt{M}}}=\sum_{j=1}^L \sum_{k=1}^K \log_2(1+ \bar{\gamma}_{jk}^{\rm z,1})
\end{align}
where $\bar{\gamma}_{jk}^{\rm m,1}=\frac{1}{\sum_{l\neq j}^L \frac{ \beta_{llk}^4 \beta_{ljk}^2 \bar{\beta}_j}{\bar{\beta}_{l} \beta_{jjk}^4  \beta_{llk}^2}+\frac{\pi^2 \sigma^4 \bar{\beta}_{j}}{4 E_t E_p \beta_{jjk}^4 K}} $, $\bar{\gamma}_{jk}^{\rm z,1}=\frac{1}{\sum_{l\neq j}^L \frac{\bar{\zeta}_j \beta_{ljk}^2}{\bar{\zeta}_l  \beta_{llk}^2}+\frac{\pi^2 \sigma^4 \bar{\zeta}_j}{4E_t E_p K }} $, $\bar{\beta}_j=\sum_{k=1}^K \beta_{jjk}^2$ and $\bar{\zeta}_j=\sum_{k=1}^K \frac{1}{\beta_{jjk}^2}$. While the terms $\overline{\text{QN}}_{jk}$ and $\overline{\text{IUI}}_{jk}$  in \eqref{SQINR_MRT} and \eqref{SQINR_ZF} decrease to zero as $M\rightarrow \infty$ in this case as well, the decrease is proportional to $\frac{1}{\sqrt{M}}$  instead of $\frac{1}{M}$ (since $t_{jjk}$ in \eqref{TT} behaves as $\frac{2 E_p K}{\pi \sigma^2 \sqrt{M}}\beta_{jjk}^2 $ as $M$ grows large in this case). Therefore the convergence of the rates to the derived asymptotic limits  in Case II will be slower than that in Case I.  For the mixed and conventional architectures,  the transmit  and training power at each BS and user respectively can also be reduced proportionally to $1/\sqrt{M}$ under both  precoders but the achieved asymptotic limits are larger than those for the one-bit MIMO system and are given as $\underset{M\rightarrow \infty}{\text{lim}} R_{\rm sum}^{\rm m, mix}|_{P_t=\frac{E_t}{\sqrt{M}}, P_p=\frac{E_p}{\sqrt{M}}}=\sum_{j=1}^L \sum_{k=1}^K \log_2(1+ \bar{\gamma}_{jk}^{\rm m,mix} )$ and $\underset{M\rightarrow \infty}{\text{lim}} R_{\rm sum}^{\rm  z, mix}|_{P_t=\frac{E_t}{\sqrt{M}}, P_p=\frac{E_p}{\sqrt{M}}}=\sum_{j=1}^L \sum_{k=1}^K \log_2(1+ \bar{\gamma}_{jk}^{\rm z, mix})$, where $\bar{\gamma}_{jk}^{\rm m,mix}= \frac{1}{\sum_{l\neq j}^L \frac{ \beta_{llk}^4 \beta_{ljk}^2 \bar{\beta}_j}{\bar{\beta}_{l} \beta_{jjk}^4  \beta_{llk}^2}+\frac{\pi \sigma^4 \bar{\beta}_{j}}{2 E_t E_p  \beta_{jjk}^4 K}}$ and $\bar{\gamma}_{jk}^{\rm z, mix}= \frac{1}{\sum_{l\neq j}^L \frac{\bar{\zeta}_j \beta_{ljk}^2}{\bar{\zeta}_l  \beta_{llk}^2}+\frac{\pi \sigma^4 \bar{\zeta}_j}{2 E_t E_p  K }} $.  For the conventional  system,  the asymptotic limits are  given as $\underset{M\rightarrow \infty}{\text{lim}} R_{\rm sum}^{\rm m, conv}|_{P_t=\frac{E_t}{\sqrt{M}}, P_p=\frac{E_p}{\sqrt{M}}}=\sum_{j=1}^L \sum_{k=1}^K \log_2(1+ \bar{\gamma}_{jk}^{\rm m,conv} )$ and $\underset{M\rightarrow \infty}{\text{lim}} R_{\rm sum}^{\rm  z, conv}|_{P_t=\frac{E_t}{\sqrt{M}}, P_p=\frac{E_p}{\sqrt{M}}}=\sum_{j=1}^L \sum_{k=1}^K \log_2(1+ \bar{\gamma}_{jk}^{\rm z, conv})$, where $\bar{\gamma}_{jk}^{\rm m,conv}= \frac{1}{\sum_{l\neq j}^L \frac{ \beta_{llk}^4 \beta_{ljk}^2 \bar{\beta}_j}{\bar{\beta}_{l} \beta_{jjk}^4  \beta_{llk}^2}+\frac{\sigma^4 \bar{\beta}_{j}}{E_t E_p \beta_{jjk}^4 K}}$ and $\bar{\gamma}_{jk}^{\rm z, conv}= \frac{1}{\sum_{l\neq j}^L \frac{\bar{\zeta}_j \beta_{ljk}^2}{\bar{\zeta}_l  \beta_{llk}^2}+\frac{\sigma^4 \bar{\zeta}_j}{E_t E_p  K }} $. Thus, \hspace{-.02in} one-bit  \hspace{-.02in} massive  \hspace{-.02in}  MIMO  \hspace{-.02in}  inherits   \hspace{-.02in} the power efficiency of conventional massive MIMO systems, since we can reduce the transmit and training powers with the same factor in $M$.

\subsection{Additional Antennas  Needed by One-Bit Massive MIMO}

 Denote the number of antennas  and  sum  rate in the one-bit, mixed, and conventional settings as $(M^{\rm 1}$, $R_{\rm sum}^{\rm 1})$, $(M^{\rm mix}$, $R_{\rm sum}^{\rm mix})$ and $(M^{\rm conv}$, $R_{\rm sum}^{\rm  conv})$ respectively. We want to find the ratios $\kappa=\frac{M^{\rm 1}}{M^{\rm conv}}$ and $\tilde{\kappa}=\frac{M^{\rm mix}}{M^{\rm conv}}$ required for the one-bit and mixed architectures  to  achieve the same sum-rate as the conventional architecture employing $M^{\rm conv}$ antennas as formulated next. \vspace{-.05in}
 \begin{subequations}
 \begin{alignat}{2} \textit{(P1)} \hspace{.15in}
&\!\min_{\kappa, \tilde{\kappa}}  \hspace{.05in}   \kappa, \tilde{\kappa} \label{P1}\\
&\text{s.t.} &  \hspace{-.15in}|R_{\rm sum}^{\rm 1}(\kappa M^{\rm conv})-R_{\rm sum}^{\rm conv}(M^{\rm conv})| \leq \epsilon \label{constraint7} \\
& &  \hspace{-.15in} |R_{\rm sum}^{\rm mix}(\tilde{\kappa} M^{\rm conv})-R_{\rm sum}^{\rm conv}(M^{\rm conv})| \leq \epsilon \label{constraint6}
\end{alignat}
\end{subequations}  
 Since \textit{(P1)} has two independent single parameters, we utilize two independent searches over $ [1, \dots, \kappa_{\rm max}]$ to numerically find  $\kappa$ and $\tilde{\kappa}$ that guarantee  \eqref{constraint7}  and \eqref{constraint6} respectively for a given $\epsilon$ under both precoders. In the corollaries that follow, we obtain  optimal values of   $\kappa^{\rm m}$ and $\tilde{\kappa}^{\rm m}$ under MRT  at any SNR, and optimal values of   $\kappa^{\rm z}$ and $\tilde{\kappa}^{\rm z}$  under ZF at low SNR.

\begin{corollary}\label{Corspec1}
The  ratios $\kappa^{\rm m}$ and $\tilde{\kappa}^{\rm m}$ required for the  one-bit and mixed massive MIMO systems respectively to achieve the sum average rate of the conventional  system that employs $M^{\rm conv}$ antennas at each BS   under MRT precoding are given as 
\begin{align}
\label{MRTsp1}
& \kappa^{\rm m}=\frac{M^{\rm m, 1}}{M^{\rm m, conv}}=\frac{\pi^2}{4}, \hspace{.15in} \tilde{\kappa}^{\rm m}=\frac{M^{\rm m, mix}}{M^{\rm m, conv}}=\frac{\pi}{2}.
\end{align}
\end{corollary}
\begin{IEEEproof}
Proof follows by finding $M^{\rm one}$ and $M^{\rm mix}$ in terms of $M^{\rm conv}$ to achieve $\gamma_{jk}^{\rm m, 1}=\gamma_{jk}^{\rm m, conv}$ and $\gamma_{jk}^{\rm m, mix}=\gamma_{jk}^{\rm m, conv}$.
\end{IEEEproof}

Therefore   $\pi^2/4=2.5\times$ more antennas are needed by the one-bit architecture with one-bit ADCs and DACs, while $\pi/2=1.57\times$ more antennas are needed by the mixed architecture with FR ADCs and one-bit DACs, to perform as well as the conventional system with FR ADCs and DACs  under MRT. The result also implies that $1.57\times$ more antennas are needed by the one-bit architecture to perform as well as the mixed architecture. The numbers  align with the discussion  in Sec. IV-C on the loss in SINR caused by the use of one-bit ADCs and DACs under MRT precoding. Next we present the ratios under ZF precoding  for  low SNR values.

\begin{corollary}\label{Corspec2}
At low SNR, i.e. small values of $\frac{P_t}{\sigma^2}$, the  ratios    $\kappa^{\rm z}$ and $\tilde{\kappa}^{\rm z}$  under ZF precoding are computed as
\begin{align}
\label{ZFsp1}
& \kappa^{\rm z}=\frac{M^{\rm z, 1}}{M^{\rm z, conv}}\approx\frac{\pi^2}{4}, \hspace{.1in} \tilde{\kappa}^{\rm z}=\frac{M^{\rm z, mix}}{M^{\rm z, conv}}\approx\frac{\pi}{2}.
\end{align}
\end{corollary}
\begin{IEEEproof}
The proof follows by simplifying the SQINR expressions under ZF in Sec. IV-D in the limit where $\frac{P_t}{\sigma^2}$ decreases, and solving for $\gamma_{jk}^{\rm z,1}=\gamma_{jk}^{\rm z, conv}$ and $\gamma_{jk}^{\rm z,mix}=\gamma_{jk}^{\rm z, conv}$.
\end{IEEEproof}

 While we can not obtain simple, analytical solutions for  $\kappa^{\rm z}$ and $\tilde{\kappa}^{\rm z}$  in the general SNR regime, we can infer from our discussion in Sec. IV-D that as the SNR increases  $\kappa^{\rm z}>2.5$ and $\tilde{\kappa}^{\rm z}>1.57$. This is because as the SNR increases,  the average  interference power to desired signal power ratio  becomes dominant  in the SQINR expressions, and is seen to increase by a factor of $\pi/2$ under MRT while it is increased by a factor $>\pi/2$ under ZF (seen by comparing the expressions of $\overline{\text{IUI}}_{jk}$ in Theorem \ref{Th_MRT} and Theorem \ref{Cor_MRTconv} for MRT,  and Theorem \ref{Th_ZF} and Theorem \ref{Cor_ZFconv} for ZF). Therefore, while  ${\kappa}^{\rm m}=2.5$ is enough to compensate for the decrease in sum rate caused by the use of one-bit ADCs and DACs instead of FR ADCs and DACs under MRT precoding,  ${\kappa}^{\rm z}>2.5$  will be required to compensate for the decrease in the sum  rate in the one-bit setting under ZF precoding as the SNR increases. Using a similar comparison  between MRT and ZF of the sum rate loss caused by using  FR ADCs and one-bit DACs instead of FR ADCs and DACs, we can see that    $\tilde{\kappa}^{\rm z}$ should be $>1.57$   as the SNR increases.
 
  Theoretically as $M^{\rm conv}$ increases to very large numbers,  $\kappa$ and $\tilde{\kappa}$  decrease to one under both precoders as outlined next.
\begin{remark}
As $M^{\rm m, conv}(M^{\rm z, conv})\rightarrow \infty$,  $\kappa^{\rm m} (\kappa^{\rm z})\rightarrow 1$ and $\tilde{\kappa}^{\rm m} (\tilde{\kappa}^{\rm z})\rightarrow 1$   for a given $\epsilon$. This is because    $R_{\rm  sum}^{\rm m, 1}$, $R_{\rm  sum}^{\rm m, mix}$ and $R_{\rm sum}^{\rm m, conv}$ all converge to $\sum_{l=1}^L \sum_{j=1}^K R^{\rm m, \infty}_{jk}$, and  $R_{\rm sum}^{\rm z, 1}$, $R_{\rm sum}^{\rm z, mix}$  and $R_{\rm sum}^{\rm z, conv}$ all converge to $\sum_{l=1}^L \sum_{j=1}^K R^{\rm z, \infty}_{jk}$ with the average PC to desired signal power ratio becoming the identical limiting factor in all cases as  shown in Theorem \ref{Cor_MRTlimit}. 
\end{remark}

\subsection{Energy Efficiency }

Since  one-bit and mixed massive MIMO cellular systems need a higher number of antennas at each BS to achieve the same sum average rate as the conventional system, it is interesting to study if we gain in terms of energy efficiency (EE) when we use one-bit ADCs and DACs. To this end, we define EE as  
\begin{align}
\label{EE}
&{\rm EE}=\frac{R_{\rm sum}}{L P_{\rm tot}} \hspace{.07in} \rm{(bits/Hz/Joule)}
\end{align}
 where $P_{\rm tot}=\frac{1}{\zeta}P_t+M(2P_{\rm ADC}+2P_{\rm DAC}+P_{\rm RF})$ is the  power consumption of each BS,  $\zeta $ is the PA efficiency, and $P_{\rm ADC}$, $P_{\rm DAC}$ and $P_{\rm RF}$ are the power consumptions of each ADC, DAC and RF chain. The latter is given as $P_{\rm RF}=P_{\rm TF}+P_{\rm LPF}+P_{\rm LNA}+2P_{\rm LO}+2P_{\rm M}$, with $P_{\rm TF}$, $P_{\rm LPF}$, $P_{\rm LNA}$, $P_{\rm LO}$ and $P_{\rm M}$ representing the power consumption of the transmit side filter, low pass filter, low noise amplifier, local oscillator, and mixer respectively. The power consumption of a $b$-bit data converter  is given as $P_{\rm ADC}=P_{\rm DAC}=cf_s 2^b$, where $c$ is the energy consumption per conversion step per Hz and $f_s$ is the sampling frequency.

Under the constraint of achieving the same sum average rate under one-bit and conventional massive MIMO settings, which can be met by finding $M^{\rm 1}=\kappa M^{\rm conv}$ by solving \textit{(P1)}, the EE for the two architectures are given as ${\rm EE}^{\rm 1}=\frac{R_{\rm sum}^{\rm 1}(\kappa M^{\rm conv})}{\frac{1}{\zeta}P_t+\kappa M^{\rm conv} (8 c f_s+P_{\rm RF})}$ and ${\rm EE}^{\rm conv}=\frac{R_{\rm sum}^{\rm conv}(M^{\rm conv})}{\frac{1}{\zeta}P_t+ M^{\rm conv} (2^{b+2} c f_s+P_{\rm RF})}$. By comparing ${\rm EE}^{\rm 1}$ and ${\rm EE}^{\rm conv}$, we can see that the decrease in the EE  with $f_s$ will be  less for  the one-bit massive MIMO system than the conventional  system that utilizes high resolution data converters (large $b$) and therefore consumes excessive amount of power as $f_s$ increases. The range of sampling frequencies where the one-bit massive MIMO system outperforms the conventional massive MIMO system in terms of EE while achieving the same sum-rate is computed to be 
\begin{align}
\label{f1}
&f_s\geq f_s^*=\frac{P_{RF} (\kappa-1)}{2^{b+2}c-8\kappa c}.
\end{align}
The value of $f_s^*$ increases with $\kappa$ and $P_{\rm RF}$, and decreases with $b$ since the power consumption of the conventional system quickly exceeds that of the one-bit system for larger values of $b$. Therefore, one-bit massive MIMO systems promise to yield  EE gains over conventional systems particularly at mmWave frequencies, while achieving the same sum-rate with higher $\kappa$. 

Next we compare the EE of the mixed and conventional massive MIMO systems under the constraint of achieving the same sum-rate by setting $M^{\rm mix}=\tilde{\kappa} M^{\rm conv}$. The EE for the mixed case is given as ${\rm EE}^{\rm mix}=\frac{R_{\rm sum}^{\rm mix}(\tilde{\kappa} M^{\rm conv})}{\frac{1}{\zeta}P_t+\tilde{\kappa} M^{\rm conv} (4 c f_s+2^{b+1}cf_s+P_{\rm RF})}$. By comparing ${\rm EE}^{\rm mix}$ and ${\rm EE}^{\rm conv}$, we can see that the range of sampling frequencies where the mixed architecture outperforms  the conventional architecture in terms of EE  is given as
\begin{align}
\label{f2}
&f_s\geq \tilde{f}_s^*=\frac{P_{RF} (\tilde{\kappa}-1)}{2^{b+2}c-4c\tilde{\kappa} -2^{b+1}c\tilde{\kappa}}.
\end{align}
Comparing \eqref{f1} and \eqref{f2}, we  observe that the range of frequencies where the one-bit architecture outperforms the conventional one is  larger than that where the mixed architecture outperforms the conventional one, i.e. $\tilde{f}_s^*> f_s^*$. This is because the mixed architecture uses high resolution ADCs resulting in a significantly larger power consumption than the one-bit architecture that uses one-bit ADCs and DACs.

Further we can see that ${\rm EE}^{\rm one}\geq {\rm EE}^{\rm mix}$ with both architectures achieving the same sum average rate when
\begin{align}
&f_s\geq \bar{f}_s^*=\frac{P_{RF} (\kappa-\tilde{\kappa})}{4c\tilde{\kappa}+2^{b+1}c \tilde{\kappa}-8c{\kappa}}.
\end{align}
Therefore the higher the resolution $b$ of the FR ADCs used in the mixed architecture, the larger will be the range of sampling frequencies where one-bit architecture outperforms the mixed architecture in terms of EE. Overall we observe in the simulations that the EE gains from using one-bit ADCs and DACs (one-bit architecture)  are more significant than the EE gains from using one-bit DACs and FR ADCs (mixed architecture) when compared to the conventional architecture. 

\section{Simulation Results}
\label{Sec:Sim}

\begin{figure}[!t]
\centering
\tikzset{every picture/.style={scale=.95}, every node/.style={scale=.8}}
%
%
\definecolor{mycolor1}{rgb}{0.00000,0.49804,0.00000}%
\definecolor{mycolor2}{rgb}{0.74902,0.00000,0.74902}%
\begin{tikzpicture}

\begin{axis}[%
width=.75\columnwidth,
height=.45\columnwidth,
scale only axis,
xmin=-20,
xmax=20,
xlabel style={font=\color{white!15!black}},
xlabel={$\rho_{p}\text{ (dB)}$},
ymin=0,
ymax=1.25,
ylabel style={font=\color{white!15!black}},
ylabel={Average NMSE per user, $\Gamma$},
axis background/.style={fill=white},
xmajorgrids,
ymajorgrids,
legend style={at={(axis cs: 20,1.25)},anchor=north east,legend cell align=left,align=left,draw=white!15!black, /tikz/column 2/.style={
                column sep=5pt,
            }},]
						
						\addplot [color=blue, line width=1.0pt, mark=o, mark size=2.0pt,mark options={solid, blue}]
  table[row sep=crcr]{%
-20	0.96963301424143\\
-14.7712125471966	0.912594628800961\\
-10	0.810653899471184\\
-4.77121254719662	0.677935621024206\\
0	0.595919602300733\\
5.22878745280338	0.555994746163245\\
10	0.543042923696357\\
15.2287874528034	0.538322954997411\\
20	0.53695578602953\\
};
\addlegendentry{\footnotesize 1-bit ADCs with PC (Lemma \ref{L1})}

\addplot [color=red, line width=1.0pt, mark size=2.0pt,mark=diamond, mark options={solid, red}]
  table[row sep=crcr]{%
-20	0.891783093143359\\
-14.7712125471966	0.741720028960221\\
-10	0.572244982346192\\
-4.77121254719662	0.444720062262007\\
0	0.393018019642083\\
5.22878745280338	0.372571080938977\\
10	0.366473618168704\\
15.2287874528034	0.364311412087369\\
20	0.36369092542274\\
};
\addlegendentry{\footnotesize 1-bit ADCs w/o PC (Lemma \ref{NoPCest})}

\addplot [color=mycolor1, line width=1.0pt, mark size=2.0pt,mark=square, mark options={solid, mycolor1}]
  table[row sep=crcr]{%
-20	0.952299650314605\\
-14.7712125471966	0.862703963978406\\
-10	0.702575840796399\\
-4.77121254719662	0.494102456513343\\
0	0.36527199556417\\
5.22878745280338	0.302558178195589\\
10	0.282213503039303\\
15.2287874528034	0.274799393544412\\
20	0.272651849551556\\
};
\addlegendentry{\footnotesize FR ADCs with PC (Lemma \ref{Cor1})}

\addplot [color=black, line width=1.0pt, mark size=2.0pt,mark=triangle, mark options={solid, rotate=270, black}]
  table[row sep=crcr]{%
-20	0.830013280212483\\
-14.7712125471966	0.594294770206022\\
-10	0.328083989501312\\
-4.77121254719662	0.127768313458262\\
0	0.0465549348230913\\
5.22878745280338	0.014436958614052\\
10	0.00485908649173947\\
15.2287874528034	0.00146270112140424\\
20	0.000488042947779414\\
};
\addlegendentry{\footnotesize FR ADCs w/o PC (Lemma \ref{Cor1noPC})}

\addplot [color=mycolor2, line width=1.0pt, mark size=2.0pt, mark=pentagon, mark options={solid, mycolor2}]
  table[row sep=crcr]{%
-20	0.636619772367582\\
-14.7712125471966	0.63661977236758\\
-10	0.636619772367581\\
-4.77121254719662	0.636619772367582\\
0	0.636619772367581\\
5.22878745280338	0.636619772367581\\
10	0.636619772367581\\
15.2287874528034	0.636619772367582\\
20	0.636619772367581\\
};
\addlegendentry{\footnotesize $\frac{\Gamma-\Gamma^{\rm s}}{\Gamma^{\rm FR}-\Gamma^{\rm FR^s}}$ (Remark 1)}

\end{axis}
\end{tikzpicture}%
\caption{Average NMSE per user in channel estimates versus effective training SNR for  $M=128$ and $K=8$. }
\label{PC}
\end{figure}
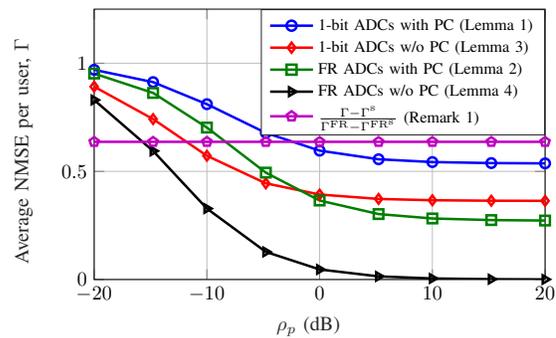

\begin{figure*}[!t]
\begin{minipage}[b]{0.47\linewidth}
\centering
\tikzset{every picture/.style={scale=.95}, every node/.style={scale=.8}}
%
%
\definecolor{mycolor1}{rgb}{0.74902,0.00000,0.74902}%
\definecolor{mycolor2}{rgb}{0.60000,0.20000,0.00000}%
\begin{tikzpicture}

\begin{axis}[%
width=.75\textwidth,
height=.48\textwidth,
scale only axis,
xmin=-30,
xmax=20,
xlabel style={font=\color{white!15!black}},
xlabel={Average transmit power $P_t$ (dB)},
ymin=0,
ymax=3.2,
ylabel style={font=\color{white!15!black}},
ylabel={Sum average rate per user (bps/Hz)},
axis background/.style={fill=white},
title style={font=\bfseries},
xmajorgrids,
ymajorgrids,
legend style={at={(axis cs: -30,3.2)},anchor=north west,legend cell align=left,align=left,draw=white!15!black, /tikz/column 2/.style={
                column sep=5pt,
            }},]
\addplot [color=blue, line width=1.0pt]
  table[row sep=crcr]{%
-30	0.0433678169648711\\
-28	0.0675211402022201\\
-26	0.104129684874094\\
-24	0.158359527844887\\
-22	0.236124176425752\\
-20	0.342834582069627\\
-18	0.481229417974631\\
-16	0.648900269672157\\
-14	0.836919256900611\\
-12	1.03094590682807\\
-10	1.21489201790045\\
-8	1.37549998401932\\
-6	1.5055067662819\\
-4	1.60406506193439\\
-2	1.67490444319304\\
0	1.72378657480917\\
2	1.75653418561876\\
4	1.77802614275782\\
6	1.79193697690371\\
8	1.80085901684704\\
10	1.80654755984772\\
12	1.8101606935468\\
14	1.81245003890659\\
16	1.81389836901441\\
18	1.8148137427041\\
20	1.81539191875771\\
};
\addlegendentry{\footnotesize One-bit (Th)}

\addplot [color=blue,only marks, mark size=2.0pt,draw=none, mark=o, mark options={solid, blue}]
  table[row sep=crcr]{%
-30	0.0433800252725379\\
-28	0.0675904903421044\\
-26	0.104064396152569\\
-24	0.158610621160517\\
-22	0.236719822227944\\
-20	0.343211251128056\\
-18	0.482725158594781\\
-16	0.651573204123188\\
-14	0.841928687730361\\
-12	1.03728635905962\\
-10	1.22487197606724\\
-8	1.38925349644017\\
-6	1.52111115760368\\
-4	1.62035562921496\\
-2	1.6944306249897\\
0	1.74335057964578\\
2	1.77820599135869\\
4	1.80098733001936\\
6	1.8137085306212\\
8	1.82051405941598\\
10	1.82959925969083\\
12	1.83285033827971\\
14	1.83505504571413\\
16	1.83686108083049\\
18	1.83679933362522\\
20	1.83734001973278\\
};
\addlegendentry{\footnotesize One-bit (MC)}

\addplot [color=mycolor1, line width=1.0pt]
  table[row sep=crcr]{%
-30	0.067309936149904\\
-28	0.104129417474258\\
-26	0.159093143081765\\
-24	0.23879365187851\\
-22	0.349880548218522\\
-20	0.497010980518855\\
-18	0.680174679167289\\
-16	0.89261041564424\\
-14	1.12081073896636\\
-12	1.34724233852543\\
-10	1.55485079993387\\
-8	1.73133312798141\\
-6	1.87132559154271\\
-4	1.97591342092794\\
-2	2.05032640699122\\
0	2.10132344729147\\
2	2.13533346053637\\
4	2.15758824212804\\
6	2.17196552555003\\
8	2.18117556274183\\
10	2.18704320138559\\
12	2.19076826698033\\
14	2.19312780363774\\
16	2.19462024730051\\
18	2.19556338497167\\
20	2.19615905101423\\
};
\addlegendentry{\footnotesize Mixed (Th)}

\addplot [color=mycolor1, only marks, mark size=2.0pt,draw=none, mark=o, mark options={solid, mycolor1}]
  table[row sep=crcr]{%
-30	0.0673864660296317\\
-28	0.104334244676397\\
-26	0.158845389746758\\
-24	0.238708217728668\\
-22	0.350307252078027\\
-20	0.49585867230634\\
-18	0.679501985244605\\
-16	0.891976517550051\\
-14	1.12091659303774\\
-12	1.34758130896752\\
-10	1.55482669242328\\
-8	1.73307230351082\\
-6	1.87180441180702\\
-4	1.97576292317157\\
-2	2.0506719298346\\
0	2.10154278719649\\
2	2.13663094621105\\
4	2.16001742974412\\
6	2.17386790660181\\
8	2.17872540503752\\
10	2.18666967979336\\
12	2.19177103009337\\
14	2.19386074720996\\
16	2.19489251699243\\
18	2.19456551046357\\
20	2.19506437397369\\
};
\addlegendentry{\footnotesize  Mixed (MC)}

\addplot [color=mycolor2, line width=1.0pt]
  table[row sep=crcr]{%
-30	0.103807910638797\\
-28	0.159087455512097\\
-26	0.239841243710154\\
-24	0.353579941420184\\
-22	0.506389625794695\\
-20	0.700237100125701\\
-18	0.930520538393386\\
-16	1.18533019857965\\
-14	1.44727868478323\\
-12	1.69741623723102\\
-10	1.91966206903734\\
-8	2.10404052947097\\
-6	2.2476895775566\\
-4	2.35364888757978\\
-2	2.42838000115062\\
0	2.47929531702329\\
2	2.51312002835044\\
4	2.53519821099477\\
6	2.54943847748107\\
8	2.55855137621001\\
10	2.56435333813622\\
12	2.56803518391719\\
14	2.57036673303698\\
16	2.57184122980177\\
18	2.57277292854045\\
20	2.57336133109697\\
};
\addlegendentry{\footnotesize  Conv. (Th)}

\addplot [color=mycolor2,only marks, mark size=2.0pt,draw=none, mark=o, mark options={solid, mycolor2}]
  table[row sep=crcr]{%
-30	0.1038532443115\\
-28	0.15922423815661\\
-26	0.239652428864463\\
-24	0.353557661817578\\
-22	0.506655633536521\\
-20	0.699298146025283\\
-18	0.930002567635843\\
-16	1.18448784274392\\
-14	1.44718935626388\\
-12	1.69757145522592\\
-10	1.91988224136579\\
-8	2.10560438875123\\
-6	2.24878224363555\\
-4	2.35388039776068\\
-2	2.42876856137803\\
0	2.48036700895365\\
2	2.51502450091711\\
4	2.53834539784019\\
6	2.55137724580914\\
8	2.55431267469307\\
10	2.56387404978834\\
12	2.56923330455543\\
14	2.57151679701556\\
16	2.5726106229746\\
18	2.57128331681088\\
20	2.57173137977191\\
};
\addlegendentry{\footnotesize Conv. (MC)}

\addplot [color=blue, dashdotted, line width=1.0pt]
  table[row sep=crcr]{%
-30	0.0352692817160206\\
-28	0.0551511742568257\\
-26	0.085606272187149\\
-24	0.131418197959933\\
-22	0.198527717833769\\
-20	0.293237647103531\\
-18	0.420386732964486\\
-16	0.580651000607481\\
-14	0.76813317880086\\
-12	0.970067615542588\\
-10	1.16963685651553\\
-8	1.35081765281272\\
-6	1.50272639622622\\
-4	1.62140676668208\\
-2	1.70880342107343\\
0	1.77023219394524\\
2	1.81193560735665\\
4	1.83955799740095\\
6	1.85754750937176\\
8	1.86913241152392\\
10	1.87653819128311\\
12	1.88124998616377\\
14	1.88423867656125\\
16	1.88613073278842\\
18	1.88732706840574\\
20	1.88808291513206\\
};

\addplot [color=blue, only marks, draw=none,mark size=2.0pt, mark=o, mark options={solid, blue}]
  table[row sep=crcr]{%
-30	0.0381287680159101\\
-28	0.0596173269948434\\
-26	0.0921946494694098\\
-24	0.141480143705633\\
-22	0.213399627773323\\
-20	0.313151243994724\\
-18	0.44737142731428\\
-16	0.613961641071849\\
-14	0.807877192607773\\
-12	1.01229382450792\\
-10	1.21404225453194\\
-8	1.39537144171217\\
-6	1.54439195900023\\
-4	1.6599429553865\\
-2	1.74691481697337\\
0	1.80526525310327\\
2	1.84641609273656\\
4	1.87465579121903\\
6	1.8907588676146\\
8	1.89888000815338\\
10	1.91050738970733\\
12	1.91430964241425\\
14	1.91792105733523\\
16	1.9194750710876\\
18	1.92004802358675\\
20	1.91941047624574\\
};

\addplot [color=mycolor1, dashdotted, line width=1.0pt]
  table[row sep=crcr]{%
-30	0.0549442295998095\\
-28	0.0855284995397771\\
-26	0.131862192269235\\
-24	0.200450787529932\\
-22	0.29871274192457\\
-20	0.433397987001321\\
-18	0.607907393196857\\
-16	0.819363273674851\\
-14	1.05717383235785\\
-12	1.30450451394517\\
-10	1.54235261881889\\
-8	1.7543581315709\\
-6	1.93031165547377\\
-4	2.0672171510028\\
-2	2.16799483759896\\
0	2.23892154288838\\
2	2.28715789041945\\
4	2.31915871893974\\
6	2.34002584303542\\
8	2.35347592859362\\
10	2.3620793141691\\
12	2.36755528627894\\
14	2.37102961159384\\
16	2.3732294834432\\
18	2.37462059996621\\
20	2.37549957022693\\
};

\addplot [color=mycolor1, only marks, mark size=2.0pt,draw=none, mark=o, mark options={solid, mycolor1}]
  table[row sep=crcr]{%
-30	0.0549234847078713\\
-28	0.0855516510610757\\
-26	0.131830843934082\\
-24	0.200402254532956\\
-22	0.298741185823839\\
-20	0.433158119652814\\
-18	0.607610931947614\\
-16	0.819056188007073\\
-14	1.05715303381106\\
-12	1.30484086795884\\
-10	1.54256505977483\\
-8	1.75384766471323\\
-6	1.93036532269081\\
-4	2.06653478334717\\
-2	2.1681448534696\\
0	2.23922139431095\\
2	2.28684905059092\\
4	2.32106233067393\\
6	2.34019457813581\\
8	2.35137890755279\\
10	2.36196190512156\\
12	2.36815777699437\\
14	2.37201858954778\\
16	2.3728254472501\\
18	2.374429243497\\
20	2.37404944509179\\
};

\addplot [color=mycolor2, dashdotted, line width=1.0pt]
  table[row sep=crcr]{%
-30	0.0906500012644358\\
-28	0.139955539681511\\
-26	0.213207524078701\\
-24	0.318723916423908\\
-22	0.464526668739237\\
-20	0.655677456668826\\
-18	0.891222698656348\\
-16	1.1624169182376\\
-14	1.45360645153059\\
-12	1.74553216605704\\
-10	2.01946571304948\\
-8	2.26068113397951\\
-6	2.46053002354825\\
-4	2.61688597579504\\
-2	2.7330440669709\\
0	2.81560978814506\\
2	2.87225754147569\\
4	2.91009997615053\\
6	2.93490139158285\\
8	2.95094350184844\\
10	2.96122903037418\\
12	2.96778577096354\\
14	2.97194995846774\\
16	2.97458832247353\\
18	2.97625740457618\\
20	2.97731227876703\\
};

\addplot [color=mycolor2,only marks, mark size=2.0pt,draw=none, mark=o, mark options={solid, mycolor2}]
  table[row sep=crcr]{%
-30	0.0906912603632858\\
-28	0.14019624013304\\
-26	0.212989874824767\\
-24	0.31849621301806\\
-22	0.464889599923225\\
-20	0.65484646227134\\
-18	0.890903172827071\\
-16	1.16158454374944\\
-14	1.4537098817616\\
-12	1.74610596708291\\
-10	2.01948646807897\\
-8	2.26103298607379\\
-6	2.46062252639827\\
-4	2.61563370523232\\
-2	2.73335338435083\\
0	2.81560132536764\\
2	2.87213109076312\\
4	2.9136235067431\\
6	2.93545373078329\\
8	2.94838020470645\\
10	2.96002204497697\\
12	2.96863686784796\\
14	2.97331011556066\\
16	2.97381293407554\\
18	2.97552337254033\\
20	2.97472244282256\\
};

\node at (axis cs: 20,0.4) [anchor = east] {\small Solid lines: MRT precoding};
\node at (axis cs: 20.8,0.17) [anchor = east] {\small Dashdotted lines: ZF precoding};

\end{axis}
\end{tikzpicture}%
\caption{Sum average rate per user versus $P_t$ for $L=4$, $M=128$ and $K=8$. }
\label{Fig2}
\end{minipage}
\hspace{.4cm}
\begin{minipage}[b]{0.47\linewidth}
\centering
\tikzset{every picture/.style={scale=.95}, every node/.style={scale=.8}}
%
%
\definecolor{mycolor1}{rgb}{0.74902,0.00000,0.74902}%
\definecolor{mycolor2}{rgb}{0.60000,0.20000,0.00000}%
\begin{tikzpicture}

\begin{axis}[%
width=.75\textwidth,
height=.48\textwidth,
scale only axis,
xmin=-30,
xmax=20,
xlabel style={font=\color{white!15!black}},
xlabel={$\text{Average transmit power P}_\text{t}\text{ (dB)}$},
ymin=0,
ymax=6.3,
ylabel style={font=\color{white!15!black}},
ylabel={Sum average rate per user (bps/Hz)},
axis background/.style={fill=white},
xmajorgrids,
ymajorgrids,
legend style={at={(axis cs: -30,6.3)},anchor=north west,legend cell align=left,align=left,draw=white!15!black, /tikz/column 2/.style={
                column sep=5pt,
            }},]
\addplot [color=blue, line width=1.0pt]
  table[row sep=crcr]{%
-30	0.0578635011059908\\
-28	0.0901963882290773\\
-26	0.139336504313492\\
-24	0.212401481156782\\
-22	0.317669943468434\\
-20	0.462897913763628\\
-18	0.652245068253943\\
-16	0.882636096777166\\
-14	1.14167666470588\\
-12	1.40924875671227\\
-10	1.66283295150456\\
-8	1.88401161472722\\
-6	2.06282003607646\\
-4	2.19821163825755\\
-2	2.29542843733682\\
0	2.36246165672432\\
2	2.40734519079798\\
4	2.43679107263932\\
6	2.45584551574972\\
8	2.46806459368169\\
10	2.47585447726979\\
12	2.48080197487705\\
14	2.48393666651353\\
16	2.48591974371296\\
18	2.48717306737479\\
20	2.48796469388628\\
};
\addlegendentry{\footnotesize One-bit (Th)}

\addplot [color=blue, only marks, mark size=2.0pt,draw=none, mark=o, mark options={solid, blue}]
  table[row sep=crcr]{%
-30	0.0578858101067664\\
-28	0.0902808565387627\\
-26	0.139299846051786\\
-24	0.212642505066262\\
-22	0.318387146802733\\
-20	0.463706199229508\\
-18	0.654250661571313\\
-16	0.886762378469965\\
-14	1.14913832587372\\
-12	1.42129031289778\\
-10	1.68012519046049\\
-8	1.90767556138584\\
-6	2.09012189115755\\
-4	2.23082554925551\\
-2	2.33116335004887\\
0	2.40116287530358\\
2	2.45022558588523\\
4	2.48170799542661\\
6	2.49909533314799\\
8	2.50985588863083\\
10	2.52073827721087\\
12	2.52661973406848\\
14	2.52889315251682\\
16	2.53117176249941\\
18	2.52935143404994\\
20	2.53353342920487\\
};
\addlegendentry{\footnotesize One-bit (MC)}

\addplot [color=mycolor1, line width=1.0pt]
  table[row sep=crcr]{%
-30	0.0898802841003823\\
-28	0.139259965900924\\
-26	0.213224392576495\\
-24	0.320948669648445\\
-22	0.471874195557694\\
-20	0.672863597498728\\
-18	0.92436429305231\\
-16	1.21736730530142\\
-14	1.53337273735738\\
-12	1.84815771934403\\
-10	2.13793219607231\\
-8	2.38522043018493\\
-6	2.5820506068112\\
-4	2.72950394161387\\
-2	2.83462754407956\\
0	2.90677344896701\\
2	2.95493371852006\\
4	2.9864677277593\\
6	3.0068480136573\\
8	3.0199069818944\\
10	3.02822813683983\\
12	3.03351137773492\\
14	3.03685812197029\\
16	3.03897508018105\\
18	3.04031291086329\\
20	3.04115787116798\\
};
\addlegendentry{\footnotesize Mixed (Th)}

\addplot [color=mycolor1, only marks, mark size=2.0pt,draw=none, mark=o, mark options={solid, mycolor1}]
  table[row sep=crcr]{%
-30	0.089893014712439\\
-28	0.139415845569443\\
-26	0.212845063127991\\
-24	0.320856803839823\\
-22	0.472181402153254\\
-20	0.671667615522986\\
-18	0.923235337027262\\
-16	1.21657161569979\\
-14	1.53326322699894\\
-12	1.84828047965084\\
-10	2.13862469569027\\
-8	2.38624340315676\\
-6	2.58210959491438\\
-4	2.72951752397163\\
-2	2.83434961909894\\
0	2.90583646835261\\
2	2.95564852500441\\
4	2.98959156027763\\
6	3.00825345771755\\
8	3.0177428446306\\
10	3.02754690544517\\
12	3.035181361281\\
14	3.03739232568873\\
16	3.0394748754982\\
18	3.03693529169702\\
20	3.04222273846071\\
};
\addlegendentry{\footnotesize Mixed (MC)}

\addplot [color=mycolor2, line width=1.0pt]
  table[row sep=crcr]{%
-30	0.138780018442435\\
-28	0.213110233625131\\
-26	0.322147772737986\\
-24	0.476501529462342\\
-22	0.685045289982478\\
-20	0.951142615482608\\
-18	1.26919742691673\\
-16	1.62361661607988\\
-14	1.99116510004676\\
-12	2.34586218377398\\
-10	2.66465252224981\\
-8	2.93207827387057\\
-6	3.14243683232294\\
-4	3.29878057588022\\
-2	3.40966227593629\\
0	3.48550245500519\\
2	3.53601852139281\\
4	3.56904897140272\\
6	3.59037747605203\\
8	3.60403634227196\\
10	3.61273665134626\\
12	3.61825937929137\\
14	3.62175733207921\\
16	3.62396973725342\\
18	3.62536780715647\\
20	3.62625078231465\\
};
\addlegendentry{\footnotesize Conv. (Th)}

\addplot [color=mycolor2, only marks, mark size=2.0pt,draw=none, mark=o, mark options={solid, mycolor2}]
  table[row sep=crcr]{%
-30	0.138786208433029\\
-28	0.213230951008771\\
-26	0.321868863641447\\
-24	0.476431633004406\\
-22	0.685288592850792\\
-20	0.950448851937676\\
-18	1.26845242159945\\
-16	1.62286880271677\\
-14	1.99109120522399\\
-12	2.34550182034058\\
-10	2.66588605083135\\
-8	2.93342628493511\\
-6	3.14295659406391\\
-4	3.30008319988613\\
-2	3.40922379146163\\
0	3.48436523172758\\
2	3.5373520334208\\
4	3.57363833406988\\
6	3.5920829456266\\
8	3.59991402373774\\
10	3.61170999870262\\
12	3.62065807676203\\
14	3.62278077204332\\
16	3.6252778177301\\
18	3.61998173439577\\
20	3.62757153654367\\
};
\addlegendentry{\footnotesize Conv. (MC)}

\addplot [color=blue, dashdotted, line width=1.0pt]
  table[row sep=crcr]{%
-30	0.051117485384265\\
-28	0.0799078439952462\\
-26	0.123975224906441\\
-24	0.190191538930863\\
-22	0.287052741637774\\
-20	0.423524453291197\\
-18	0.606473062587488\\
-16	0.836984329596974\\
-14	1.10718669372184\\
-12	1.3998840267325\\
-10	1.6920931708642\\
-8	1.96115822812568\\
-6	2.19056202546087\\
-4	2.37291381872084\\
-2	2.50935236727473\\
0	2.60653202428447\\
2	2.67318328629863\\
4	2.71765722553435\\
6	2.74676992212117\\
8	2.76558225460507\\
10	2.77763530610219\\
12	2.78531500343688\\
14	2.7901907775768\\
16	2.79327932879448\\
18	2.79523293813797\\
20	2.7964675267896\\
};

\addplot [color=blue, only marks, mark size=2.0pt,draw=none, mark=o, mark options={solid, blue}]
  table[row sep=crcr]{%
-30	0.0552503385024064\\
-28	0.0862935656644238\\
-26	0.133509477600471\\
-24	0.204617290877722\\
-22	0.308172061392931\\
-20	0.452250647147537\\
-18	0.644526490757112\\
-16	0.884669126703944\\
-14	1.16344081287723\\
-12	1.46200955705105\\
-10	1.75747236044259\\
-8	2.0275420276696\\
-6	2.25330291727869\\
-4	2.43345006459074\\
-2	2.56805164135989\\
0	2.6610973707694\\
2	2.72855133827714\\
4	2.7729145527775\\
6	2.80013222516199\\
8	2.81733403448496\\
10	2.82733397234741\\
12	2.83720524416756\\
14	2.84137728018893\\
16	2.84486675599882\\
18	2.84570407164733\\
20	2.8473947096655\\
};

\addplot [color=mycolor1, dashdotted, line width=1.0pt]
  table[row sep=crcr]{%
-30	0.0796095099333697\\
-28	0.123867392460271\\
-26	0.190849699858794\\
-24	0.289884104491198\\
-22	0.431593996065891\\
-20	0.625737218650913\\
-18	0.877679061029226\\
-16	1.18469105562716\\
-14	1.53414941549859\\
-12	1.9050137182025\\
-10	2.27209752838642\\
-8	2.61141148632173\\
-6	2.90486563609158\\
-4	3.14306023369832\\
-2	3.32546778101765\\
0	3.45825993500664\\
2	3.55101696797433\\
4	3.61378446896439\\
6	3.65528911899005\\
8	3.68229613120301\\
10	3.6996800602085\\
12	3.7107900940767\\
14	3.71785762624224\\
16	3.72234017806566\\
18	3.72517781104063\\
20	3.72697197311345\\
};

\addplot [color=mycolor1, only marks, mark size=2.0pt,draw=none, mark=o, mark options={solid, mycolor1}]
  table[row sep=crcr]{%
-30	0.0796044071936076\\
-28	0.12387570192381\\
-26	0.190850593976204\\
-24	0.289889334462446\\
-22	0.43155929889569\\
-20	0.625792077586189\\
-18	0.877684790911813\\
-16	1.18469118953498\\
-14	1.5341157434481\\
-12	1.90514424859443\\
-10	2.27219170961094\\
-8	2.61180308342298\\
-6	2.90526011344625\\
-4	3.14345083902576\\
-2	3.32620204592661\\
0	3.45729183195881\\
2	3.55189312052094\\
4	3.61450657094581\\
6	3.65540761384563\\
8	3.68207361396457\\
10	3.69953265766878\\
12	3.71073889937816\\
14	3.71794504204291\\
16	3.72370857995035\\
18	3.72523546113576\\
20	3.72704953758317\\
};

\addplot [color=mycolor2, dashdotted, line width=1.0pt]
  table[row sep=crcr]{%
-30	0.131278726228986\\
-28	0.202555856820808\\
-26	0.308337007424166\\
-24	0.460591059324653\\
-22	0.671083164450516\\
-20	0.9480727123444\\
-18	1.29278903392389\\
-16	1.69758816280943\\
-14	2.14689486732588\\
-12	2.62018326658354\\
-10	3.09535595549701\\
-8	3.5515246805824\\
-6	3.97106873701828\\
-4	4.34103848877153\\
-2	4.65384797122118\\
0	4.90727347361552\\
2	5.103904464206\\
4	5.25012663979104\\
6	5.35466054015674\\
8	5.42689176284686\\
10	5.47546001754317\\
12	5.50745616502685\\
14	5.52823024862595\\
16	5.54158453442334\\
18	5.55011239837258\\
20	5.55553464079426\\
};

\addplot [color=mycolor2, only marks, mark size=2.0pt,draw=none, mark=o, mark options={solid, mycolor2}]
  table[row sep=crcr]{%
-30	0.13128391601019\\
-28	0.20265858327388\\
-26	0.308066688831027\\
-24	0.460507454868237\\
-22	0.67138554724425\\
-20	0.947397071693028\\
-18	1.29195484970295\\
-16	1.69704907240453\\
-14	2.14691308411734\\
-12	2.62055995212991\\
-10	3.09572437157074\\
-8	3.55284027726887\\
-6	3.97140103779077\\
-4	4.3405482767521\\
-2	4.65525007370123\\
0	4.90502990011417\\
2	5.10677003852887\\
4	5.25318855195033\\
6	5.35565203853743\\
8	5.42786957635106\\
10	5.47489582306388\\
12	5.50621035054999\\
14	5.5272583462615\\
16	5.54460443370005\\
18	5.55019955811257\\
20	5.55498276515372\\
};

\node at (axis cs: 20,0.8) [anchor = east] {\small Solid lines: MRT precoding};
\node at (axis cs: 20.8,0.4) [anchor = east] {\small Dashdotted lines: ZF precoding};

\end{axis}
\end{tikzpicture}%
\caption{Sum average rate per user versus $P_t$ for $L=4$, $M=128$ and $K=8$ for special case without PC. }
\label{Fig2woPC}
\end{minipage}
\end{figure*}

We consider $L=4$ cells  with  Cartesian coordinates of the BSs set as $(0,0,0)$, $(525, 0, 0)$, $(0, 525, 0)$, and $(525, 525, 0)$ (all in metres). The BS in each cell has $M$ antennas serving $K$ users distributed uniformly on a circle  of radius $250$ metres around it \cite{massiveMIMOO}. Moreover  $\sigma^2=-80$\rm{dBm}, $\rho_p=\frac{1}{\sigma^2}$,  $\beta_{jlk}=\frac{10^{-3}}{d_{jlk}^\alpha}$,  $\alpha=3$, and $d_{jlk}$ is the distance between BS $j$ and user $k$ in cell $l$. Remaining parameters are stated under each figure. 

We first study the average NMSE per user in the channel estimates, defined as $\Gamma=\frac{1}{KL}\sum_{l=1}^L \sum_{j=1}^K \Gamma_{jk}$,  for the scenario where the BS has one-bit ADCs considering cases with PC (Lemma \ref{L1}) and without PC (Lemma \ref{NoPCest}), and for the scenario where the BS has FR ADCs under cases with PC (Lemma \ref{Cor1}) and without PC (Lemma \ref{Cor1noPC}). The results are plotted in Fig. \ref{PC} against the effective training SNR $\rho_p$, with the NMSE  seen to be  larger when we have one-bit ADCs instead of FR ADCs. As $\rho_p$ increases, the average NMSE for the conventional system with FR ADCs  goes to zero when there is no PC, and saturates at a non-zero value when there is PC. For the one-bit scenario, the NMSE saturates at  $1-\frac{2}{\pi}=0.36$ when there is no PC, with this number representing the effect of quantization noise on the NMSE, while it saturates at a larger value when there is PC. We also plot  $\frac{\Gamma-\Gamma^{\rm s}}{\Gamma^{\rm FR}-\Gamma^{\rm FR^s}}$, which represents the ratio of the impact of PC when the BS has one-bit ADCs to when the BS has FR ADCs, which is shown to be constant at $\frac{2}{\pi}\approx 0.64$. The reduced impact of PC on the NMSE in the one-bit case compared to the conventional case was explained in Remark 1 by noting that when the BS has one-bit ADCs,  the received training signals from the interfering users from other cells that are using the same pilot sequence are quantized reducing their impact on the derived estimates. 

\begin{figure*}[!t]
\begin{minipage}[b]{0.47\linewidth}
\centering
\tikzset{every picture/.style={scale=.95}, every node/.style={scale=.8}}
%
%
\definecolor{mycolor1}{rgb}{0.00000,0.44706,0.74118}%
\definecolor{mycolor2}{rgb}{0.00000,0.49804,0.00000}%

\begin{tikzpicture}

\begin{axis}[%
width=.75\textwidth,
height=.375\textwidth,
scale only axis,
xmin=-30,
xmax=20,
xlabel style={font=\color{white!15!black}},
xlabel={Average transmit power $P_t\text{ (dB)}$},
ymin=0.45,
ymax=0.72,
ylabel style={font=\color{white!15!black}},
ylabel={$\gamma^{\rm 1}/\gamma^{\rm mix}$},
axis background/.style={fill=white},
xmajorgrids,
ymajorgrids,
legend style={at={(axis cs: -30,0.45)},anchor=south west,legend cell align=left,align=left,draw=white!15!black, /tikz/column 2/.style={
                column sep=5pt,
            }},]

\addplot [color=mycolor1, line width=1.0pt,mark=diamond, mark size=2pt,mark options={solid, mycolor1}]
  table[row sep=crcr]{%
-30	0.638218485573124\\
-28	0.639117112136303\\
-26	0.640489503898759\\
-24	0.642543273863904\\
-22	0.645524850286427\\
-20	0.649667298284378\\
-18	0.655082743034638\\
-16	0.661622278010474\\
-14	0.668798620984428\\
-12	0.675888419024221\\
-10	0.682200537372388\\
-8	0.687321175433641\\
-6	0.691172035937946\\
-4	0.693907028223415\\
-2	0.695772209010899\\
0	0.697009640825632\\
2	0.697815856025602\\
4	0.698335014875618\\
6	0.698666836526625\\
8	0.698877917138889\\
10	0.699011788274212\\
12	0.699096530609836\\
14	0.699150109370943\\
16	0.699183959145633\\
18	0.699205334392684\\
20	0.699218828227318\\
};
\addlegendentry{\footnotesize MRT with PC}

\addplot [color=mycolor2, line width=1.0pt,mark=x, mark size=2pt, mark options={solid, mycolor2}]
  table[row sep=crcr]{%
-30	0.636619772367581\\
-28	0.636619772367581\\
-26	0.636619772367581\\
-24	0.636619772367581\\
-22	0.636619772367581\\
-20	0.636619772367581\\
-18	0.636619772367581\\
-16	0.636619772367581\\
-14	0.636619772367581\\
-12	0.636619772367581\\
-10	0.636619772367581\\
-8	0.636619772367581\\
-6	0.636619772367581\\
-4	0.636619772367581\\
-2	0.636619772367581\\
0	0.636619772367581\\
2	0.636619772367581\\
4	0.636619772367581\\
6	0.636619772367581\\
8	0.636619772367581\\
10	0.636619772367581\\
12	0.636619772367582\\
14	0.636619772367581\\
16	0.636619772367581\\
18	0.636619772367581\\
20	0.636619772367581\\
};
\addlegendentry{\footnotesize MRT without PC}

\addplot [color=mycolor1, line width=1.0pt,dashdotted, mark size=2pt, mark=diamond, mark options={solid, mycolor1}]
  table[row sep=crcr]{%
-30	0.637524953296386\\
-28	0.63801739441451\\
-26	0.638746295848444\\
-24	0.639783593106497\\
-22	0.641171160064582\\
-20	0.642852758544527\\
-18	0.644579186760425\\
-16	0.645846960934774\\
-14	0.64597779545003\\
-12	0.644382045035577\\
-10	0.640867721792948\\
-8	0.635776868580986\\
-6	0.629864409487521\\
-4	0.62399940298807\\
-2	0.618856632847527\\
0	0.614762563136868\\
2	0.611732839996355\\
4	0.609605983096897\\
6	0.608166645562751\\
8	0.607216320096697\\
10	0.60659898098185\\
12	0.606202159606748\\
14	0.605948807922603\\
16	0.605787752856561\\
18	0.605685651565715\\
20	0.605621036862009\\
};
\addlegendentry{\footnotesize ZF with PC}

\addplot [color=mycolor2, line width=1.0pt,dashdotted, mark=x,mark size=2pt, mark options={solid, mycolor2}]
  table[row sep=crcr]{%
-30	0.635746030179972\\
-28	0.635240112804354\\
-26	0.634445896800981\\
-24	0.6332058449967\\
-22	0.631285866720983\\
-20	0.628350956903316\\
-18	0.623949610096867\\
-16	0.617530009319466\\
-14	0.608523472004037\\
-12	0.596525336321681\\
-10	0.581554667088833\\
-8	0.564282068281606\\
-6	0.546043296152424\\
-4	0.528514742213966\\
-2	0.513162192433739\\
0	0.500800861265874\\
2	0.491522262575227\\
4	0.48492547207936\\
6	0.48041701806872\\
8	0.47741928181593\\
10	0.475462551988494\\
12	0.47420076799567\\
14	0.473393507138252\\
16	0.472879648447555\\
18	0.472553608687193\\
20	0.472347162918335\\
};
\addlegendentry{\footnotesize ZF without PC}

\end{axis}
\end{tikzpicture}%
\caption{Ratio of average SQINR per user in one-bit setting to average SQINR per user in mixed setting. }
\label{Fig2ratio2}
\end{minipage}
\hspace{.4cm}
\begin{minipage}[b]{0.47\linewidth}
\centering
\tikzset{every picture/.style={scale=.95}, every node/.style={scale=.8}}
%
%
\definecolor{mycolor1}{rgb}{0.00000,0.44700,0.74100}%
\definecolor{mycolor2}{rgb}{0.00000,0.49804,0.00000}%
\begin{tikzpicture}

\begin{axis}[%
width=.75\textwidth,
height=.375\textwidth,
scale only axis,
xmin=-30,
xmax=20,
xlabel style={font=\color{white!15!black}},
xlabel={Average transmit power $P_t\text{ (dB)}$},
ymin=0.09,
ymax=0.51,
ylabel style={font=\color{white!15!black}},
ylabel={$\gamma^{\rm 1}/\gamma^{\rm conv}$},
axis background/.style={fill=white},
xmajorgrids,
ymajorgrids,
legend style={at={(axis cs: -30,0.09)},anchor=south west,legend cell align=left,align=left,draw=white!15!black, /tikz/column 2/.style={
                column sep=5pt,
            }},]

\addplot [color=mycolor1, line width=1.0pt, mark size=2pt, mark=diamond, mark options={solid, mycolor1}]
  table[row sep=crcr]{%
-30	0.407894235254937\\
-28	0.409354937839627\\
-26	0.411577403561642\\
-24	0.414884763835682\\
-22	0.419647581210431\\
-20	0.426191261421631\\
-18	0.434622843556346\\
-16	0.444630582454623\\
-14	0.455412312628048\\
-12	0.465880521457817\\
-10	0.475068087907648\\
-8	0.482444776685341\\
-6	0.487955274424388\\
-4	0.49185344114075\\
-2	0.494505983674958\\
0	0.496263733204659\\
2	0.497408281399442\\
4	0.498145101596952\\
6	0.498615979261773\\
8	0.498915498040208\\
10	0.499105452336556\\
12	0.499225694208931\\
14	0.499301717089963\\
16	0.499349746297752\\
18	0.499380075405007\\
20	0.499399221631131\\
};
\addlegendentry{\footnotesize MRT with PC}

\addplot [color=mycolor2, line width=1.0pt, mark size=2pt, mark=x, mark options={solid, mycolor2}]
  table[row sep=crcr]{%
-30	0.405284734569351\\
-28	0.405284734569351\\
-26	0.405284734569351\\
-24	0.405284734569351\\
-22	0.405284734569351\\
-20	0.405284734569351\\
-18	0.405284734569351\\
-16	0.405284734569351\\
-14	0.405284734569351\\
-12	0.405284734569351\\
-10	0.405284734569351\\
-8	0.405284734569351\\
-6	0.405284734569351\\
-4	0.405284734569351\\
-2	0.405284734569351\\
0	0.405284734569351\\
2	0.405284734569351\\
4	0.405284734569351\\
6	0.405284734569351\\
8	0.405284734569351\\
10	0.405284734569351\\
12	0.405284734569351\\
14	0.405284734569351\\
16	0.405284734569351\\
18	0.405284734569351\\
20	0.405284734569351\\
};
\addlegendentry{\footnotesize MRT without PC}

\addplot [color=mycolor1, dashdotted, line width=1.0pt, mark size=2pt, mark=diamond, mark options={solid, mycolor1}]
  table[row sep=crcr]{%
-30	0.381618777374259\\
-28	0.382512993182327\\
-26	0.383821622201276\\
-24	0.385651850104332\\
-22	0.388036662707101\\
-20	0.390814239984796\\
-18	0.393489277125822\\
-16	0.395194391691319\\
-14	0.394889663334598\\
-12	0.391771118657325\\
-10	0.385644334953758\\
-8	0.377053720875053\\
-6	0.367136003891918\\
-4	0.357243604140763\\
-2	0.348486157439494\\
0	0.341446142631958\\
2	0.33619307892773\\
4	0.332482067246851\\
6	0.329959283497115\\
8	0.328288445491768\\
10	0.327200817485595\\
12	0.326500756836333\\
14	0.326053413471455\\
16	0.325768880551906\\
18	0.325588436439858\\
20	0.325474216970666\\
};
\addlegendentry{\footnotesize ZF with PC}

\addplot [color=mycolor2, dashdotted, line width=1.0pt, mark size=2pt, mark=x, mark options={solid, mycolor2}]
  table[row sep=crcr]{%
-30	0.378601649752098\\
-28	0.377818241615871\\
-26	0.376588239991045\\
-24	0.374667364143326\\
-22	0.371692264532957\\
-20	0.367142068758849\\
-18	0.360312600524726\\
-16	0.350337843625606\\
-14	0.336311398579221\\
-12	0.317550526625172\\
-10	0.29397022630615\\
-8	0.26640810106737\\
-6	0.236669216802024\\
-4	0.207155174345454\\
-2	0.180191149651779\\
0	0.157395506432035\\
2	0.13940834421026\\
4	0.126020247739698\\
6	0.11651346580399\\
8	0.110002499910307\\
10	0.105660075385847\\
12	0.102817633898141\\
14	0.100980662002033\\
16	0.0998035472442738\\
18	0.0990534484746433\\
20	0.0985771744417534\\
};
\addlegendentry{\footnotesize ZF without PC}

\end{axis}
\end{tikzpicture}%
\caption{Ratio of average SQINR per user in one-bit setting to average SINR per user in conventional setting. }
\label{Fig2ratio1}
\end{minipage}
\end{figure*}

We now validate the derived closed-form ergodic rate expressions in Fig. \ref{Fig2} and Fig. \ref{Fig2woPC}   considering cases with and without PC respectively, where we plot the   sum average rate per user given as $\frac{1}{KL}R_{\rm sum}$ versus $P_t$.  The theoretical (Th) results for MRT precoding are plotted  using   Theorem \ref{Th_MRT}, Theorem \ref{Cor_MRT}, and Theorem \ref{Cor_MRTconv} for the scenarios with one-bit ADCs and DACs (one-bit), FR ADCs and one-bit DACs (mixed), and FR ADCs and DACs (conventional) respectively. The theoretical results for ZF are plotted  using the expressions in  Theorem \ref{Th_ZF}, Theorem \ref{Cor_ZF}, and Theorem \ref{Cor_ZFconv} for the three scenarios just discussed.  The  theoretical expressions for the case without PC under one-bit, mixed and conventional implementations are given in Corollaries \ref{Th_MRTs},  \ref{Cor_MRTs}, and  \ref{Cor_MRTconvs} respectively for MRT, and in Corollaries \ref{Th_ZFs},  \ref{Cor_ZFs}, and  \ref{Cor_ZFconvs} respectively for ZF. The Monte-Carlo (MC) curves are plotted using the SQINR definition in \eqref{SQINR} to compute the rates  for all  scenarios. The match between the Monte-Carlo simulated results and the closed-form theoretical results is excellent, even for moderate system dimensions.

As expected, the sum average rate is lower when we use one-bit ADCs and DACs compared to when we use FR ADCs and/or DACs.  We also observe that the MRT precoder slightly outperforms the ZF precoder at low transmit power (or SNR) levels, i.e. in noise-limited scenarios, while  ZF  outperforms  MRT  at high transmit powers where interference is dominant \cite{emilref}. The performance improvement of ZF over MRT  at high SNR levels is more noticeable in conventional settings than in one-bit settings, since the one-bit quantization operation significantly affects the interference cancellation ability of ZF, reducing its gain over  MRT.  In fact, while using one-bit ADCs and DACs increases the  interference by a factor of $\frac{\pi}{2}$ under MRT, it is increased by a factor $\gg \frac{\pi}{2}$ under ZF as shown in Sec. IV-D. Therefore the use of one-bit ADCs and DACs has a more significant adverse impact on the performance of ZF  at high SNR. Moreover comparing the results in Fig. \ref{Fig2} and \ref{Fig2woPC},  the sum average rate is lower under PC for all the cases.

Next we study in Fig. \ref{Fig2ratio2} the ratio of average SQINR per user in the one-bit setting to that in the mixed setting, i.e. $\frac{\gamma^{\rm 1}}{\gamma^{\rm mix}}=\frac{\frac{1}{KL}\sum_{j=1}^L \sum_{k=1}^K\gamma_{jk}^{\rm 1}}{\frac{1}{KL}\sum_{j=1}^L \sum_{k=1}^K\gamma_{jk}^{\rm mix}}$ under MRT and ZF for the cases with and without PC. The ratio is $\frac{2}{\pi}=0.64$ for small  SNR values in all cases as predicted by our analysis. Under MRT, the ratio stays constant at $\frac{2}{\pi}$ for all SNR values when there is no PC, and improves  under PC  as characterized in \eqref{M1}. This is because PC has a smaller adverse impact on the performance of one-bit architecture than the mixed architecture as shown in Fig. \ref{PC}. Under ZF precoding without PC, the ratio decreases as the SNR increases as highlighted in \eqref{Z1} because the interference to desired signal power ratio becomes dominant in the denominator of the SQINR, and is significantly increased by the use of  one-bit ADCs as employed in the one-bit setting instead of FR ADCs as employed in the mixed setting. Finally even under ZF, the ratio is better when we have PC.

\begin{figure*}[!t]
\begin{minipage}[b]{0.48\linewidth}
\centering
\tikzset{every picture/.style={scale=.95}, every node/.style={scale=.8}}
%
%
\definecolor{mycolor1}{rgb}{0.00000,0.49804,0.00000}%
\definecolor{mycolor2}{rgb}{0.60000,0.20000,0.00000}%
\begin{tikzpicture}

\begin{axis}[%
width=.75\textwidth,
height=.5\textwidth,
scale only axis,
xmin=0,
xmax=800,
xlabel style={font=\color{white!15!black}},
xlabel={$M$},
ymin=1,
ymax=4.85,
ylabel style={font=\color{white!15!black}},
ylabel={Sum average rate per user (bps/Hz)},
axis background/.style={fill=white},
xmajorgrids,
ymajorgrids,
legend style={at={(axis cs: 800,1)},anchor=south east,legend cell align=left,align=left,draw=white!15!black, /tikz/column 2/.style={
                column sep=5pt,
            }},]
\addplot [color=blue, line width=0.8pt, mark=o, mark size=2.0pt,  mark options={solid, blue}]
  table[row sep=crcr]{%
50	1.08090977540877\\
100	1.60342239441729\\
200	2.18256774558941\\
300	2.52204886227306\\
400	2.75551188121175\\
600	3.06777931616999\\
800	3.2745900014559\\
};
\addlegendentry{\small One-bit}

\addplot [color=mycolor1, line width=0.8pt, mark=square,  mark size=2.0pt, mark options={solid, mycolor1}]
  table[row sep=crcr]{%
50	1.41167348374425\\
100	1.97836731299874\\
200	2.56000636001637\\
300	2.88440221817223\\
400	3.10181456803482\\
600	3.38629919000052\\
800	3.57100119375295\\
};
\addlegendentry{\small Mixed}

\addplot [color=red,  line width=0.8pt, mark=triangle,  mark size=2.0pt, mark options={solid, rotate=270, red}]
  table[row sep=crcr]{%
50	1.77588837577302\\
100	2.35944564437583\\
200	2.92004492836961\\
300	3.22007473748651\\
400	3.41689607672698\\
600	3.66954514950059\\
800	3.83048738675464\\
};
\addlegendentry{\small Conv.}

\addplot [color=mycolor2, line width=0.8pt, mark=diamond,  mark size=2.0pt, mark options={solid, mycolor2}]
  table[row sep=crcr]{%
50	4.74801258444337\\
100	4.74801258444337\\
200	4.74801258444337\\
300	4.74801258444337\\
400	4.74801258444337\\
600	4.74801258444337\\
800	4.74801258444337\\
};
\addlegendentry{\small $M\rightarrow\infty$}


\addplot [color=blue, dotted, line width=0.8pt, mark=o, mark size=2.0pt,  mark options={solid, blue}]
  table[row sep=crcr]{%
50	1.15260358760981\\
100	1.67066338747953\\
200	2.25374147938167\\
300	2.59356293782519\\
400	2.82018394723678\\
600	3.11834738187651\\
800	3.32784226873997\\
};


\addplot [color=mycolor1, dotted, line width=1.0pt, mark=square, mark size=2.0pt,  mark options={solid, mycolor1}]
  table[row sep=crcr]{%
50	1.47861093776894\\
100	2.04204637346917\\
200	2.61989022882755\\
300	2.94327518018042\\
400	3.15675891030774\\
600	3.42332427315047\\
800	3.60958828187423\\
};


\addplot [color=red, dotted, line width=1.0pt, mark=triangle,  mark size=2.0pt, mark options={solid, rotate=270, red}]
  table[row sep=crcr]{%
50	1.90029263927536\\
100	2.47229154486308\\
200	3.02339172374289\\
300	3.31962169599923\\
400	3.51289041797107\\
600	3.73774932226613\\
800	3.89698155860646\\
};

\node at (axis cs: 0,4.4) [anchor = west] {\small Solid lines: Hardening Bound \eqref{rate}};
\node at (axis cs: 0,4.1) [anchor = west] {\small Dotted lines: Perfect CE \eqref{rateperfect}};

\end{axis}
\end{tikzpicture}%
\caption{Sum average rate per user versus $M$, for $L=4$, $K=8$ and $P_t=10\rm{dB}$ under MRT precoding.}
\label{Fig3}
\end{minipage}
\hspace{.4cm}
\begin{minipage}[b]{0.48\linewidth}
\centering
\tikzset{every picture/.style={scale=.95}, every node/.style={scale=.8}}
%
%
\definecolor{mycolor1}{rgb}{0.00000,0.49804,0.00000}%
\definecolor{mycolor2}{rgb}{0.60000,0.20000,0.00000}%
\begin{tikzpicture}

\begin{axis}[%
width=.75\textwidth,
height=.5\textwidth,
scale only axis,
xmin=0,
xmax=800,
xlabel style={font=\color{white!15!black}},
xlabel={$M$},
ymin=1,
ymax=4.6,
ylabel style={font=\color{white!15!black}},
ylabel={Sum average rate per user (bps/Hz)},
axis background/.style={fill=white},
xmajorgrids,
ymajorgrids,
legend style={at={(axis cs: 800,1)},anchor=south east,legend cell align=left,align=left,draw=white!15!black, /tikz/column 2/.style={
                column sep=5pt,
            }},]
\addplot [color=blue, line width=0.8pt, mark=o, mark size=2.0pt, mark options={solid, blue}]
  table[row sep=crcr]{%
50	1.05411898247782\\
100	1.6598699852299\\
200	2.2576286058852\\
300	2.58442091135611\\
400	2.80179831974469\\
600	3.08502271532078\\
800	3.26856323619637\\
};
\addlegendentry{\small One-bit}

\addplot [color=mycolor1, line width=0.8pt, mark=square, mark size=2.0pt, mark options={solid, mycolor1}]
  table[row sep=crcr]{%
50	1.48988179972267\\
100	2.14631361131749\\
200	2.7240388397646\\
300	3.02004310449515\\
400	3.21084512769607\\
600	3.4528774937624\\
800	3.60575812249063\\
};
\addlegendentry{\small Mixed}

\addplot [color=red, line width=0.8pt, mark=triangle, mark size=2.0pt, mark options={solid, rotate=270, red}]
  table[row sep=crcr]{%
50	2.19635046526431\\
100	2.77933533589162\\
200	3.25813470983622\\
300	3.49392384636214\\
400	3.64237004460928\\
600	3.82571517755935\\
800	3.93781566387151\\
};
\addlegendentry{\small Conv.}

\addplot [color=mycolor2,  line width=0.8pt, mark=diamond,  mark size=2.0pt,mark options={solid, mycolor2}]
  table[row sep=crcr]{%
50	4.46712632793422\\
100	4.46712632793422\\
200	4.46712632793422\\
300	4.46712632793422\\
400	4.46712632793422\\
600	4.46712632793422\\
800	4.46712632793422\\
};
\addlegendentry{\small $M\rightarrow\infty$}


\addplot [color=blue, dotted, line width=0.8pt, mark=o, mark size=2.0pt, mark options={solid, blue}]
  table[row sep=crcr]{%
50	1.19774750427563\\
100	1.75601315942888\\
200	2.33384831648156\\
300	2.6508409199919\\
400	2.86279532396407\\
600	3.13109069380185\\
800	3.31683821829149\\
};


\addplot [color=mycolor1, dotted, line width=1.0pt, mark=square, mark size=2.0pt, mark options={solid, mycolor1}]
  table[row sep=crcr]{%
50	1.60235603179535\\
100	2.20280196761457\\
200	2.7585899087913\\
300	3.05333316674849\\
400	3.2398685553841\\
600	3.4760655360083\\
800	3.63102470906462\\
};


\addplot [color=red, dotted, line width=1.0pt, mark=triangle,  mark size=2.0pt,mark options={solid, rotate=270, red}]
  table[row sep=crcr]{%
50	2.31650378589836\\
100	2.88280527989552\\
200	3.34173301415772\\
300	3.57603077707463\\
400	3.71650925860154\\
600	3.88484996965168\\
800	3.99317314761513\\
};

\node at (axis cs: 0,4.25) [anchor = west] {\small Solid lines: Hardening Bound \eqref{rate}};
\node at (axis cs: 0,4) [anchor = west] {\small Dotted lines: Perfect CE \eqref{rateperfect}};

\end{axis}
\end{tikzpicture}%
\caption{Sum average rate per user versus $M$, for $L=4$, $K=8$ and $P_t=10\rm{dB}$ under ZF precoding.}
\label{Fig3b}
\end{minipage}
\end{figure*}

In Fig. \ref{Fig2ratio1}, we study the ratio of average SQINR  in the one-bit setting to the average SINR  in the conventional setting under both precoders. The ratio is around $\frac{4}{\pi^2}=0.4$ for small  SNR values in all cases as predicted by our analysis. Under MRT, the ratio stays constant at $\frac{4}{\pi^2}$  when there is no PC, and improves  when there is PC  as characterized in \eqref{M3}. Under ZF precoding without PC, the ratio decreases as the SNR increases as highlighted in \eqref{Z3} since the interference suppression capability of ZF is significantly impacted by the use of one-bit ADCs and DACs. The ratio does get better when we include PC even under ZF.

Next in Fig. \ref{Fig3} and \ref{Fig3b}, we plot the sum average rate per user against the number of antennas under MRT  and ZF precoding respectively, using the achievable rate expressions in:  (i)  \eqref{rate} derived exploiting channel hardening and (ii)  \eqref{rateperfect} for a genie-aided user that has perfect CSI. The first expression is used for theoretical analysis  in this work, while the latter is computed numerically as a benchmark. As expected, the performance gap between the considered achievable rate expression that assumes statistical CSI at the users and the achievable rate that assumes perfect  CSI at the users is very small, as Rayleigh fading channels considered in this work always harden \cite{No}, \cite[Fig. 2]{cff}.  We also plot on these figures  the  asymptotic sum average rate for the scenario where $M\rightarrow \infty$    using Theorem \ref{Cor_MRTlimit}. The  asymptotic performance  is the same under one-bit, mixed and conventional implementations for both precoders, because theoretically the impact of quantization noise due to the use of one-bit ADCs and DACs goes to zero as $M\rightarrow \infty$. Numerically we  observe that when $M=800$, MRT precoding under one-bit, mixed and conventional settings can achieve $70\%$, $75\%$ and $81\%$ of the asymptotic sum  rate, while ZF precoding under one-bit, mixed and  conventional  settings can achieve $73\%$, $81\%$ and $88\%$ of the asymptotic sum  rate.

Next Fig. \ref{Fig4} and Fig. \ref{Fig4b}  study the power efficiency of using larger antenna arrays in one-bit, mixed  and conventional  MIMO scenarios under Case I and Case II respectively, described in Sec. V-A. In Case I,  the training power at users is fixed at $P_p=1$W and the transmit power at the BSs scales as $P_t=\frac{E_t}{M}$, while in Case II we have $P_p=\frac{E_p}{\sqrt{M}} $ and $P_t=\frac{E_t}{\sqrt{M}}$, where $E_p=1$W and $E_t=10$W. Looking at Case I first, we see that even with scaling down the transmit power with $M$, the sum  rate increases and converges to the asymptotic limits derived in Sec. V-A, with the limits being lower for the one-bit architecture than those for the mixed and conventional architectures. Moreover as $P_t$ scales down with $M$, the  thermal noise term become dominant  in the derived SQINR expressions,  leading to MRT outperforming ZF.

Even in Case II, where we are scaling both the uplink training and downlink transmit powers proportionally to $1/\sqrt{M}$, the sum  rate  increases with $M$ and will eventually converge to constant values for both precoders. However, convergence in this case is much slower than that in Case I, because the terms $\overline{\text{QN}}_{jk}$ and $\overline{\text{IUI}}_{jk}$ in the SQINR expressions for all three architectures scale down proportionally to $1/\sqrt{M}$ instead of $1/M$ as discussed after \eqref{limZFPEII}. In this case ZF  is performing  better than MRT,  since the transmit power is not reduced as aggressively as it was in Case I. Overall one-bit massive MIMO  inherits the power efficiency of mixed and conventional massive MIMO architectures since we can decrease the transmit  and training powers by same factor in $M$ in all three architectures and reach given asymptotic limits.

\begin{figure*}[!t]
\begin{minipage}[b]{0.47\linewidth}
\centering
\tikzset{every picture/.style={scale=.95}, every node/.style={scale=.8}}
%
%
\definecolor{mycolor1}{rgb}{0.74902,0.00000,0.74902}%
\definecolor{mycolor2}{rgb}{0.00000,0.49804,0.00000}%
\definecolor{mycolor3}{rgb}{0.60000,0.20000,0.00000}%
\begin{tikzpicture}

\begin{axis}[%
width=.75\textwidth,
height=.55\textwidth,
scale only axis,
xmin=0,
xmax=800,
xlabel style={font=\color{white!15!black}},
xlabel={$M$},
ymin=0.35,
ymax=2.4,
ylabel style={font=\color{white!15!black}},
ylabel={Sum average rate per user (bps/Hz)},
axis background/.style={fill=white},
xmajorgrids,
ymajorgrids,
legend style={at={(axis cs: 800,0.35)},anchor=south east,legend cell align=left,align=left,draw=white!15!black, /tikz/column 2/.style={
                column sep=5pt,
            }},]
\addplot [color=blue, line width=1.0pt,  mark size=2.0pt,mark=diamond, mark options={solid, blue}]
  table[row sep=crcr]{%
50	0.806849660601885\\
100	1.04515040719689\\
200	1.23043303313452\\
300	1.30887766806399\\
400	1.3522817237412\\
600	1.39891475721047\\
800	1.4235654803201\\
};
\addlegendentry{\footnotesize MRT, One-bit}

\addplot [color=mycolor1, line width=1.0pt, mark=+, mark size=2.0pt, mark options={solid, mycolor1}]
  table[row sep=crcr]{%
50	1.08908509265636\\
100	1.3648823000195\\
200	1.56950290051118\\
300	1.65381638280358\\
400	1.69990876610896\\
600	1.74899912824154\\
800	1.77477181457328\\
};
\addlegendentry{\footnotesize MRT, Mixed}

\addplot [color=mycolor2, line width=1.0pt, mark=square, mark size=2.0pt, mark options={solid, mycolor2}]
  table[row sep=crcr]{%
50	1.41682498032699\\
100	1.71816976071957\\
200	1.93222180036496\\
300	2.01830863501063\\
400	2.06487880723226\\
600	2.1141067785598\\
800	2.13980126987487\\
};
\addlegendentry{\footnotesize MRT, Conv.}

\addplot [color=mycolor3, line width=1.0pt,  mark size=2.0pt,mark=o, mark options={solid, mycolor3}]
  table[row sep=crcr]{%
50	0.707197229569147\\
100	0.985037135051507\\
200	1.17926924303577\\
300	1.25641191050041\\
400	1.29788602303906\\
600	1.34152152087944\\
800	1.36421074878263\\
};
\addlegendentry{\footnotesize ZF, One-bit}

\addplot [color=black, line width=1.0pt, mark=triangle, mark size=2.0pt, mark options={solid, rotate=270, black}]
  table[row sep=crcr]{%
50	1.01919260273673\\
100	1.33411997281864\\
200	1.53377407773558\\
300	1.60900078949847\\
400	1.64855282271718\\
600	1.68951440616697\\
800	1.7105546311106\\
};
\addlegendentry{\footnotesize ZF, Mixed}

\addplot [color=red, line width=1.0pt, mark=x, mark size=2.0pt, mark options={solid, red}]
  table[row sep=crcr]{%
50	1.55928602441069\\
100	1.8137258263269\\
200	1.96008396942497\\
300	2.01253760870963\\
400	2.03954869476718\\
600	2.06711663680722\\
800	2.08111809520312\\
};
\addlegendentry{\footnotesize ZF, Conv.}

\addplot [color=blue, dashdotted, line width=1.0pt, mark size=2.0pt, mark=diamond, mark options={solid, blue}]
  table[row sep=crcr]{%
50	1.5035562801265\\
100	1.5035562801265\\
200	1.5035562801265\\
300	1.5035562801265\\
400	1.5035562801265\\
600	1.5035562801265\\
800	1.5035562801265\\
};

\addplot [color=mycolor1, dashdotted, line width=1.0pt, mark size=2.0pt, mark=+, mark options={solid, mycolor1}]
  table[row sep=crcr]{%
50	1.85757977574884\\
100	1.85757977574884\\
200	1.85757977574884\\
300	1.85757977574884\\
400	1.85757977574884\\
600	1.85757977574884\\
800	1.85757977574884\\
};

\addplot [color=mycolor2, dashdotted, line width=1.0pt,  mark size=2.0pt,mark=square, mark options={solid, mycolor2}]
  table[row sep=crcr]{%
50	2.2216732591953\\
100	2.2216732591953\\
200	2.2216732591953\\
300	2.2216732591953\\
400	2.2216732591953\\
600	2.2216732591953\\
800	2.2216732591953\\
};

\addplot [color=mycolor3, dashdotted, line width=1.0pt, mark size=2.0pt, mark=o, mark options={solid, mycolor3}]
  table[row sep=crcr]{%
50	1.43609858843493\\
100	1.43609858843493\\
200	1.43609858843493\\
300	1.43609858843493\\
400	1.43609858843493\\
600	1.43609858843493\\
800	1.43609858843493\\
};

\addplot [color=black, dashdotted, line width=1.0pt,  mark size=2.0pt,mark=triangle, mark options={solid, rotate=270, black}]
  table[row sep=crcr]{%
50	1.77607841627065\\
100	1.77607841627065\\
200	1.77607841627065\\
300	1.77607841627065\\
400	1.77607841627065\\
600	1.77607841627065\\
800	1.77607841627065\\
};

\addplot [color=red, dashdotted, line width=1.0pt, mark size=2.0pt, mark=x, mark options={solid, red}]
  table[row sep=crcr]{%
50	2.12403737798124\\
100	2.12403737798124\\
200	2.12403737798124\\
300	2.12403737798124\\
400	2.12403737798124\\
600	2.12403737798124\\
800	2.12403737798124\\
};

\node at (axis cs: 300,2.32) [anchor = west] {\small Dashdotted lines: Limit $M\rightarrow \infty$};

\end{axis}
\end{tikzpicture}%
\caption{Sum average rate for  $P_p=1$W and $P_t=\frac{E_t}{M}$ (Case I).}
\label{Fig4}
\end{minipage}
\hspace{.4cm}
\begin{minipage}[b]{0.47\linewidth}
\centering
\tikzset{every picture/.style={scale=.95}, every node/.style={scale=.8}}
%
%
\definecolor{mycolor1}{rgb}{0.74902,0.00000,0.74902}%
\definecolor{mycolor2}{rgb}{0.00000,0.49804,0.00000}%
\definecolor{mycolor3}{rgb}{0.60000,0.20000,0.00000}%
\begin{tikzpicture}

\begin{axis}[%
width=.75\textwidth,
height=.55\textwidth,
scale only axis,
xmin=0,
xmax=800,
xlabel style={font=\color{white!15!black}},
xlabel={$M$},
ymin=0.8,
ymax=4.4,
ylabel style={font=\color{white!15!black}},
ylabel={Sum average rate per user (bps/Hz)},
axis background/.style={fill=white},
xmajorgrids,
ymajorgrids,
legend style={at={(axis cs: 800,0.8)},anchor=south east,legend cell align=left,align=left,draw=white!15!black, /tikz/column 2/.style={
                column sep=5pt,
            }},]
\addplot [color=blue, line width=1.0pt, mark size=2.0pt,mark=diamond, mark options={solid, blue}]
  table[row sep=crcr]{%
50	0.981003791778951\\
100	1.42781870036569\\
200	1.91224746867473\\
300	2.19119641815019\\
400	2.38069311226279\\
600	2.63077980590417\\
800	2.7940033431467\\
};
\addlegendentry{\footnotesize MRT, One-bit}

\addplot [color=mycolor1, line width=1.0pt, mark=+, mark size=2.0pt,mark options={solid, mycolor1}]
  table[row sep=crcr]{%
50	1.29649237881861\\
100	1.79087302280416\\
200	2.28868582473827\\
300	2.56165069516265\\
400	2.74237498108245\\
600	2.97573306421725\\
800	3.12513359223534\\
};
\addlegendentry{\footnotesize MRT, Mixed}

\addplot [color=mycolor2, line width=1.0pt, mark=square,mark size=2.0pt, mark options={solid, mycolor2}]
  table[row sep=crcr]{%
50	1.6500232674971\\
100	2.16795640382013\\
200	2.65670381782015\\
300	2.9138946593169\\
400	3.08059363715439\\
600	3.29194985528359\\
800	3.42502533927541\\
};
\addlegendentry{\footnotesize MRT, Conv.}

\addplot [color=mycolor3, line width=1.0pt, mark=o, mark size=2.0pt,mark options={solid, mycolor3}]
  table[row sep=crcr]{%
50	0.922165848770752\\
100	1.44018301631503\\
200	1.94414830459776\\
300	2.24700498935814\\
400	2.42749439847371\\
600	2.66147633824889\\
800	2.82247946134109\\
};
\addlegendentry{\footnotesize ZF, One-bit}

\addplot [color=black, line width=1.0pt, mark=triangle,mark size=2.0pt, mark options={solid, rotate=270, black}]
  table[row sep=crcr]{%
50	1.32519601168927\\
100	1.89696867677915\\
200	2.39519733929661\\
300	2.64865731165673\\
400	2.81145326192179\\
600	3.0176384204876\\
800	3.14815063954992\\
};
\addlegendentry{\footnotesize ZF, Mixed}

\addplot [color=red, line width=1.0pt, mark=x,mark size=2.0pt, mark options={solid, red}]
  table[row sep=crcr]{%
50	1.96455883284148\\
100	2.47627867126999\\
200	2.89397365401247\\
300	3.09947364121402\\
400	3.22945414753408\\
600	3.39204908852951\\
800	3.49385384849505\\
};
\addlegendentry{\footnotesize ZF, Conv.}

\addplot [color=blue, dashdotted, line width=1.0pt, mark size=2.0pt,mark=diamond, mark options={solid, blue}]
  table[row sep=crcr]{%
50	4.05296499652377\\
100	4.05296499652377\\
200	4.05296499652377\\
300	4.05296499652377\\
400	4.05296499652377\\
600	4.05296499652377\\
800	4.05296499652377\\
};

\addplot [color=mycolor1, dashdotted, line width=1.0pt, mark size=2.0pt,mark=+, mark options={solid, mycolor1}]
  table[row sep=crcr]{%
50	4.17276146600506\\
100	4.17276146600506\\
200	4.17276146600506\\
300	4.17276146600506\\
400	4.17276146600506\\
600	4.17276146600506\\
800	4.17276146600506\\
};

\addplot [color=mycolor2, dashdotted, line width=1.0pt, mark size=2.0pt,mark=square, mark options={solid, mycolor2}]
  table[row sep=crcr]{%
50	4.26331747267177\\
100	4.26331747267177\\
200	4.26331747267177\\
300	4.26331747267177\\
400	4.26331747267177\\
600	4.26331747267177\\
800	4.26331747267177\\
};

\addplot [color=mycolor3, dashdotted, line width=1.0pt,mark size=2.3pt, mark=o, mark options={solid, mycolor3}]
  table[row sep=crcr]{%
50	4.05296499652377\\
100	4.05296499652377\\
200	4.05296499652377\\
300	4.05296499652377\\
400	4.05296499652377\\
600	4.05296499652377\\
800	4.05296499652377\\
};

\addplot [color=black, dashdotted, line width=1.0pt, mark=triangle,mark size=2.75pt, mark options={solid, rotate=270, black}]
  table[row sep=crcr]{%
50	4.17276146600506\\
100	4.17276146600506\\
200	4.17276146600506\\
300	4.17276146600506\\
400	4.17276146600506\\
600	4.17276146600506\\
800	4.17276146600506\\
};

\addplot [color=red, dashdotted, line width=1.0pt, mark=x, mark size=2.0pt,mark options={solid, red}]
  table[row sep=crcr]{%
50	4.26331747267177\\
100	4.26331747267177\\
200	4.26331747267177\\
300	4.26331747267177\\
400	4.26331747267177\\
600	4.26331747267177\\
800	4.26331747267177\\
};

\node at (axis cs: 300,3.8) [anchor = west] {\small Dashdotted lines: Limit $M\rightarrow \infty$};

\end{axis}
\end{tikzpicture}%
\caption{Sum \hspace{-.01in}average \hspace{-.01in}rate \hspace{-.01in}for \hspace{-.01in}$P_p\hspace{-.04in}=\hspace{-.04in}\frac{E_p}{\sqrt{M}} $ and $P_t\hspace{-.04in}=\hspace{-.04in}\frac{E_t}{\sqrt{M}}$ \hspace{-.03in}(Case II).}
\label{Fig4b}
\end{minipage}
\end{figure*}

The relationship between the number of antennas  $M^{\rm 1}$ and $M^{\rm mix}$ needed for the one-bit and mixed  architectures to perform as well as the conventional architecture with  $M^{\rm conv}$ antennas is illustrated in Fig. \ref{Fig5} and Fig. \ref{Fig5b} respectively. We numerically solve problem \textit{(P1)} for $\epsilon=10^{-3}$ using a simple search  to find $\kappa=\frac{M^{\rm 1}}{M^{\rm conv}}$ and $\tilde{\kappa}=\frac{M^{\rm mix}}{M^{\rm conv}}$ for different values of $M^{\rm conv}$. We can see that the ratios $\kappa$ and $\tilde{\kappa}$ are constant at $2.5$ and $1.57$ respectively for MRT precoding at all SNR values in accordance with Corollary \ref{Corspec1}. For ZF precoding,  $\kappa=2.5$  at low SNR values in accordance with Corollary \ref{Corspec2}, while  it increases to $3.91$ for $M^{\rm conv}=100$  as  $P_t$ increases to $20\rm {dB}$, because the interference becomes dominant and  conventional ZF precoder better  suppresses  interference than  one-bit quantized ZF precoder, and therefore the latter requires a higher number of antennas to get similar performance.  Similarly, $\tilde{\kappa}=1.57$  at low SNR values under ZF precoding in accordance with Corollary \ref{Corspec2}, while  it increases   as  $P_t$ increases.  We also see that  as $M^{\rm conv}$ increases to very large numbers, $\kappa$ and $\tilde{\kappa}$ eventually start to decrease and approach one for both precoders  as discussed in Remark 2, since  the effect of quantization, interference and noise  decreases with $M$ and the sum  rate of one-bit, mixed and conventional architectures approach the same limit dictated by normalized PC term. 

\begin{figure*}[!t]
\begin{minipage}[b]{0.47\linewidth}
\centering
\tikzset{every picture/.style={scale=.95}, every node/.style={scale=.8}}
%
%
\definecolor{mycolor1}{rgb}{0.85098,0.32549,0.09804}%
\definecolor{mycolor2}{rgb}{0.74902,0.00000,0.74902}%
\definecolor{mycolor3}{rgb}{0.60000,0.20000,0.00000}%
\definecolor{mycolor4}{rgb}{0.46667,0.67451,0.18824}%

\begin{tikzpicture}

\begin{axis}[%
width=.75\textwidth,
height=.5\textwidth,
scale only axis,
xmin=-30,
xmax=20,
xlabel style={font=\color{white!15!black}},
xlabel={Average transmit power $P_t$ (dB)},
ymin=1,
ymax=4.5,
ylabel style={ font=\color{white!15!black}},
ylabel={$\kappa$},
axis background/.style={fill=white},
title style={font=\bfseries},
title={},
xmajorgrids,
ymajorgrids,
legend style={at={(axis cs: -30,4.5)},anchor=north west,legend cell align=left,align=left,draw=white!15!black, /tikz/column 2/.style={
                column sep=5pt,
            }},]
\addplot [color=black, line width=1.9pt]
  table[row sep=crcr]{%
-30	2.47\\
-29	2.47\\
-28	2.47\\
-27	2.47\\
-26	2.47\\
-25	2.47\\
-24	2.47\\
-23	2.47\\
-22	2.47\\
-21	2.47\\
-20	2.47\\
-19	2.47\\
-18	2.47\\
-17	2.47\\
-16	2.47\\
-15	2.47\\
-14	2.47\\
-13	2.47\\
-12	2.47\\
-11	2.47\\
-10	2.47\\
-9	2.47\\
-8	2.47\\
-7	2.47\\
-6	2.47\\
-5	2.47\\
-4	2.47\\
-3	2.47\\
-2	2.47\\
-1	2.47\\
0	2.47\\
1	2.47\\
2	2.47\\
3	2.47\\
4	2.47\\
5	2.47\\
6	2.47\\
7	2.47\\
8	2.47\\
9	2.47\\
10	2.47\\
11	2.47\\
12	2.47\\
13	2.47\\
14	2.47\\
15	2.47\\
16	2.47\\
17	2.47\\
18	2.47\\
19	2.47\\
20	2.47\\
};
\addlegendentry{\footnotesize  $M^{\rm conv}=10^2$}

\addplot [color=red, line width=1.2pt]
  table[row sep=crcr]{%
-30	2.47\\
-29	2.47\\
-28	2.47\\
-27	2.47\\
-26	2.47\\
-25	2.47\\
-24	2.47\\
-23	2.47\\
-22	2.47\\
-21	2.47\\
-20	2.47\\
-19	2.47\\
-18	2.47\\
-17	2.47\\
-16	2.47\\
-15	2.47\\
-14	2.47\\
-13	2.47\\
-12	2.47\\
-11	2.47\\
-10	2.47\\
-9	2.47\\
-8	2.47\\
-7	2.47\\
-6	2.47\\
-5	2.47\\
-4	2.47\\
-3	2.47\\
-2	2.47\\
-1	2.47\\
0	2.47\\
1	2.47\\
2	2.47\\
3	2.47\\
4	2.47\\
5	2.47\\
6	2.47\\
7	2.47\\
8	2.47\\
9	2.47\\
10	2.47\\
11	2.47\\
12	2.47\\
13	2.47\\
14	2.47\\
15	2.47\\
16	2.47\\
17	2.47\\
18	2.47\\
19	2.47\\
20	2.47\\
};
\addlegendentry{\footnotesize $M^{\rm conv}=10^3$}

\addplot [color=mycolor4, line width=1.2pt]
  table[row sep=crcr]{%
-30	2.47\\
-29	2.47\\
-28	2.47\\
-27	2.47\\
-26	2.47\\
-25	2.47\\
-24	2.47\\
-23	2.47\\
-22	2.47\\
-21	2.47\\
-20	2.47\\
-19	2.47\\
-18	2.47\\
-17	2.47\\
-16	2.47\\
-15	2.47\\
-14	2.47\\
-13	2.46\\
-12	2.46\\
-11	2.46\\
-10	2.46\\
-9	2.46\\
-8	2.46\\
-7	2.46\\
-6	2.46\\
-5	2.45\\
-4	2.45\\
-3	2.45\\
-2	2.45\\
-1	2.45\\
0	2.45\\
1	2.45\\
2	2.45\\
3	2.45\\
4	2.45\\
5	2.45\\
6	2.45\\
7	2.45\\
8	2.45\\
9	2.45\\
10	2.45\\
11	2.45\\
12	2.45\\
13	2.45\\
14	2.45\\
15	2.45\\
16	2.45\\
17	2.45\\
18	2.45\\
19	2.45\\
20	2.45\\
};
\addlegendentry{\footnotesize \footnotesize $M^{\rm conv}=10^5$}

\addplot [color=mycolor2, line width=1.2pt]
  table[row sep=crcr]{%
-30	2.47\\
-29	2.47\\
-28	2.47\\
-27	2.47\\
-26	2.47\\
-25	2.47\\
-24	2.46\\
-23	2.46\\
-22	2.46\\
-21	2.46\\
-20	2.45\\
-19	2.45\\
-18	2.45\\
-17	2.44\\
-16	2.43\\
-15	2.42\\
-14	2.42\\
-13	2.41\\
-12	2.39\\
-11	2.38\\
-10	2.37\\
-9	2.36\\
-8	2.34\\
-7	2.33\\
-6	2.32\\
-5	2.31\\
-4	2.3\\
-3	2.29\\
-2	2.28\\
-1	2.27\\
0	2.27\\
1	2.26\\
2	2.26\\
3	2.25\\
4	2.25\\
5	2.25\\
6	2.25\\
7	2.25\\
8	2.25\\
9	2.24\\
10	2.24\\
11	2.24\\
12	2.24\\
13	2.24\\
14	2.24\\
15	2.24\\
16	2.24\\
17	2.24\\
18	2.24\\
19	2.24\\
20	2.24\\
};
\addlegendentry{\footnotesize  $M^{\rm conv}=10^6$}

\addplot [color=blue, line width=1.2pt]
  table[row sep=crcr]{%
-30	2.45\\
-29	2.45\\
-28	2.44\\
-27	2.44\\
-26	2.43\\
-25	2.42\\
-24	2.4\\
-23	2.38\\
-22	2.36\\
-21	2.34\\
-20	2.31\\
-19	2.27\\
-18	2.23\\
-17	2.18\\
-16	2.13\\
-15	2.07\\
-14	2.01\\
-13	1.94\\
-12	1.86\\
-11	1.79\\
-10	1.72\\
-9	1.65\\
-8	1.59\\
-7	1.53\\
-6	1.48\\
-5	1.44\\
-4	1.4\\
-3	1.37\\
-2	1.34\\
-1	1.32\\
0	1.3\\
1	1.28\\
2	1.27\\
3	1.26\\
4	1.25\\
5	1.24\\
6	1.24\\
7	1.23\\
8	1.23\\
9	1.23\\
10	1.23\\
11	1.22\\
12	1.22\\
13	1.22\\
14	1.22\\
15	1.22\\
16	1.22\\
17	1.22\\
18	1.22\\
19	1.22\\
20	1.22\\
};
\addlegendentry{\footnotesize $M^{\rm conv}=10^7$}

\addplot [color=black, dashdotted, line width=1.5pt]
  table[row sep=crcr]{%
-30	2.43\\
-29	2.44\\
-28	2.44\\
-27	2.44\\
-26	2.44\\
-25	2.45\\
-24	2.45\\
-23	2.46\\
-22	2.47\\
-21	2.48\\
-20	2.49\\
-19	2.5\\
-18	2.52\\
-17	2.54\\
-16	2.57\\
-15	2.6\\
-14	2.64\\
-13	2.68\\
-12	2.73\\
-11	2.79\\
-10	2.86\\
-9	2.93\\
-8	3\\
-7	3.08\\
-6	3.17\\
-5	3.25\\
-4	3.33\\
-3	3.41\\
-2	3.48\\
-1	3.55\\
0	3.61\\
1	3.66\\
2	3.71\\
3	3.75\\
4	3.78\\
5	3.8\\
6	3.83\\
7	3.84\\
8	3.86\\
9	3.87\\
10	3.88\\
11	3.89\\
12	3.89\\
13	3.9\\
14	3.9\\
15	3.91\\
16	3.91\\
17	3.91\\
18	3.91\\
19	3.91\\
20	3.91\\
};

\addplot [color=red, dashdotted, line width=1.2pt]
  table[row sep=crcr]{%
-30	2.47\\
-29	2.47\\
-28	2.47\\
-27	2.48\\
-26	2.48\\
-25	2.49\\
-24	2.49\\
-23	2.5\\
-22	2.5\\
-21	2.51\\
-20	2.53\\
-19	2.54\\
-18	2.56\\
-17	2.58\\
-16	2.6\\
-15	2.63\\
-14	2.67\\
-13	2.71\\
-12	2.76\\
-11	2.82\\
-10	2.88\\
-9	2.95\\
-8	3.03\\
-7	3.11\\
-6	3.2\\
-5	3.28\\
-4	3.37\\
-3	3.45\\
-2	3.53\\
-1	3.59\\
0	3.66\\
1	3.71\\
2	3.76\\
3	3.8\\
4	3.83\\
5	3.86\\
6	3.89\\
7	3.91\\
8	3.92\\
9	3.93\\
10	3.94\\
11	3.95\\
12	3.96\\
13	3.96\\
14	3.97\\
15	3.97\\
16	3.97\\
17	3.98\\
18	3.98\\
19	3.98\\
20	3.98\\
};

\addplot [color=mycolor4, dashdotted, line width=1.2pt]
  table[row sep=crcr]{%
-30	2.47\\
-29	2.47\\
-28	2.48\\
-27	2.48\\
-26	2.48\\
-25	2.48\\
-24	2.49\\
-23	2.49\\
-22	2.5\\
-21	2.51\\
-20	2.52\\
-19	2.53\\
-18	2.55\\
-17	2.57\\
-16	2.59\\
-15	2.62\\
-14	2.65\\
-13	2.69\\
-12	2.74\\
-11	2.79\\
-10	2.86\\
-9	2.92\\
-8	3\\
-7	3.08\\
-6	3.16\\
-5	3.24\\
-4	3.32\\
-3	3.4\\
-2	3.48\\
-1	3.54\\
0	3.6\\
1	3.66\\
2	3.7\\
3	3.74\\
4	3.77\\
5	3.8\\
6	3.82\\
7	3.84\\
8	3.85\\
9	3.87\\
10	3.88\\
11	3.88\\
12	3.89\\
13	3.9\\
14	3.9\\
15	3.9\\
16	3.91\\
17	3.91\\
18	3.91\\
19	3.91\\
20	3.91\\
};

\addplot [color=mycolor2, dashdotted, line width=1.2pt]
  table[row sep=crcr]{%
-30	2.47\\
-29	2.47\\
-28	2.48\\
-27	2.48\\
-26	2.48\\
-25	2.48\\
-24	2.48\\
-23	2.49\\
-22	2.49\\
-21	2.5\\
-20	2.5\\
-19	2.51\\
-18	2.52\\
-17	2.53\\
-16	2.55\\
-15	2.56\\
-14	2.59\\
-13	2.61\\
-12	2.64\\
-11	2.67\\
-10	2.7\\
-9	2.74\\
-8	2.79\\
-7	2.83\\
-6	2.87\\
-5	2.92\\
-4	2.96\\
-3	3\\
-2	3.03\\
-1	3.07\\
0	3.09\\
1	3.12\\
2	3.14\\
3	3.16\\
4	3.17\\
5	3.18\\
6	3.19\\
7	3.2\\
8	3.21\\
9	3.21\\
10	3.22\\
11	3.22\\
12	3.23\\
13	3.23\\
14	3.23\\
15	3.23\\
16	3.23\\
17	3.23\\
18	3.23\\
19	3.23\\
20	3.23\\
};

\addplot [color=blue, dashdotted, line width=1.2pt]
  table[row sep=crcr]{%
-30	2.46\\
-29	2.45\\
-28	2.45\\
-27	2.44\\
-26	2.43\\
-25	2.42\\
-24	2.41\\
-23	2.4\\
-22	2.38\\
-21	2.36\\
-20	2.33\\
-19	2.3\\
-18	2.27\\
-17	2.22\\
-16	2.17\\
-15	2.12\\
-14	2.06\\
-13	1.99\\
-12	1.92\\
-11	1.84\\
-10	1.77\\
-9	1.7\\
-8	1.63\\
-7	1.56\\
-6	1.5\\
-5	1.45\\
-4	1.41\\
-3	1.37\\
-2	1.33\\
-1	1.31\\
0	1.28\\
1	1.26\\
2	1.25\\
3	1.24\\
4	1.23\\
5	1.22\\
6	1.21\\
7	1.21\\
8	1.2\\
9	1.2\\
10	1.2\\
11	1.19\\
12	1.19\\
13	1.19\\
14	1.19\\
15	1.19\\
16	1.19\\
17	1.19\\
18	1.19\\
19	1.19\\
20	1.19\\
};

\node at (axis cs: 20,1.8) [anchor = east] {\small Solid  lines: MRT};
\node at (axis cs: 20,1.5) [anchor = east] {\small Dashdotted  lines: ZF};

\end{axis}
\end{tikzpicture}%
\caption{Ratio  $\kappa=\frac{M^{\rm 1}}{M^{\rm conv}}$ versus $P_t$ for $L=4$ and $K = 8$.}
\label{Fig5}
\end{minipage}
\hspace{.4cm}
\begin{minipage}[b]{0.47\linewidth}
\centering
\tikzset{every picture/.style={scale=.95}, every node/.style={scale=.8}}
%
%
\definecolor{mycolor1}{rgb}{0.85098,0.32549,0.09804}%
\definecolor{mycolor2}{rgb}{0.74902,0.00000,0.74902}%
\definecolor{mycolor3}{rgb}{0.60000,0.20000,0.00000}%
\definecolor{mycolor4}{rgb}{0.46667,0.67451,0.18824}%

\begin{tikzpicture}

\begin{axis}[%
width=.75\textwidth,
height=.5\textwidth,
scale only axis,
xmin=-30,
xmax=20,
xlabel style={font=\color{white!15!black}},
xlabel={Average transmit power $P_t$ (dB)},
ymin=1,
ymax=2.4,
ylabel style={font=\color{white!15!black}},
ylabel={$\tilde{\kappa}$},
axis background/.style={fill=white},
title style={font=\bfseries},
title={},
xmajorgrids,
ymajorgrids,
legend style={at={(axis cs: -30,2.4)},anchor=north west,legend cell align=left,align=left,draw=white!15!black, /tikz/column 2/.style={
                column sep=5pt,
            }},]

\addplot [color=black, line width=1.9pt]
  table[row sep=crcr]{%
-30	1.57\\
-29	1.57\\
-28	1.57\\
-27	1.57\\
-26	1.57\\
-25	1.57\\
-24	1.57\\
-23	1.57\\
-22	1.57\\
-21	1.57\\
-20	1.57\\
-19	1.57\\
-18	1.57\\
-17	1.57\\
-16	1.57\\
-15	1.57\\
-14	1.57\\
-13	1.57\\
-12	1.57\\
-11	1.57\\
-10	1.57\\
-9	1.57\\
-8	1.57\\
-7	1.57\\
-6	1.57\\
-5	1.57\\
-4	1.57\\
-3	1.57\\
-2	1.57\\
-1	1.57\\
0	1.57\\
1	1.57\\
2	1.57\\
3	1.57\\
4	1.57\\
5	1.57\\
6	1.57\\
7	1.57\\
8	1.57\\
9	1.57\\
10	1.57\\
11	1.57\\
12	1.57\\
13	1.57\\
14	1.57\\
15	1.57\\
16	1.57\\
17	1.57\\
18	1.57\\
19	1.57\\
20	1.57\\
};
\addlegendentry{\footnotesize  $M^{\rm conv}=10^2$}

\addplot [color=red, line width=1.2pt]
  table[row sep=crcr]{%
-30	1.57\\
-29	1.57\\
-28	1.57\\
-27	1.57\\
-26	1.57\\
-25	1.57\\
-24	1.57\\
-23	1.57\\
-22	1.57\\
-21	1.57\\
-20	1.57\\
-19	1.57\\
-18	1.57\\
-17	1.57\\
-16	1.57\\
-15	1.57\\
-14	1.57\\
-13	1.57\\
-12	1.57\\
-11	1.57\\
-10	1.57\\
-9	1.57\\
-8	1.57\\
-7	1.57\\
-6	1.57\\
-5	1.57\\
-4	1.57\\
-3	1.57\\
-2	1.57\\
-1	1.57\\
0	1.57\\
1	1.57\\
2	1.57\\
3	1.57\\
4	1.57\\
5	1.57\\
6	1.57\\
7	1.57\\
8	1.57\\
9	1.57\\
10	1.57\\
11	1.57\\
12	1.57\\
13	1.57\\
14	1.57\\
15	1.57\\
16	1.57\\
17	1.57\\
18	1.57\\
19	1.57\\
20	1.57\\
};
\addlegendentry{\footnotesize $M^{\rm conv}=10^3$}

\addplot [color=mycolor4, line width=1.2pt]
  table[row sep=crcr]{%
-30	1.57\\
-29	1.57\\
-28	1.57\\
-27	1.57\\
-26	1.57\\
-25	1.57\\
-24	1.57\\
-23	1.57\\
-22	1.57\\
-21	1.57\\
-20	1.57\\
-19	1.57\\
-18	1.57\\
-17	1.57\\
-16	1.57\\
-15	1.57\\
-14	1.57\\
-13	1.57\\
-12	1.57\\
-11	1.57\\
-10	1.57\\
-9	1.57\\
-8	1.57\\
-7	1.57\\
-6	1.56\\
-5	1.56\\
-4	1.56\\
-3	1.56\\
-2	1.56\\
-1	1.56\\
0	1.56\\
1	1.56\\
2	1.56\\
3	1.56\\
4	1.56\\
5	1.56\\
6	1.56\\
7	1.56\\
8	1.56\\
9	1.56\\
10	1.56\\
11	1.56\\
12	1.56\\
13	1.56\\
14	1.56\\
15	1.56\\
16	1.56\\
17	1.56\\
18	1.56\\
19	1.56\\
20	1.56\\
};
\addlegendentry{\footnotesize $M^{\rm conv}=10^5$}

\addplot [color=mycolor2, line width=1.2pt]
  table[row sep=crcr]{%
-30	1.57\\
-29	1.57\\
-28	1.57\\
-27	1.57\\
-26	1.57\\
-25	1.57\\
-24	1.57\\
-23	1.57\\
-22	1.57\\
-21	1.57\\
-20	1.56\\
-19	1.56\\
-18	1.56\\
-17	1.55\\
-16	1.55\\
-15	1.55\\
-14	1.54\\
-13	1.53\\
-12	1.53\\
-11	1.52\\
-10	1.51\\
-9	1.5\\
-8	1.49\\
-7	1.48\\
-6	1.48\\
-5	1.47\\
-4	1.46\\
-3	1.46\\
-2	1.45\\
-1	1.45\\
0	1.44\\
1	1.44\\
2	1.44\\
3	1.44\\
4	1.44\\
5	1.43\\
6	1.43\\
7	1.43\\
8	1.43\\
9	1.43\\
10	1.43\\
11	1.43\\
12	1.43\\
13	1.43\\
14	1.43\\
15	1.43\\
16	1.43\\
17	1.43\\
18	1.43\\
19	1.43\\
20	1.43\\
};
\addlegendentry{\footnotesize $M^{\rm conv}=10^6$}

\addplot [color=blue, line width=1.2pt]
  table[row sep=crcr]{%
-30	1.56\\
-29	1.56\\
-28	1.56\\
-27	1.55\\
-26	1.55\\
-25	1.54\\
-24	1.53\\
-23	1.52\\
-22	1.51\\
-21	1.49\\
-20	1.47\\
-19	1.45\\
-18	1.42\\
-17	1.39\\
-16	1.36\\
-15	1.32\\
-14	1.28\\
-13	1.23\\
-12	1.19\\
-11	1.14\\
-10	1.1\\
-9	1.06\\
-8	1.01\\
-7	1\\
-6	1\\
-5	1\\
-4	1\\
-3	1\\
-2	1\\
-1	1\\
0	1\\
1	1\\
2	1\\
3	1\\
4	1\\
5	1\\
6	1\\
7	1\\
8	1\\
9	1\\
10	1\\
11	1\\
12	1\\
13	1\\
14	1\\
15	1\\
16	1\\
17	1\\
18	1\\
19	1\\
20	1\\
};
\addlegendentry{\footnotesize $M^{\rm conv}=10^7$}

\addplot [color=black, dashdotted, line width=1.5pt]
  table[row sep=crcr]{%
-30	1.6\\
-29	1.6\\
-28	1.61\\
-27	1.61\\
-26	1.61\\
-25	1.61\\
-24	1.61\\
-23	1.61\\
-22	1.62\\
-21	1.62\\
-20	1.62\\
-19	1.63\\
-18	1.64\\
-17	1.65\\
-16	1.66\\
-15	1.67\\
-14	1.68\\
-13	1.7\\
-12	1.72\\
-11	1.74\\
-10	1.76\\
-9	1.79\\
-8	1.82\\
-7	1.85\\
-6	1.88\\
-5	1.91\\
-4	1.94\\
-3	1.97\\
-2	2\\
-1	2.03\\
0	2.05\\
1	2.07\\
2	2.09\\
3	2.1\\
4	2.11\\
5	2.12\\
6	2.13\\
7	2.14\\
8	2.14\\
9	2.15\\
10	2.15\\
11	2.15\\
12	2.16\\
13	2.16\\
14	2.16\\
15	2.16\\
16	2.16\\
17	2.16\\
18	2.16\\
19	2.16\\
20	2.16\\
};

\addplot [color=red, dashdotted, line width=1.2pt]
  table[row sep=crcr]{%
-30	1.58\\
-29	1.58\\
-28	1.58\\
-27	1.58\\
-26	1.58\\
-25	1.58\\
-24	1.58\\
-23	1.59\\
-22	1.59\\
-21	1.59\\
-20	1.6\\
-19	1.6\\
-18	1.61\\
-17	1.62\\
-16	1.63\\
-15	1.64\\
-14	1.65\\
-13	1.67\\
-12	1.69\\
-11	1.71\\
-10	1.74\\
-9	1.76\\
-8	1.79\\
-7	1.83\\
-6	1.86\\
-5	1.89\\
-4	1.92\\
-3	1.96\\
-2	1.99\\
-1	2.01\\
0	2.04\\
1	2.06\\
2	2.08\\
3	2.09\\
4	2.11\\
5	2.12\\
6	2.13\\
7	2.13\\
8	2.14\\
9	2.14\\
10	2.15\\
11	2.15\\
12	2.15\\
13	2.16\\
14	2.16\\
15	2.16\\
16	2.16\\
17	2.16\\
18	2.16\\
19	2.16\\
20	2.16\\
};

\addplot [color=mycolor4, dashdotted, line width=1.2pt]
  table[row sep=crcr]{%
-30	1.57\\
-29	1.57\\
-28	1.57\\
-27	1.58\\
-26	1.58\\
-25	1.58\\
-24	1.58\\
-23	1.58\\
-22	1.58\\
-21	1.59\\
-20	1.59\\
-19	1.6\\
-18	1.6\\
-17	1.61\\
-16	1.62\\
-15	1.63\\
-14	1.64\\
-13	1.66\\
-12	1.68\\
-11	1.7\\
-10	1.72\\
-9	1.75\\
-8	1.78\\
-7	1.81\\
-6	1.84\\
-5	1.87\\
-4	1.9\\
-3	1.93\\
-2	1.96\\
-1	1.98\\
0	2\\
1	2.02\\
2	2.04\\
3	2.06\\
4	2.07\\
5	2.08\\
6	2.09\\
7	2.09\\
8	2.1\\
9	2.1\\
10	2.11\\
11	2.11\\
12	2.11\\
13	2.12\\
14	2.12\\
15	2.12\\
16	2.12\\
17	2.12\\
18	2.12\\
19	2.12\\
20	2.12\\
};

\addplot [color=mycolor2, dashdotted, line width=1.2pt]
  table[row sep=crcr]{%
-30	1.57\\
-29	1.57\\
-28	1.57\\
-27	1.58\\
-26	1.58\\
-25	1.58\\
-24	1.58\\
-23	1.58\\
-22	1.58\\
-21	1.58\\
-20	1.58\\
-19	1.58\\
-18	1.59\\
-17	1.59\\
-16	1.59\\
-15	1.6\\
-14	1.6\\
-13	1.61\\
-12	1.61\\
-11	1.62\\
-10	1.63\\
-9	1.64\\
-8	1.65\\
-7	1.66\\
-6	1.67\\
-5	1.68\\
-4	1.69\\
-3	1.7\\
-2	1.71\\
-1	1.71\\
0	1.72\\
1	1.73\\
2	1.73\\
3	1.74\\
4	1.74\\
5	1.74\\
6	1.74\\
7	1.75\\
8	1.75\\
9	1.75\\
10	1.75\\
11	1.75\\
12	1.75\\
13	1.75\\
14	1.75\\
15	1.75\\
16	1.75\\
17	1.75\\
18	1.75\\
19	1.75\\
20	1.75\\
};

\addplot [color=blue, dashdotted, line width=1.2pt]
  table[row sep=crcr]{%
-30	1.56\\
-29	1.56\\
-28	1.56\\
-27	1.55\\
-26	1.55\\
-25	1.54\\
-24	1.53\\
-23	1.52\\
-22	1.51\\
-21	1.49\\
-20	1.48\\
-19	1.45\\
-18	1.43\\
-17	1.4\\
-16	1.36\\
-15	1.32\\
-14	1.27\\
-13	1.23\\
-12	1.17\\
-11	1.12\\
-10	1.07\\
-9	1.01\\
-8	1\\
-7	1\\
-6	1\\
-5	1\\
-4	1\\
-3	1\\
-2	1\\
-1	1\\
0	1\\
1	1\\
2	1\\
3	1\\
4	1\\
5	1\\
6	1\\
7	1\\
8	1\\
9	1\\
10	1\\
11	1\\
12	1\\
13	1\\
14	1\\
15	1\\
16	1\\
17	1\\
18	1\\
19	1\\
20	1\\
};
\node at (axis cs: 20,1.2) [anchor = east] {\small Solid  lines: MRT};
\node at (axis cs: 20,1.1) [anchor = east] {\small Dashdotted  lines: ZF};

\end{axis}
\end{tikzpicture}%
\caption{Ratio  $\tilde{\kappa}=\frac{M^{\rm mix}}{M^{\rm conv}}$  versus $P_t$ for $L=4$ and $K = 8$.}
\label{Fig5b}
\end{minipage}
\end{figure*}

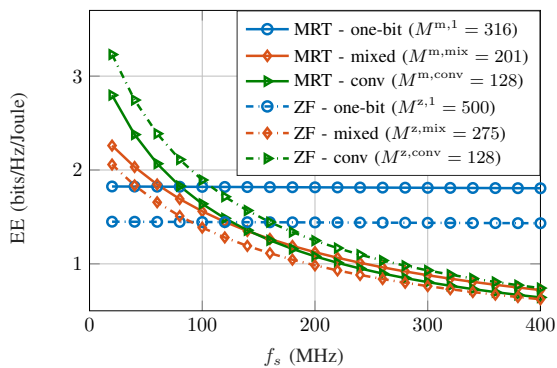
\begin{figure}[!t]
\centering
\tikzset{every picture/.style={scale=.95}, every node/.style={scale=.8}}
%
%
\definecolor{mycolor1}{rgb}{0.00000,0.44700,0.74100}%
\definecolor{mycolor2}{rgb}{0.85000,0.32500,0.09800}%
\definecolor{mycolor3}{rgb}{0.00000,0.49804,0.00000}%
\definecolor{mycolor4}{rgb}{0.00000,0.44706,0.74118}%
\begin{tikzpicture}

\begin{axis}[%
width=.75\columnwidth,
height=.5\columnwidth,
scale only axis,
xmin=0,
xmax=400,
xlabel style={font=\color{white!15!black}},
xlabel={$f_s$ (MHz)},
ymin=0.5,
ymax=3.7,
ylabel style={font=\color{white!15!black}},
ylabel={EE (bits/Hz/Joule)},
axis background/.style={fill=white},
xmajorgrids,
ymajorgrids,
legend style={at={(axis cs: 400,3.7)},anchor=north east,legend cell align=left,align=left,draw=white!15!black, /tikz/column 2/.style={
                column sep=5pt,
            }},]
\addplot [color=mycolor1, line width=1.0pt, mark=o,mark size=2.0pt,  mark options={solid, mycolor1}]
  table[row sep=crcr]{%
20	1.82446723881453\\
40	1.82345446905468\\
60	1.82244282305689\\
80	1.82143229895182\\
100	1.82042289487427\\
120	1.81941460896318\\
140	1.8184074393616\\
160	1.8174013842167\\
180	1.81639644167975\\
200	1.81539260990609\\
220	1.81438988705516\\
240	1.81338827129044\\
260	1.81238776077948\\
280	1.81138835369386\\
300	1.81039004820921\\
320	1.80939284250514\\
340	1.80839673476532\\
360	1.80740172317739\\
380	1.80640780593297\\
400	1.80541498122766\\
};
\addlegendentry{\small MRT - one-bit ($M^{\rm m,1}=316$)}

\addplot [color=mycolor2, line width=1.0pt, mark size=2.0pt, mark=diamond, mark options={solid, mycolor2}]
  table[row sep=crcr]{%
20	2.26019578338921\\
40	2.03211867917842\\
60	1.84585303843244\\
80	1.69086681903009\\
100	1.5598912522682\\
120	1.44774782951875\\
140	1.35064733315577\\
160	1.26575322137564\\
180	1.19089993459622\\
200	1.12440557268789\\
220	1.06494400896846\\
240	1.01145554383624\\
260	0.963083196070882\\
280	0.91912644077459\\
300	0.879007063198129\\
320	0.842243581318747\\
340	0.808431829230546\\
360	0.777230036891869\\
380	0.748347236599571\\
400	0.721534161820239\\
};
\addlegendentry{\small MRT - mixed ($M^{\rm m,mix}=201$)}

\addplot [color=mycolor3, line width=1.0pt, mark size=2.0pt, mark=triangle, mark options={solid, rotate=270, mycolor3}]
  table[row sep=crcr]{%
20	2.79714909226063\\
40	2.37737771591534\\
60	2.06715707924077\\
80	1.82855193773573\\
100	1.63932952359121\\
120	1.48559682808775\\
140	1.35822545057434\\
160	1.25097036271045\\
180	1.15941477213903\\
200	1.08034672587823\\
220	1.01137450952579\\
240	0.950680529011594\\
260	0.896858794636295\\
280	0.848804651229679\\
300	0.805638155363007\\
320	0.766649701289442\\
340	0.731260706513189\\
360	0.698994705571067\\
380	0.669455773129971\\
400	0.64231219619268\\
};
\addlegendentry{\small MRT - conv ($M^{\rm m,conv}=128$)}

\addplot [color=mycolor1, dashdotted, line width=1.0pt,mark size=2.0pt,  mark=o, mark options={solid, mycolor1}]
  table[row sep=crcr]{%
20	1.44936830071807\\
40	1.4484920651686\\
60	1.44761688845952\\
80	1.44674276867272\\
100	1.44586970389475\\
120	1.44499769221677\\
140	1.44412673173453\\
160	1.44325682054839\\
180	1.44238795676325\\
200	1.4415201384886\\
220	1.44065336383847\\
240	1.43978763093143\\
260	1.43892293789055\\
280	1.43805928284342\\
300	1.43719666392212\\
320	1.43633507926321\\
340	1.43547452700772\\
360	1.43461500530112\\
380	1.43375651229334\\
400	1.43289904613871\\
};
\addlegendentry{\small ZF - one-bit ($M^{\rm z,1}=500$)}

\addplot [color=mycolor2, dashdotted, line width=1.0pt, mark size=2.0pt, mark=diamond, mark options={solid, mycolor2}]
  table[row sep=crcr]{%
20	2.05818997723805\\
40	1.83562651946628\\
60	1.65650001719787\\
80	1.50922486690042\\
100	1.38599924371255\\
120	1.28137700764391\\
140	1.19144101141344\\
160	1.11330171425185\\
180	1.04478094963942\\
200	0.984205701831826\\
220	0.930269690598697\\
240	0.881938103638536\\
260	0.838380559696006\\
280	0.798923001829616\\
300	0.763012564648992\\
320	0.730191508986272\\
340	0.700077606344792\\
360	0.672349184901871\\
380	0.646733593873542\\
400	0.622998208054296\\
};
\addlegendentry{\small ZF - mixed ($M^{\rm z,mix}=275$)}

\addplot [color=mycolor3, dashdotted, line width=1.0pt, mark size=2.0pt, mark=triangle, mark options={solid, rotate=270, mycolor3}]
  table[row sep=crcr]{%
20	3.23005374146571\\
40	2.7453158672938\\
60	2.38708350458454\\
80	2.11155030823696\\
100	1.89304262537237\\
120	1.71551727655542\\
140	1.56843309157896\\
160	1.44457852165111\\
180	1.33885320343494\\
200	1.24754808160153\\
220	1.16790128476014\\
240	1.09781391637968\\
260	1.03566238681977\\
280	0.980171077424449\\
300	0.930323859102804\\
320	0.885301302992825\\
340	0.844435281478268\\
360	0.807175588259338\\
380	0.773065022070979\\
400	0.741720531895019\\
};
\addlegendentry{\small ZF - conv ($M^{\rm z,conv}=128$)}

\end{axis}
\end{tikzpicture}%
\caption{EE versus sampling frequency of ADCs/DACs for $L=4$, $K=8$, $P_t=10$\rm{dB}, and $c=494$fJ/step/Hz. }
\label{Fig5c}
\end{figure}

Next we study if we gain in terms of EE when we use one-bit ADCs and DACs instead of FR ADCs and DACs under the constraint of achieving the same sum average rate. The power consumption of each RF chain is computed as outlined after \eqref{EE} with parameters set as $P_{\rm TF}=14$mW, $P_{\rm LPF}=14$mW, $P_{\rm LNA}=39$mW, $P_{\rm LO}=5$mW and $P_{\rm M}=16.8$mW  \cite{addref3}. Moreover, $P_{\rm ADC}=P_{\rm DAC}=cf_s2^b$ where $c=494$fJ/step/Hz,  $b=1$ for one-bit ADCs/DACs, and $b=10$ for high resolution ADCs/DACs.  We find by solving  \textit{(P1)} that $M^{\rm m, 1}=316$ and $M^{\rm z, 1}=500$ antennas are needed at each BS of the one-bit system to achieve the same sum  rate as the conventional system that has $128$ antennas at each BS under MRT and ZF precoding respectively, while $M^{\rm m, mix}=201$ and $M^{\rm z, mix}=275$ antennas are needed at each BS of the mixed system to achieve the same sum  rate as the conventional system. With these values, we achieve $R_{\rm sum}^{\rm m, 1}=R_{\rm sum}^{\rm m, mix}=R_{\rm sum}^{\rm m, conv}$ under MRT and $R_{\rm sum}^{\rm z, 1}=R_{\rm sum}^{\rm z, mix}=R_{\rm sum}^{\rm z, conv}$ under ZF, and compute ${\rm EE}^{\rm m, 1}$, ${\rm EE}^{\rm m, mix}$, ${\rm EE}^{\rm m, conv}$, and ${\rm EE}^{\rm z, 1}$, ${\rm EE}^{\rm z, mix}$,  ${\rm EE}^{\rm z, conv}$ using the results in Sec. V-C.  The results are plotted against $f_s$ in Fig. \ref{Fig5c}. 

First we observe that in contrast to the conventional case where ${\rm EE}^{\rm z, conv}>{\rm EE}^{\rm m,conv}$ due to the larger sum  rate under ZF than MRT, the EE under one-bit quantized ZF is lower than that under one-bit quantized MRT  because it needs much larger numbers of antennas $M^{\rm z, 1}$ and $M^{\rm z, mix}$ to achieve $R_{\rm sum}^{\rm z, 1}=R_{\rm sum}^{\rm z, mix}=R_{\rm sum}^{\rm z, conv}$ than the numbers needed by MRT, therefore consuming more power. Overall we observe a significant decrease in the EE of the mixed and conventional systems with $f_s$ under both precoders, because the power consumption of FR ADCs (and FR DACs in case of conventional system) becomes quite dominant for large values of $b$ and $f_s$. The EE achieved by the one-bit MIMO system exceeds that achieved by the mixed system for $f_s > 62\rm{MHz}$ and $f_s>90\rm{MHz} $  under MRT and ZF precoding respectively, and that achieved by the conventional system for $f_s > 80\rm{MHz}$ and $f_s>160\rm{MHz} $ under MRT and ZF precoding respectively, thanks to the huge power savings the use of one-bit ADCs and DACs brings. On the other hand, the mixed architecture only achieves a larger EE than the conventional architecture when  $f_s > 150\rm{MHz}$ under MRT, while under ZF the conventional system has a better EE than the mixed setting for the range of sampling frequencies considered. The mixed architecture is not seen to yield large (or any) EE gains compared to conventional system because it still uses power hungry FR ADCs  while requiring a larger number of antennas than that utilized by the conventional system to overcome the impact of one-bit DACs and achieve the same sum rate. Overall we conclude that one-bit massive MIMO architecture employing one-bit ADCs and DACs is a potential energy efficient solution for mmWave  systems, that utilize larger bandwidths and higher sampling rates, even under  linear MRT and ZF precoders.

\section{Conclusion}
\label{Sec:Con}

In this work, we studied a multi-cell one-bit massive MIMO system employing one-bit ADCs and DACs at each BS, and derived closed-form expressions of the ergodic achievable downlink rates at the users under one-bit quantized MRT and ZF precoding schemes implemented using imperfect CSI. We also simplified these  results for the mixed (FR ADCs, one-bit DACs) and conventional (FR ADCs and DACs) architectures under both precoders, and considered cases with and without PC. The results revealed that the decrease in the SQINR due to the use of one-bit data converters  is more pronounced under ZF precoding than that under MRT precoding  especially at high SNR values. Interestingly for both precoders, the decrease in the SQINR with the introduction of PC was seen to be more dominant when the BSs employ FR ADCs and DACs as compared to when the BSs employ one-bit ADCs and DACs.  The developed expressions were utilized to study  the number of antennas needed by the one-bit, mixed and conventional architectures to achieve the same sum rate, and the EE gains of the one-bit  architecture over the mixed and conventional architectures at higher sampling frequencies. Interesting future research directions include studying the multi-cell one-bit massive MIMO system under M-ZF and M-MMSE precoders with arbitrary or optimized pilot re-use factors as well as under non-linear precoder designs.

\appendices

\section{Proof of Theorem \ref{Th_MRT}}
\label{App:Th_MRT}
Under MRT precoding in \eqref{MRT},  $\mathbf{w}^{\rm m}_{jk}=\hat{\mathbf{h}}_{jjk}$ and   $\mathbf{A}_j^{\rm m}=\sqrt{\frac{2}{\pi \bar{t}_{j}}} \mathbf{I}_M$  from Lemma \ref{LemmaMRT}, where $\bar{t}_{j}=\sum_{k=1}^K t_{jjk}$. Using these results, we compute  $\text{DS}_{jk}$, $\text{CU}_{jk}$, $\text{QN}_{jk}$, and $\text{IUI}_{jk}$  to express the SQINR in \eqref{SQINR} in a closed-form for large $(M,K)$ values.

\subsubsection{Computation of $\text{DS}_{jk}$} First using  $\eta_j=\frac{P_t}{M}$, we write $\text{DS}_{jk}=\eta_{j}|\mathbb{E}[\mathbf{h}_{jjk}^H \mathbf{A}^{\rm m}_j \mathbf{w}^{\rm m}_{jk}]|^2=\frac{P_t}{M} \frac{2}{\pi \bar{t}_{j}}|\mathbb{E}[\mathbf{h}_{jjk}^H  \hat{\mathbf{h}}_{jjk}]|^2\overset{(a)}{=}\frac{P_t}{M} \frac{2}{\pi \bar{t}_{j}}|\mathbb{E}[||\hat{\mathbf{h}}_{jjk}||^2]|^2\overset{(b)}{=} \frac{2 P_t M t_{jjk}^2}{\pi \bar{t}_{j}}$ where (a) follows by writing $\mathbf{h}_{jjk}$ as $ \hat{\mathbf{h}}_{jjk}+ \tilde{\mathbf{h}}_{jjk}$, where $ \hat{\mathbf{h}}_{jjk}$ and $ \tilde{\mathbf{h}}_{jjk}$ are uncorrelated, and (b) follows using  $\hat{\mathbf{h}}_{jjk}\sim \mathcal{CN}(\mathbf{0}, t_{jjk} \mathbf{I}_M)$ from  Lemma \ref{L1}.

\subsubsection{Computation of  $\text{CU}_{jk}$} We compute  $ \text{CU}_{jk}=\eta_j \times \\ \text{Var}[\mathbf{h}_{jjk}^H \mathbf{A}^{\rm m}_j \mathbf{w}^{\rm m}_{jk}]\hspace{-.04in}=\hspace{-.04in}\frac{P_t}{M} \frac{2}{\pi \bar{t}_{j}}(\mathbb{E}[|\mathbf{h}_{jjk}^H  \hat{\mathbf{h}}_{jjk}|^2]-(\mathbb{E}[\mathbf{h}_{jjk}^H  \hat{\mathbf{h}}_{jjk}])^2)$ as
\begin{align}
&\text{CU}_{jk}=\frac{P_t}{M} \frac{2}{\pi \bar{t}_{j}}\left(\mathbb{E}[||\hat{\mathbf{h}}_{jjk}||^4]+\mathbb{E}[|\tilde{\mathbf{h}}^H_{jjk}\hat{\mathbf{h}}_{jjk}|^2]-(M t_{jjk})^2 \right) \nonumber \\
\label{step}
&\hspace{-.04in}\overset{(a)}{=}\hspace{-.04in}\frac{2 P_t}{M \pi \bar{t}_{j}}\hspace{-.02in}\left(M(M\hspace{-.03in}+\hspace{-.03in}1)t_{jjk}^2\hspace{-.02in}+\hspace{-.02in}Mt_{jj}\tilde{t}_{jjk}\hspace{-.03in}-\hspace{-.02in}M^2t_{jjk}^2\right)\hspace{-.04in}=\hspace{-.03in}\frac{2 P_t \beta_{jjk} t_{jjk}}{\pi \bar{t}_{j}}. 
\end{align}
where (a) follows by applying \cite[Lemma 4]{append}  and  $\hat{\mathbf{h}}_{jjk}\sim \mathcal{CN}(\mathbf{0}, t_{jjk} \mathbf{I}_M)$  to compute $\mathbb{E}[||\hat{\mathbf{h}}_{jjk}||^4$. Also $\mathbb{E}[|\tilde{\mathbf{h}}^H_{jjk}\hat{\mathbf{h}}_{jjk}|^2]$ is computed as $\mathbb{E}[|\tilde{\mathbf{h}}^H_{jjk}\hat{\mathbf{h}}_{jjk}|^2]=\mathbb{E}[\tilde{\mathbf{h}}^H_{jjk}\hat{\mathbf{h}}_{jjk}\hat{\mathbf{h}}_{jjk}^H \tilde{\mathbf{h}}_{jjk}]=\mathbb{E}[\tilde{\mathbf{h}}^H_{jjk}t_{jjk}\mathbf{I}_M \tilde{\mathbf{h}}_{jjk}]=M \tilde{t}_{jjk} t_{jjk}$, where $\tilde{t}_{jjk}=\beta_{jjk}-t_{jjk}$ as defined after Lemma \ref{L1}.

\subsubsection{Computation of  $\text{QN}_{jk}$} Using $\mathbf{C}_{\mathbf{q}^{\rm m}_l \mathbf{q}^{\rm m}_l}=\left( 1-\frac{2}{\pi}\right)\mathbf{I}_M$ from  Lemma \ref{LemmaMRT}, we get $\text{QN}_{jk}=\sum_{l=1}^L \eta_l \mathbb{E}[\mathbf{h}_{ljk}^H \mathbf{C}_{\mathbf{q}^{\rm m}_l \mathbf{q}^{\rm m}_l} \mathbf{h}_{ljk}]=\sum_{l=1}^L \frac{P_t}{M} \left( 1-\frac{2}{\pi}\right) \mathbb{E}[||\mathbf{h}_{ljk} ||^2]=\sum_{l=1}^L P_t\left( 1-\frac{2}{\pi}\right) \beta_{ljk}$.

\subsubsection{Computation of  $\text{IUI}_{jk}$} The calculation of $\text{IUI}_{jk}=\sum_{(l,m)\neq (j,k)} \eta_l\mathbb{E}[|\mathbf{h}_{ljk}^H \mathbf{A}^{\rm m}_l \mathbf{w}^{\rm m}_{lm}|^2] $ is split into three cases:

\vspace{.05in}
(i) For $j=l, m \neq k$, we write 
\begin{align}
&\hspace{-.06in}\mathbb{E}[|\mathbf{h}_{jjk}^H \mathbf{A}^{\rm m}_j \mathbf{w}^{\rm m}_{jm}|^2]\hspace{-.04in}=\hspace{-.04in}\frac{2}{\pi \bar{t}_{j}} (\mathbb{E}[|\hat{\mathbf{h}}_{jjk}^H \hat{\mathbf{h}}_{jjm}|^2]+\mathbb{E}[|\tilde{\mathbf{h}}_{jjk}^H \hat{\mathbf{h}}_{jjm}|^2]) \nonumber \\
& \overset{(a)}{=}\frac{2}{\pi \bar{t}_{j}}(t_{jjk}t_{jjm}M+\tilde{t}_{jjk}t_{jjm}M) =\frac{2}{\pi \bar{t}_{j}} M \beta_{jjk}t_{jjm}
\end{align}
where (a) follows using similar steps outlined after \eqref{step}, and the distributions of $ \hat{\mathbf{h}}_{jjk}$ and  $ \tilde{\mathbf{h}}_{jjk}$ from Sec. III-B.

(ii) For  $j \neq l, m\neq k$, we write $
\mathbb{E}[|\mathbf{h}_{ljk}^H \mathbf{A}^{\rm m}_l \mathbf{w}^{\rm m}_{lm}|^2]=\frac{2}{\pi \bar{t}_{l}} (\mathbb{E}[|\hat{\mathbf{h}}_{ljk}^H \hat{\mathbf{h}}_{llm}|^2]+\mathbb{E}[|\tilde{\mathbf{h}}_{ljk}^H \hat{\mathbf{h}}_{llm}|^2]) \overset{(a)}{=}\frac{2}{\pi \bar{t}_{l}}(t_{ljk}t_{llm}M+(\beta_{ljk}-t_{ljk}) t_{llm}M) =\frac{2}{\pi \bar{t}_{l}} M \beta_{ljk}t_{llm}$, where (a) follows using similar steps outlined after \eqref{step} and considering  $\hat{\mathbf{h}}_{ljk}\sim \mathcal{CN}(\mathbf{0}, t_{ljk} \mathbf{I}_M)$ and  $\tilde{\mathbf{h}}_{ljk}\sim \mathcal{CN}(\mathbf{0}, (\beta_{ljk}-t_{ljk}) \mathbf{I}_M)$.

(iii) For $j \neq l, m=k$, we first make the following remark.
\begin{remark}
The channel estimates at BS $l$ with respect to user $k$ in cell $l$ and user $k$ in cell $j$ (if computed) will be correlated since the two users use the same pilot sequence. Mathematically, we can see that  $\hat{\mathbf{h}}_{ljk}=\frac{\beta_{ljk}}{\beta_{llk}}\hat{\mathbf{h}}_{llk}$, with the variances of their elements related as $\frac{{t}_{ljk}}{t_{llk}}=\frac{\beta^2_{ljk}}{\beta^2_{llk}}$. Thus we have
\end{remark}
\begin{align}
&\mathbb{E}[|\mathbf{h}_{ljk}^H \mathbf{A}^{\rm m}_l \mathbf{w}^{\rm m}_{lk}|^2]\hspace{-.03in}=\hspace{-.03in}\frac{2}{\pi \bar{t}_{l}} (\mathbb{E}[|\hat{\mathbf{h}}_{ljk}^H \hat{\mathbf{h}}_{llk}|^2]\hspace{-.03in}+\hspace{-.03in}\mathbb{E}[|\tilde{\mathbf{h}}_{ljk}^H \hat{\mathbf{h}}_{llk}|^2]) \nonumber \\
&\overset{(a)}{=} \frac{2}{\pi \bar{t}_{l}} \Big(\mathbb{E}\left[\Big|\frac{\beta_{ljk}}{\beta_{llk}}\hat{\mathbf{h}}_{llk}^H \hat{\mathbf{h}}_{llk}\Big|^2\right]+\mathbb{E}[|\tilde{\mathbf{h}}_{ljk}^H \hat{\mathbf{h}}_{llk}|^2]\Big) \nonumber \\ 
&\overset{(b)}{=}\frac{2}{\pi \bar{t}_{l}}\left(\frac{\beta_{ljk}^2}{\beta_{llk}^2}(M^2+M) t_{llk}^2+(\beta_{ljk}-t_{ljk}) t_{llk}M\right) \\ \label{PCeq}
&\overset{(c)}{=}\frac{2}{\pi \bar{t}_{l}}\left(\frac{\beta_{ljk}^2}{\beta_{llk}^2}M^2 t_{llk}^2+M\beta_{ljk}t_{llk}\right) 
\end{align}
where (a) follows using Remark 3,  (b) follows by utilizing  \cite[Lemma 4]{append} and similar steps listed after \eqref{step}, and (c) follows from  the result $\frac{{t}_{ljk}}{t_{llk}}=\frac{\beta^2_{ljk}}{\beta^2_{llk}}$ in Remark 3 and simplifying.  

\vspace{.02in}
Combining these three results  yields $\text{IUI}_{jk}=\sum_{l=1}^L \sum_{m\neq k}^K \frac{2P_t}{\pi \bar{t}_{l}} \beta_{ljk}t_{llm}+\sum_{l\neq j}^L \frac{2P_t}{\pi \bar{t}_{l}  } \beta_{ljk}t_{llk}+\sum_{l\neq j}^L \frac{2P_t M}{\pi \bar{t}_{l}  }\left(\frac{\beta_{ljk}t_{llk}}{\beta_{llk}}  \right)^2$, where the last term   captures the effect of PC. Combining the expressions of these four terms in the SQINR definition in \eqref{SQINR}, and simplifying by dividing all the terms in the denominator  with the expression of $\text{DS}_{jk}$ in the numerator yields \eqref{SQINR_MRT}, and completes the proof of Theorem \ref{Th_MRT}.

\section{Proof of Theorem \ref{Th_ZF}}
\label{App:Th_ZF}
For ZF precoding,  $\mathbf{A}_j^{\rm z}=\sqrt{\frac{2 K (c-1)^2}{\pi \zeta_{j}}} \mathbf{I}_M$ from  Lemma \ref{LemmaZF}.  Using this, we compute  $\text{DS}_{jk}$, $\text{CU}_{jk}$, $\text{QN}_{jk}$, and $\text{IUI}_{jk}$  to express the SQINR in \eqref{SQINR} in a closed-form as follows.

\subsubsection{Computation of $\text{DS}_{jk}$} First using   $\eta_j=\frac{P_t}{M}$ and $\mathbf{A}_j^{\rm z}=\sqrt{\frac{2 K (c-1)^2}{\pi \zeta_{j}}} \mathbf{I}_M$, we write $\text{DS}_{jk}=\eta_{j}|\mathbb{E}[\mathbf{h}_{jjk}^H \mathbf{A}^{\rm z}_j \mathbf{w}^{\rm z}_{jk}]|^2=\frac{P_t}{M}\frac{2 K (c-1)^2}{\pi \zeta_{j}}|\mathbb{E}[\mathbf{h}_{jjk}^H \mathbf{w}^{\rm z}_{jk}]|^2$. We then use the following result:
\begin{align}
\label{step5}
&\mathbf{H}_{jj}^H \mathbf{W}_j^{\rm z}=\mathbf{H}_{jj}^H \hat{\mathbf{H}}_{jj}( \hat{\mathbf{H}}^H_{jj} \hat{\mathbf{H}}_{jj})^{-1} \nonumber \\
&\hspace{.1in}=(\hat{\mathbf{H}}_{jj}+\widetilde{\mathbf{H}}_{jj})^H \hat{\mathbf{H}}_{jj}( \hat{\mathbf{H}}^H_{jj} \hat{\mathbf{H}}_{jj})^{-1}=\mathbf{I}_K+\tilde{\mathbf{H}}_{jj}^H \mathbf{W}_j^{\rm z},
\end{align}
where $\widetilde{\mathbf{H}}_{jj}={\mathbf{H}}_{jj}-\hat{\mathbf{H}}_{jj}$. Therefore $\mathbf{h}_{jjk}^H \mathbf{w}^{\rm z}_{jk}=1+\tilde{\mathbf{h}}^H_{jjk} \mathbf{w}^{\rm z}_{jk}$ and $\mathbb{E}[\mathbf{h}_{jjk}^H \mathbf{w}^{\rm z}_{jk}]=1+ \mathbb{E}[\tilde{\mathbf{h}}_{jjk}^H \mathbf{w}^{\rm z}_{jk}]=1$ since $\tilde{\mathbf{h}}_{jjk}$ and $\mathbf{w}^{\rm z}_{jk}$ are uncorrelated. Consequently we have $\text{DS}_{jk}=\frac{2 K P_t (c-1)^2}{\pi M \zeta_j}$.

\vspace{.04in}
\subsubsection{Computation of  $\text{CU}_{jk}$} To compute $ \text{CU}_{jk}=\eta_j\text{Var}[\mathbf{h}_{jjk}^H \mathbf{A}^{\rm z}_j \mathbf{w}^{\rm z}_{jk}]$, we utilize \eqref{step5} to write $\text{CU}_{jk}=\frac{P_t}{M} \frac{2 K (c-1)^2}{\pi \zeta_{j}}\text{Var}[\mathbf{h}_{jjk}^H  \mathbf{w}^{\rm z}_{jk}]=\frac{P_t}{M} \frac{2 K (c-1)^2}{\pi \zeta_{j}}\mathbb{E}[||\tilde{\mathbf{h}}_{jjk}^H  {\mathbf{w}}^{\rm z}_{jk}||^2]$. It then follows that
\begin{align}
& \text{CU}_{jk}\overset{(a)}{=}\frac{P_t}{M} \frac{2 K (c-1)^2}{\pi \zeta_{j}} (\beta_{jjk}-t_{jjk})\mathbb{E}[||{\mathbf{w}}^{\rm z}_{jk}||^2]\nonumber \\
&\overset{(b)}{=}\frac{P_t}{M} \frac{2 K (c-1)^2}{\pi \zeta_{j}} (\beta_{jjk}-t_{jjk})\mathbb{E}[[(\hat{\mathbf{H}}^H_{jj} \hat{\mathbf{H}}_{jj})^{-1}]_{k,k}], \nonumber \\
&\overset{(c)}{=}\frac{P_t}{M} \frac{2 K (c-1)^2}{\pi \zeta_{j}}\frac{\beta_{jjk}-t_{jjk}}{t_{jjk}(M-K)}
\end{align}
where  (a) follows by writing $\mathbb{E}[||\tilde{\mathbf{h}}_{jjk}^H  {\mathbf{w}}^{\rm z}_{jk}||^2]=\mathbb{E}[  {\mathbf{w}}^{\rm z^H}_{jk} \tilde{\mathbf{h}}_{jjk} \tilde{\mathbf{h}}_{jjk}^H  {\mathbf{w}}^{\rm z}_{jk}]=\mathbb{E}[  {\mathbf{w}}^{\rm z^H}_{jk}\tilde{t}_{jjk}\mathbf{I}_M  {\mathbf{w}}^{\rm z}_{jk}]$  where $\tilde{t}_{jjk}=\beta_{jjk}-t_{jjk}$ from Lemma \ref{L1}. Next (b) follows  by noting that $||{\mathbf{w}}^{\rm z}_{jk}||^2=[\mathbf{W}_j^{\rm z^H} \mathbf{W}^{\rm z}_j]_{k,k}=[(\hat{\mathbf{H}}^H_{jj} \hat{\mathbf{H}}_{jj})^{-1}]_{k,k}$ using \eqref{ZF}, and (c) follows by utilizing \cite[(29)]{lit12}.

\subsubsection{Computation of  $\text{QN}_{jk}$} The computation of $\text{QN}_{jk}$ is the same as done for the  MRT precoder  in Appendix  A and will yield the expression $\text{QN}_{jk}=\sum_{l=1}^L P_t\left( 1-\frac{2}{\pi}\right) \beta_{ljk}$.

\subsubsection{Computation of  $\text{IUI}_{jk}$} The calculation of $\text{IUI}_{jk}=\sum_{(l,m)\neq (j,k)} \eta_l\mathbb{E}[|\mathbf{h}_{ljk}^H \mathbf{A}^{\rm z}_l \mathbf{w}^{\rm z}_{lm}|^2] $ is divided into three cases: 

(i) For $j=l, m \neq k$, we write  $\mathbb{E}[|\mathbf{h}_{jjk}^H \mathbf{A}^{\rm z}_j \mathbf{w}^{\rm z}_{jm}|^2]\overset{(a)}{=}\frac{2 K (c-1)^2}{\pi \zeta_{j}} \mathbb{E}[|\tilde{\mathbf{h}}_{jjk}^H {\mathbf{w}}^{\rm z}_{jm}|^2]=\frac{2 K (c-1)^2}{\pi \zeta_{j}}(\beta_{jjk}-t_{jjk}) \mathbb{E}[||\mathbf{w}^{\rm z}_{jm}||^2]\overset{(b)}{=}\frac{2 K (c-1)^2}{\pi \zeta_{j}}\frac{\beta_{jjk}-t_{jjk}}{t_{jjm}(M-K)}$, where (a) and (b) follow using \eqref{step5} and  \cite[(29)]{lit12} respectively.

(ii) For  $j \neq l, m\neq k$, we write \vspace{-.05in} \small
\begin{align}
&\mathbb{E}[|\mathbf{h}_{ljk}^H \mathbf{A}^{\rm z}_l \mathbf{w}^{\rm z}_{lm}|^2]\overset{(a)}{=}\frac{2 K (c-1)^2}{\pi \zeta_{l}} \mathbb{E}[|\tilde{\mathbf{h}}_{ljk}^H {\mathbf{w}}^{\rm z}_{lm}|^2]\overset{(b)}{=}\frac{2 K (c-1)^2}{\pi \zeta_{l}}\nonumber \\
&\times (\beta_{ljk}-t_{ljk}) \mathbb{E}[||\mathbf{w}^{\rm z}_{lm}||^2]\overset{(c)}{=}\frac{2 K (c-1)^2}{\pi \zeta_{l}} \frac{\beta_{ljk}-t_{ljk}}{t_{llm}(M-K)}\nonumber \\ 
& \overset{(d)}{=}\frac{2 K (c-1)^2}{\pi \zeta_{l}}\left(\frac{\beta_{ljk}}{t_{llm}(M-K)} - \frac{\beta_{ljk}^2 t_{llk}}{\beta_{llk}^2 t_{llm}(M-K)}\right)
\end{align}\normalsize
where (a) follows using Remark 3 in Appendix A and \eqref{step5}, (b) and (c) follow using similar steps shown in the derivation of $\text{CU}_{jk}$, and (d) follows from  definition of $t_{ljk}$ in Remark 3.

(iii) For $j \neq l, m=k$, we utilize the observation from Remark 3 in Appendix A, and write $\mathbb{E}[|\mathbf{h}_{ljk}^H \mathbf{A}^{\rm z}_l \mathbf{w}^{\rm z}_{lk}|^2]$ \vspace{-.05in}
\begin{align}
&=\frac{2 K (c-1)^2}{\pi \zeta_{l}}\left(\mathbb{E}[|\hat{\mathbf{h}}_{ljk}^H \mathbf{w}_{lk}^{\rm z}|^2] +\mathbb{E}[|\tilde{\mathbf{h}}_{ljk}^H \mathbf{w}_{lk}^{\rm z}|^2] \right) \nonumber \\
&\overset{(a)}{=}\frac{2 K (c-1)^2}{\pi \zeta_{l}} \left(\mathbb{E}\left[\left|\frac{\beta_{ljk}}{\beta_{llk}}\hat{\mathbf{h}}_{llk}^H {\mathbf{w}}^{\rm z}_{lk}\right|^2\right]+\mathbb{E}[|\tilde{\mathbf{h}}_{ljk}^H {\mathbf{w}}^{\rm z}_{lk}|^2]\right)\nonumber \\
&\overset{(b)}{=}\frac{2 K (c-1)^2}{\pi \zeta_{l}}\left(\frac{\beta_{ljk}^2}{\beta_{llk}^2}+\frac{(\beta_{ljk}-t_{ljk})}{ t_{llk}(M-K)}\right) \\
&\overset{(c)}{=}\frac{2 K (c-1)^2}{\pi \zeta_{l}}\left(\frac{\beta_{ljk}^2}{\beta_{llk}^2}-\frac{\beta_{ljk}^2}{\beta_{llk}^2 (M-K)}+\frac{\beta_{ljk}}{t_{llk} (M-K)}\right)\nonumber
\end{align}
where (a) and (c) follows by using Remark 3, and (b) follows by utilizing \eqref{step5} to get $\hat{\mathbf{h}}_{llk}^H {\mathbf{w}}^{\rm z}_{lk}=1$ and $\mathbb{E}[||\tilde{\mathbf{h}}_{ljk}^H  {\mathbf{w}}^{\rm z}_{lk}||^2]=\tilde{t}_{ljk}\mathbb{E}[||{\mathbf{w}}^{\rm z}_{lk}||^2]=\frac{\tilde{t}_{ljk}}{t_{llk}(M-K)}$  where $\tilde{t}_{ljk}=\beta_{ljk}-t_{ljk}$.  Combining these three terms will yield an expression for $\text{IUI}_{jk}$. 

Combining the expressions of these four terms in the SQINR definition in \eqref{SQINR} and simplifying  by dividing all the terms in the denominator with the expression of $\text{DS}_{jk}$ in the numerator yields \eqref{SQINR_ZF}, and completes the proof of Theorem \ref{Th_ZF}.

\bibliographystyle{IEEEtran}
\bibliography{bib}

\begin{IEEEbiography}[{\includegraphics[width=1 in, height=2 in,clip,keepaspectratio ]{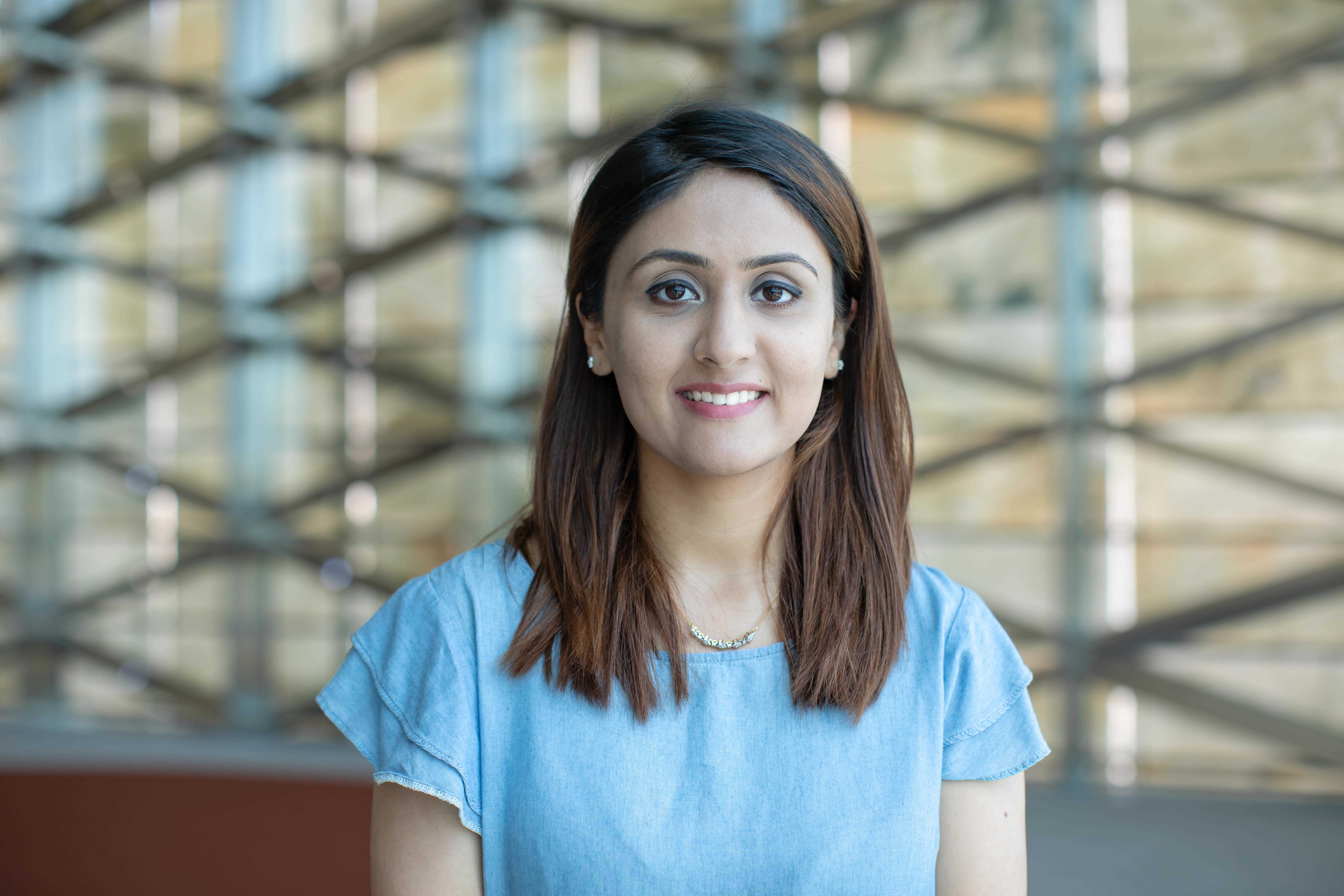}}]{Qurrat-Ul-Ain Nadeem}
(S'15, M'19)  received the B.S. degree in electrical engineering from Lahore University of Management Sciences (LUMS), Pakistan in 2013 and the M.S. and Ph.D. degrees in electrical engineering from King Abdullah University of Science and Technology (KAUST), Saudi Arabia in 2015 and 2018 respectively. She is currently a Post-Doctoral Research Fellow with the School of Engineering at the University of British Columbia, Canada. She received the Paul Baron Young Scholar Award by The Marconi Society in 2018, and the Postdoctoral Fellowship Award by the Natural Sciences and Engineering Research Council of Canada (NSERC) in 2021. Her research interests lie in the areas of communication theory, signal processing, and electromagnetics and antenna theory.
\end{IEEEbiography}

\begin{IEEEbiography}[{\includegraphics[width=1 in, height=2 in,clip,keepaspectratio ]{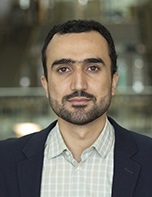}}]{Anas Chaaban} (S'09 - M'14 - SM'17) received the Ma{\^i}trise {\`e}s Sciences degree in electronics from Lebanese University, Lebanon, in 2006, the M.Sc. degree in communications technology and the Dr. Ing. (Ph.D.) degree in electrical engineering and information technology from the University of Ulm and the Ruhr-University of Bochum, Germany, in 2009 and 2013, respectively. From 2008 to 2009, he was with the Daimler AG Research Group On Machine Vision, Ulm, Germany. He was a Research Assistant with the Emmy-Noether Research Group on Wireless Networks located at the University of Ulm, Germany, from 2009 to 2011, and at the Ruhr-University of Bochum from 2011 to 2013. He was a Postdoctoral Researcher with the Ruhr-University of Bochum from 2013 to 2014, and with King Abdullah University of Science and Technology from 2015 to 2017. He joined the School of Engineering at the University of British Columbia as an Assistant Professor in 2018. His research interests are in the areas of information theory and wireless communications.

\end{IEEEbiography}

\end{document}